\documentclass[sigconf,screen]{acmart}

\usepackage{xcolor, soul}
\usepackage{xspace}
\usepackage{multirow}
\usepackage{tabularx}
\usepackage{tikz}
\usepackage[bottom]{footmisc}
\newcommand*\circled[1]{\tikz[baseline=(char.base)]{\node[shape=circle,fill,inner sep=0.5pt] (char) {\textcolor{white}{#1}};}}
\usepackage[most]{tcolorbox}
\usepackage{amsmath}
\usepackage{hyperref}

\usepackage{subcaption}
\usepackage{fancyhdr}
\usepackage{mathtools}

\definecolor{darkspringgreen}{rgb}{0.09, 0.45, 0.27}
\definecolor{denim}{rgb}{0.08, 0.38, 0.74}
\definecolor{darkolivegreen}{rgb}{0.33, 0.42, 0.18}
\definecolor{tangerine}{rgb}{0.95, 0.52, 0.0}
\definecolor{mahogany}{rgb}{0.75, 0.25, 0.0}
\definecolor{coolblack}{rgb}{0.0, 0.18, 0.39}

\definecolor{seagreen}{rgb}{0.18, 0.55, 0.34}

\newcommand{\atb}[1]{\textcolor{black}{#1}}

\definecolor{darkpink}{rgb}{0.88, 0.28, 0.54}
\definecolor{forestgreen}{rgb}{0.0, 0.27, 0.13}
\definecolor{amber}{rgb}{1.0, 0.49, 0.0}

\newcommand{\damlaa}[1]{{\color{black}#1}}

\newcommand{\sect}[1]{{Section~#1}\xspace} %
\newcommand{\head}[1]{{\noindent\textbf{#1.}\xspace}} %
\newcommand{\fig}[1]{{Figure~#1}\xspace} %

\newcommand\proposal{GenStore\xspace}
\newcommand\proposals{GenStore}
\newcommand\base{\textsf{Base}\xspace}
\newcommand\swf{\textsf{SW-filter}\xspace}
\newcommand\acc{\textsf{ACC}\xspace}
\newcommand\isf{\textsf{Ideal-ISF}\xspace}
\newcommand\iof{\textsf{Ideal-OSF}\xspace}
\newcommand\isfacc{\textsf{Ideal-ISF+ACC}\xspace}
\newcommand\ssdl{\texttt{SSD-L}\xspace}
\newcommand\ssdm{\texttt{SSD-M}\xspace}
\newcommand\ssdh{\texttt{SSD-H}\xspace}
\newcommand\dram{\texttt{\omciv{DRAM}}\xspace}

\newcommand\gs{\textsf{GS}\xspace}
\newcommand\gsos{\textsf{GS-Ext}\xspace}
\newcommand\simd{\textsf{SIMD}\xspace}

\newcolumntype{Y}{>{\centering\arraybackslash}X}
\usepackage{tikz}

\newcommand{\squishlist}{
 \begin{list}{$\circ$}
  { \setlength{\itemsep}{0pt}
     \setlength{\parsep}{0pt}
     \setlength{\topsep}{3pt}
     \setlength{\partopsep}{0pt}
     \setlength{\leftmargin}{1em}
     \setlength{\labelwidth}{1em}
     \setlength{\labelsep}{0.5em} } }

\newcommand{\squishend}{
  \end{list}  }

\DeclareRobustCommand\wcirc[1]{\tikz[baseline=(char.base)]{           \node[shape=circle,draw,inner sep=0pt,fill=white, text=black] (char) {#1};}}

\usepackage[colorinlistoftodos,prependcaption,textsize=scriptsize]{todonotes}
\usepackage{todonotes}

\usepackage{blindtext,graphicx}
\usepackage[absolute]{textpos}
\usepackage{datetime}

\newcommand\revmark[1]{\todo[linecolor=blue,backgroundcolor=blue!15,bordercolor=blue]{\textbf{#1}}}

\definecolor{seagreen}{rgb}{0.18, 0.55, 0.34}
\definecolor{ballblue}{rgb}{0.13, 0.67, 0.8}
\renewcommand{\revmark}[1]{}
\newcommand\omc[1]{{{#1}}}
\newcommand\omcc[1]{{{#1}}}
\newcommand\omccc[1]{{{#1}}}
\newcommand\omciv[1]{{{#1}}}
\newcommand\omcv[1]{{{#1}}}
\newcommand\omcvi[1]{{{#1}}}
\newcommand\omcvii[1]{{{#1}}}
\newcommand\omcviii[1]{{{#1}}}
\newcommand\omcix[1]{{{#1}}}
\newcommand\omcx[1]{{{#1}}}
\newcommand\jsr[1]{{{#1}}}
 \newcommand\hm[1]{{{#1}}}
 \newcommand\rev[1]{{{#1}}}
 \newcommand\revp[1]{{{#1}}}

 \newcommand\jsiv[1]{{{#1}}}
 \newcommand\jk[1]{{{#1}}}

\AtBeginDocument{%
  \providecommand\BibTeX{{%
    \normalfont B\kern-0.5em{\scshape i\kern-0.25em b}\kern-0.8em\TeX}}}

\setcopyright{acmlicensed}
\acmPrice{15.00}
\acmDOI{10.1145/3503222.3507702}
\acmYear{2022}
\copyrightyear{2022}
\acmSubmissionID{asplos22main-p39-p}
\acmISBN{978-1-4503-9205-1/22/02}
\acmConference[ASPLOS '22]{Proceedings of the 27th ACM International Conference on Architectural Support for Programming Languages and Operating Systems}{February 28 -- March 4, 2022}{Lausanne, Switzerland}
\acmBooktitle{Proceedings of the 27th ACM International Conference on Architectural Support for Programming Languages and Operating Systems (ASPLOS '22), February 28 -- March 4, 2022, Lausanne, Switzerland}

\begin{document}

\title[GenStore: A High-Performance and Energy-Efficient\\ In-Storage Computing System for Genome Sequence Analysis]{GenStore: A High-Performance and Energy-Efficient\\ In-Storage
Computing System for Genome Sequence Analysis\vspace{-0.2em}}

\newcommand{\tsc}[1]{\textsuperscript{#1}} 
\newcommand{\affilETH}{\tsc{1}}
\newcommand{\affilBNG}{\tsc{2}}
\newcommand{\affilKMU}{\tsc{3}}
\newcommand{\affilUofT}{\tsc{4}}

\author{%
 {%
     Nika Mansouri Ghiasi\affilETH \quad 
     Jisung Park\affilETH \quad 
     Harun Mustafa\affilETH \quad 
     Jeremie Kim\affilETH \quad
     Ataberk Olgun\affilETH
 }
}
\author{
 {%
     Arvid Gollwitzer\affilETH \quad
     Damla Senol Cali\affilBNG \quad 
     Can Firtina\affilETH \quad 
     Haiyu Mao\affilETH \quad 
     Nour Almadhoun Alserr\affilETH
 }
}
\author{
 {%
     Rachata Ausavarungnirun\affilKMU \quad
     Nandita Vijaykumar\affilUofT \quad
     Mohammed Alser\affilETH \quad
     Onur Mutlu\affilETH 
 }
}
\affiliation{
\institution{
      \vspace{8pt}
      \affilETH ETH Zürich\hspace{0.5em}
      \affilBNG Bionano Genomics\hspace{0.5em}
      \affilKMU KMUTNB\hspace{0.5em} 
      \affilUofT University of Toronto
  }
  \country{}
  }

\renewcommand{\shortauthors}{N. Mansouri Ghiasi, et al.}

\renewcommand{\authors}{Nika Mansouri Ghiasi, Jisung Park, Harun Mustafa, Jeremie Kim, Ataberk Olgun, Arvid Gollwitzer, Damla Senol Cali, Can Firtina, Haiyu Mao, Nour Almadhoun Alserr, Rachata Ausavarungnirun, Nandita Vijaykumar, Mohammed Alser, and Onur Mutlu}
\renewcommand{\title}{GenStore: A High-Performance In-Storage Processing System for Genome Sequence Analysis}

\begin{abstract}
\rev{Read mapping is a fundamental \omc{step} in many genomics applications.} \omccc{It is used} to identify potential matches and differences between \omccc{fragments (\omcvii{called}~\emph{reads}) of} a \omccc{sequenced} genome and an already known genome (called a \emph{reference genome}).
Read mapping is  costly because it needs to perform \emph{approximate string matching (ASM)} on large amounts
of data.
To address the computational challenges in genome analysis, many prior works  propose various approaches such as \omciv{accurate \emph{filters} that select the reads within \omccc{a dataset of genomic reads (called a \emph{read set})} that \emph{must} undergo expensive computation}, efficient  heuristics, and hardware acceleration. 
While effective at reducing  the amount of expensive computation, all such approaches still require the costly movement of \rev{a} large amount of data from storage to the rest of the system, which can significantly \omccc{lower} the end-to-end performance of read mapping \omc{in} conventional and emerging genomics systems. 

\omc{W}e propose \emph{\proposal}, the first in-storage processing system \omciv{designed} for genome sequence analysis %
that \omccc{greatly} reduces both data movement and computational overhead\hm{s} \rev{of \omccc{genome sequence analysis}} by exploiting low-cost and accurate \emph{in-storage filters}. 
\proposal leverages hardware/software co-design to address the challenges of in-storage processing\hm{, }%
supporting reads with  
\omciv{1)~different properties such as  read lengths and error rates, which highly depend on the sequencing technology, and 2)~different \omcvii{degrees of} genetic variation compared to the reference genome, which highly depends on the genomes that are being compared.}
Through rigorous analysis of read mapping processes \omciv{of reads with different properties and \omcvii{degrees of} genetic variation}, we meticulously design low-cost hardware accelerators and data/computation flows inside a \omc{\omcvii{NAND flash-based} solid-state drive (SSD)}. %
Our evaluation using a wide range of real genomic datasets shows that \proposal, \omciv{when implemented in three \omcvii{modern} NAND flash-based SSDs,} significantly improves the read mapping performance of  state-of-the-art software (hardware) baseline\omc{s} by \rev{2.07-6.05}$\times$ (\rev{1.52-3.32}$\times$) for \omccc{read sets with high similarity to the reference genome} and \rev{1.45-33.63}$\times$ (2.70-19.2$\times$) for \omccc{read sets with low similarity to the reference genome}.

\end{abstract}

\ccsdesc[500]{Computer systems organization~Special purpose systems}

\ccsdesc[500]{Hardware~External storage}

\keywords{Read Mapping, Filtering, Genomics, Storage, Near-Data Processing}

\settopmatter{printfolios=true, printacmref=true}
\maketitle

\pagenumbering{gobble}

\section{Introduction}
\label{sec:introduction}
Genome sequence analysis, which analyzes the DNA sequences of organisms, is important for many applications in personalized medicine~\cite{clark2019diagnosis,farnaes2018rapid,sweeney2021rapid,alkan2009personalized,flores2013p4,ginsburg2009genomic,chin2011cancer,Ashley2016}, outbreak tracing~\cite{bloom2021massively,yelagandula2021multiplexed,le2013selected,nikolayevskyy2016whole,qiu2015whole,gilchrist2015whole}, and evolutionary studies~\cite{hoban2016finding,romiguier2014comparative,ellegren2016determinants,prohaska2019human,ellegren2014genome,Prado-Martinez2013,Prohaska2019}. 
The information of an organism's DNA is converted to digital data via a process called \emph{sequencing}.
A sequencing machine extracts the sequences of DNA molecules from the organism's sample in the form of strings consisting of four \emph{base pairs (bps)}, denoted by A, C, G, and T.
No current sequencing technology has the capability to read a human DNA molecule in its entirety.
Instead, state-of-the-art sequencing machines generate randomly sampled, \omciv{inexact} sub-strings of the original genome, called \emph{reads}. 
The information about the corresponding location of each read in the complete genome is lost during sequencing in most technologies.
State-of-the-art sequencing machines produce one of two kinds of reads. 
1) Short read sequencing technologies, such as Illumina~\cite{reuter2015high,van2014ten}, 
produce reads that are \omciv{highly} accurate (99-99.9\%)\omciv{~\cite{glenn2011field,goodwin2016coming,quail2012tale}}, but short (e.g., up to a few hundred DNA \emph{base pairs}~\cite{glenn2011field,kchouk2017generations,pfeiffer2018systematic}).  
2) Long read sequencing technologies, such as Pacific Biosciences (PacBio)~\cite{amarasinghe2020opportunities} and Oxford Nanopore Technologies (ONT)~\cite{cali2017nanopore}, produce reads that are \omciv{less accurate (85-90\%\omcv{)}}~\cite{ardui2018single,kchouk2017generations,weirather2017comprehensive,van2018third}, but long (e.g., lengths ranging from thousands to millions of base pairs~\cite{wang2021nanopore}). 

Many genomics applications that involve the comparison of the genomic reads to a reference genome require a fundamental initial process, called \emph{read mapping}.
Read mapping identifies potential matching locations of reads against a reference genome~\cite{alser2020accelerating, alser2020technology} and is a \emph{very} computationally-costly \jsiv{process}~\cite{xin2013accelerating,xin2015shifted,turakhia2018darwin, cali2020genasm, alser2020sneakysnake, alser2017gatekeeper, nag2019gencache, kim2019airlift,kim2018grim,firtina2020apollo} due to two key challenges.
First, it uses \emph{\omcvii{computationally-}expensive algorithms} involving approximate string matching (ASM)~\cite{cali2020genasm,vsovsic2017edlib,alser2017gatekeeper,alser2017magnet,alser2019shouji,alser2020sneakysnake,kim2018grim,kim2019airlift,needleman1970general,smith1981identification,gotoh1982improved}.
A read is \emph{aligned} to a reference genome if the read is sufficiently similar to \omcvii{one or more subsequences in} that reference genome.
Reads generated by a sequencing machine might have differences compared to the reference genome due to either errors in the sequencing process or genetic variations~\cite{xin2013accelerating,levy2016advancements,hu2021next,10002015global}. 
ASM is widely used in existing read mappers to accurately account for such potential differences when determining the similarity between each read and the reference genome.
Second, read mapping performs \emph{large amounts of expensive ASM computation} because the genomic read datasets contain many reads (e.g., millions of reads),
\omciv{%
and each read requires ASM computation on multiple subsequences in the reference genome (see \sect{\ref{sec:background-readmapping}} for more detail\omcvii{s}).
}

Since read mapping is a key performance bottleneck in genome sequence analysis applications, there has been significant effort into improving read mapping performance via both algorithmic and system optimizations.
\omcv{Many prior works propose} efficient heuristics for ASM~\cite{zhang2000greedy,slater2005automated, li2018minimap2,myers1999fast,marco2021fast}, hardware accelerators\omciv{~\cite{alser2020accelerating, huangfu2018radar, cali2020genasm, turakhia2018darwin, fujiki2018genax, fujiki2020seedex, banerjee2018asap, khatamifard2021genvom, gupta2019rapid, li2021pim, angizi2019aligns, zokaee2018aligner, madhavan2014race, cheng2018bitmapper2, houtgast2018hardware,houtgast2017efficient, goyal2017ultra, chen2016spark, chen2014accelerating, chen2021high, zeni2020logan, ahmed2019gasal2, nishimura2017accelerating, de2016cudalign, liu2015gswabe, liu2013cudasw++, wilton2015arioc, fei2018fpgasw, waidyasooriya2015hardware, chen2015novel, rucci2018swifold, haghi2021fpga, li2021pipebsw, ham2020genesis, ham2021accelerating, wu2019fpga}}, and various \emph{filters} that try to efficiently and accurately 
\omcv{prune reads} that \omcvii{do not require} expensive computation~\cite{alser2020technology,kim2018grim,alser2020accelerating,cali2020genasm,singh2021fpga,nag2019gencache, kim20111, alser2017gatekeeper, alser2017magnet, alser2019shouji, alser2020sneakysnake, bingol2021gatekeeper, hameed2021alpha, guo2019hardware,xin2015shifted,xin2013accelerating}. 
For example, filters can be used to quickly %
prune reads that have exact matches in the reference genome. 
Pruned reads do not go through the expensive ASM process,  \omciv{which improves} read mapping performance and efficiency~\cite{alser2020accelerating,singh2021fpga,kim2018grim,alser2020technology,alser2017gatekeeper,alser2017magnet,alser2019shouji,alser2020sneakysnake,cali2020genasm,bingol2021gatekeeper,nag2019gencache}. 

While prior works improve read mapping performance, to our knowledge, \emph{none} of them consider the I/O cost that most systems must pay to read the large amount of data from the storage system to main memory and computation units.
Read mapping incurs unnecessary data movement from the storage system for large amounts of \emph{low-reuse} data. 
For example, while existing filters prune many reads to avoid expensive computation, they still need to first read the \emph{entire} read set from the storage system, even though a large fraction of the reads would be filtered out and  \emph{not} be reused \omcvii{in the} later \omcvii{stages of} the read mapping process.
The unnecessary data movement from the storage system can bottleneck read mapping performance in both conventional (software-based) and emerging (hardware-accelerated) genomics systems, while having a larger impact on emerging systems that greatly reduce the computation bottlenecks of ASM (e.g.,~\cite{alser2020accelerating,huangfu2018radar, cali2020genasm,fujiki2018genax,fujiki2020seedex, banerjee2018asap,turakhia2018darwin,nag2019gencache}). 

\emph{In-storage filtering} can be a fundamental solution for reducing the cost of the unnecessary data movement in read mapping.
Our motivational study using an ideal in-storage filter for  read mapping (Section~\ref{sec:motivation}) demonstrates that in-storage processing can greatly \omciv{accelerate} the end-to-end read mapping process.
This is because in-storage filtering not only avoids  unnecessary data movement from \omciv{storage}, but also \emph{eliminates} the computational burden of the filtering process from the rest of the system.

\textbf{Our goal} in this work is to improve the performance of genome sequence analysis by effectively reducing unnecessary data movement from \omciv{the} storage system.
To this end, we propose \emph{\proposal}, the first in-storage processing system \omciv{designed} for genome sequence analysis.
\textbf{The key idea} of \proposal is to exploit low-cost \emph{in-storage accelerators} to accurately \emph{filter out} the reads that do not require the expensive \omciv{ASM \jsiv{computation}} in read mapping, thereby significantly reducing unnecessary data movement from the storage system to main memory and processors. 

We identify two key challenges in designing an efficient in-storage system for read mapping.
First, read mapping workloads exhibit fundamentally different behavior due to \omciv{1)~}the varying read properties such as  read length and error rates, which highly depend on the sequencing technology, \omciv{and 2)} the genetic variation of reads compared to the reference genome, \omciv{which highly depends on the genomes that are being compared}.
Second, existing filtering methods incur a large number of random accesses to large datasets, which is challenging for a modern NAND flash-based solid-state drive (SSD)\footnote{
\omciv{In this work, we focus on SSDs based on NAND flash memory, the prevalent memory technology in modern storage systems.
We} expect that \proposal \omcv{ would also} provide performance \omcv{and energy} benefits with \omcv{storage devices that are built \omcvi{using}} emerging non-volatile memory technologies.} 
to cope with due to \omciv{its} poor random-access performance
and limited size of internal DRAM. 

We address these challenges with hardware/software co-design in three key directions.
First, based on our detailed analysis of read mapping, we design two different accelerators that can accelerate a wide range of read mapping applications for \omciv{reads with different properties (lengths and error rates) and genetic variations}. Each accelerator filters a large fraction of genomic read datasets using simple operations.
Second, we develop storage technology-aware algorithmic optimizations to replace expensive random accesses with more efficient sequential accesses to storage devices (e.g.,~NAND flash-based SSDs).
Third, we carefully design an efficient technique for data placement inside the storage device \omciv{that takes} full advantage of the high internal SSD bandwidth to concurrently access large amounts of genomic data.  

We design \proposal to support two in-storage filtering mechanisms in a single SSD: 1) \proposal-EM and 2) \proposal-NM.

\noindent\textbf{\proposal-EM} filters \emph{\underline{e}xactly-\underline{m}atching} reads, i.e., reads that exactly match subsequences of a reference genome.
Due to the low error rates of short reads, a large fraction of short reads map \emph{exactly} to the reference genome~\cite{10002015global,uk10k2015uk10k,nag2019gencache}. For example, on average ~80\% of human short reads map exactly to the human reference genome~\cite{10002015global,uk10k2015uk10k,nag2019gencache}.
However, finding exactly\omcvii{-}matching reads in the SSD is challenging\omcv{, as}
it incurs a number of random accesses per read to \omcv{a} large index structure \omcv{that stores unique subsequences of length $k$ (called \emph{k-mers}) and their positions in the reference genome.} 
Since each read consists of many k-mers, filtering each read requires several random accesses to the index.
\omciv{To avoid such random access\omcv{es}, we introduce} a new \emph{sorted, read-sized} k-mer index structure\omcv{, which}
enables \emph{sequentially} scanning of the read set and the new index, with only one index lookup per read during filtering. 

\noindent\textbf{\proposal-NM} filters \omciv{most of the} \emph{\underline{n}on-\underline{m}atching} reads, i.e., reads that \omciv{would not} align to any subsequence in the reference genome.
In read mapping, a significant fraction of reads might not align to the reference genome due to 1) the high sequencing error rate (in long reads) and/or 2) high genetic variation (in both short and long reads).
For example, both short and long read sets sequenced from rapidly-evolving viral samples (such as SARS-CoV-2) can have high genetic variations compared to the reference genome, leading to, on average, 36\% (up to 99.9\%) of reads not aligning \omciv{to the reference}~\cite{bhoyar2021high}.
To avoid expensive ASM operations for such
\omcv{non-matching} reads\jsiv{,} 
state-of-the-art read mappers commonly employ a step called \emph{chaining}, which calculates \omcvii{a} similarity score \omcvii{for each read} (called \emph{chaining score}) to the reference 
and filters out reads with a low score.
\omciv{\proposal-NM uses this basic idea of chaining to build an in-storage filter.}

\omciv{Calculating a chaining score for a read inside the SSD is challenging since it requires performing an expensive dynamic programming algorithm on the read's k-mers that exactly match the reference. This is particularly challenging for long reads since they have a large number of k-mers per read.
To avoid such expensive \omcv{computation},  we \emph{selectively} perform chaining only on reads with a small number of exactly-matching k-mers and send other reads to the host system for full read mapping (including chaining). 
\omcv{S}elective chaining is effective because 1) a read with many exactly-matching k-mers most likely aligns to the reference genome and \omcvii{thus} does not require in-storage filtering, and 2) selective chaining can filter many non-aligning reads, without requiring costly hardware resources in the SSD.}

To evaluate \proposal, we use a combination of synthesized Verilog models of our in-storage accelerators and state-of-the-art simulation tools that are widely used for DRAM and SSD research, Ramulator~\cite{kim2016ramulator} and MQSim~\cite{tavakkol2018mqsim}. 
To assess the performance impact of the storage system, we evaluate \omciv{three \jsiv{\proposal-enabled} systems} with \omciv{different}  SSD configurations (low-end, medium-end, and high-end).
We integrate \proposal into a state-of-the-art software read mapper (Minimap2~\cite{li2018minimap2}) and two state-of-the-art hardware read mappers (GenCache~\cite{nag2019gencache} for short reads and Darwin~\cite{turakhia2018darwin} for long reads). 
Our results show that \proposal-EM and \proposal-NM improve the performance of Minimap2 by 2.07-6.05$\times$ and 1.45-33.63$\times$, respectively, with \emph{no} accuracy loss.
\proposal-EM improves the performance of GenCache by 1.52-3.32$\times$, and \proposal-NM improves the performance Darwin by 2.70-19.2$\times$, with \emph{no} accuracy loss.

This work makes the following \textbf{key contributions}:
\vspace{-.2em}
\squishlist
    \item We introduce \proposal, the first in-storage processing system designed for genome sequence analysis. 
    \proposal fundamentally addresses the high I/O cost of reading low-reuse genomic data from storage systems. 
    \item We address the challenges of in-storage filtering for \omciv{genome sequence analysis} by analyzing the read mapping process and performing hardware/software co-design to develop in-storage filtering mechanisms and accelerators for \omciv{genomic} reads with various lengths, error rates, and genetic variations.
    \item We introduce two in-storage accelerators, 1)~\proposal-EM for filtering exactly-matching reads and 2)~\proposal-NM for filtering \omciv{\omcv{most reads that would not align to any \omcvi{subsequence} in the reference genome.} 
    \proposal filters out a large fraction of reads with  lightweight hardware accelerators and \emph{no} loss of accuracy, thereby improving the end-to-end performance and energy efficiency of genome sequence analysis.}
\squishend
\vspace{-0.3em}
\section{Background}
\label{sec:background}

\rev{We provide brief background on read mapping and NAND flash-based SSDs, necessary to understand the rest of the paper.}

\subsection{Read Mapping}
\label{sec:background-readmapping}

\revp{\head{End-to-End Workflow of Genome Sequence analysis}\revmark{CQ5}
There are three key initial steps in a standard \omc{genome} sequencing and analysis workflow\omc{~\cite{alser2020technology,alser2020accelerating}}. The first step is the collection, preparation, and sequencing of a 
DNA sample  
in the laboratory. 
Modern sequencing machines are unable to read an organism’s genome 
as a single complete sequence\omc{;} instead they generate shorter subsequences sampled randomly from the genome sequence\omc{~\cite{syed2009next,shendure2008next}}. The second step is \omciv{\emph{basecalling}}, which convert\omciv{s} the representation of 
the subsequences generated by the sequencing machine (e.g., images or electric current, \omcvii{depending} on the sequencing technology\omc{~\cite{cali2017nanopore,levy2016advancements}})
into \emph{reads}, which are sequences of \emph{nucleotides} (i.e., A, C, G, \omcvii{and} T in \omc{the} DNA alphabet).  
In order to reproduce the complete genome sequence from the shorter read sequences, \omcvii{the} third step\omcvii{,} called \omciv{\emph{read mapping,}} identif\omciv{ies} potential matching locations of each read with respect to a known reference genome (\omciv{e.g., }a representative genome sequence for a particular species)\omc{~\cite{li2018minimap2,alser2020technology,cali2020genasm,alkan2009personalized}}.
Genomic read sets can be obtained by, for example, 1) sequencing a DNA sample and storing the generated read set into the SSD of a sequencing machine\omc{~\cite{gamaarachchi2022fast,minion21}} or 2) downloading read sets from publicly available repositories\omc{~\cite{leinonen2010sequence}} 
and storing them into an SSD.

In this work, we focus on optimizing the performance of read mapping because sequencing and basecalling are performed only once per read set, \hm{whereas} read mapping can be performed \omciv{\emph{\omc{many} times}} for the same read set. \omc{This is  \omcv{common} in many genomic applications for two reasons. 
First, some applications require analyzing the genetic differences between a read set belonging to an individual and \omciv{\emph{many}} reference genomes of other individuals\omciv{. E}xamples \omciv{of such applications} include measuring the genetic diversity in a population~\cite{kostlbacher2021pangenomics, van2020emergence} and determining the donor of \omciv{a} sample by quantifying the reads that \omcvii{have a match in} each reference genome~\cite{lapierre2020metalign,meyer2021critical}.
Second, some other application\jsr{s} require repeating the read mapping step \omciv{many times to improve the outcome of read mapping. Examples of such applications are 1)~mapping with} new, more updated reference genomes~\cite{kim2019airlift}, or \omciv{2) }using different mapping parameter values (such as the maximum number of allowed differences between a read and a subseqeuence in the reference genome so that they are considered similar)~\cite{khayat2021hidden}.
}

Improving read mapping performance\revmark{E11} is critical since it is a fundamental step used in almost all genomic analyses that use sequencing data\omc{~\cite{alser2020accelerating,cali2020genasm,turakhia2018darwin,kim2018grim}}. The contribution of read mapping to the entire analysis pipeline varies depending on the application.
    \omc{For example}, read mapping takes up \omciv{to} \omc{1)}~45\% of the execution time when discovering sequence variants in cancer genomics studies~\cite{luo2014balsa}, and 2) 60\% of the \omciv{execution} time when profiling the species composition of a \omc{multi-species (i.e., \emph{metagenomic})} read set~\cite{lapierre2020metalign}.
}

\head{Read Mapping Process}
Since \omcc{the sequencing process does not provide location information for} both short and long reads \omcc{in most technologies},  \emph{read mapping} is a fundamental initial process for many genomics applications.  
\omciv{The read mapping process identifies subsequences in the reference genome to which the \omcvii{input} read\omcvii{s} match.}
\rev{For each \jsiv{\emph{matching location}, i.e., the location of each matching subsequence in the reference genome}, \omcc{a} read mapper compute\omc{s} an \emph{alignment score}, indicating the degree of similarity between the read and the region of the reference  to which the read aligns. \omcc{Matching base pairs between the read and the reference increase the alignment score, whereas \emph{edits} (i.e., base pair mismatches, insertions, or deletions relative to the reference) decrease this score.}}

Since \omcc{each} read \omcc{is} much shorter than the reference genome (e.g.,~the human reference genome contains $\sim$3.2~billion base pairs), a~read mapp\rev{er typically uses an index of the reference genome} to reduce the search space for each read. The index is a dictionary, i.e.,~a~key-value store, \revp{where the keys are unique $k$-length \omcc{subsequences} (called \emph{$k$-mers}) extracted from the reference genome, and the values are \jsiv{the} exactly-matching locations of these $k$-mers \omciv{in} the reference genome\omc{~\cite{xin2013accelerating,xin2016optimal}}. The value of $k$ is fixed during indexing and used for all subsequent steps.\footnote{\omc{$k$ is typically between 11 and 31~\cite{altschul1990basic,li2018minimap2,wood2019improved}, depending on the application.}} 
To greatly reduce the storage overhead of the index and speed up queries against it, \omc{without significantly changing the final outcome of read mapping}, some read mappers index only a subset of reference genome $k$-mers called \emph{minimizers}\omc{~\cite{schleimer2003winnowing,roberts2004reducing,marccais2017improving}}. 
\omc{A minimizer is a \omcv{representative} $k$-mer 
of a set of $k$-mers according to a scoring
mechanism. For example, some read mappers~\cite{li2016minimap, li2018minimap2} calculate hash values for all $k$-mers in a window of $w$ consecutive $k$-mers from an input sequence, and mark the $k$-mer with the smallest hash value as the minimizer $k$-mer.}}

Read mapping \omcc{is} a three-step process. In the first step (\emph{seeding}), \rev{the read mapper} queries the index structure to determine  \omciv{potential locations in the reference genome where the read could map. 
To do so, the read mapper looks up every minimizer k-mer fetched from a read in the reference index.
If the minimizer k-mer hits in the reference index, the read mapper marks the  locations
of such \omcv{a} k-mer in the reference genome as \omcv{the read's} \emph{potential matching locations}\omcv{,} also called \emph{seeds}.}
\rev{In} the second step (\emph{seed filtering} and/or \emph{chaining}), the read mapper \omciv{prunes those potential matching locations} in the reference to which the read \omcv{would not} align. \omcv{If all of the potential matching locations of the read get filtered, the read mapper discards the read from further analysis.}
\omcc{The read mapper uses a} dynamic programming (DP) algorithm  to 1) merge overlapping seeds into  longer \omcv{regions}, \rev{called} \emph{chains}~\cite{li2018minimap2}, and 2) calculate their corresponding \emph{chain\omcvii{ing} scores,}\omcc{ which refers to the approximation of the entire read's alignment score \omcv{\omcvi{in} these regions}}. 
\omcc{If the read mapper finds one or more chains with a sufficiently high chain\omcvii{ing} score (indicating a high degree of similarity to the reference genome), then the read mapper performs the third step. 
\omcv{If the read has no chain with a sufficiently high score, the read mapper prunes the read and skips the third step.}
} 
\rev{In the third step (\emph{sequence alignment})}, the read mapper \rev{determines the exact differences between \omciv{a} read and the reference genome \omciv{at the potential matching locations}}.
\rev{Sequence alignment} is done with a computationally\omcvii{-}expensive \omcc{\omcvii{DP}} algorithm\omc{~\cite{alser2020accelerating, alser2017gatekeeper, alser2017magnet, alser2019shouji, alser2020technology,cali2020genasm,kim2018grim,needleman1970general, smith1981identification, alser2020sneakysnake}} to perform approximate string matching (ASM). Finally, \omcc{the read mapper returns} the locations in the reference genome with the best alignment scores for each read.

\head{Pre-Alignment Read \omc{Filtering}} To mitigate the high performance overhead of alignment, read \omc{filtering} approaches are widely used. \omcc{\omcix{Read} filters can be incorporated at any stage of the process before alignment. }There are two \omcviii{\omcix{main} filter types}. The first \omcviii{filter} type\omc{~\cite{xin2013accelerating,xin2015shifted,alser2017gatekeeper, alser2017magnet, kim2018grim, alser2019shouji, bingol2021gatekeeper}} aims to efficiently filter  \omciv{potential matching} locations in the reference genome that lead to a large number of edits (larger than a user-defined threshold) between the read and the reference genome at \omciv{those} locations. \omcc{Doing so avoids} a costly alignment \omciv{step for potential locations at which the read \omcv{would not match the subsequences of the reference genome}.}
\omcc{T}he second \omcviii{filter} type~\cite{nag2019gencache} aims to detect if a read matches a \omcc{subsequence} of the reference genome with \emph{no} \omcc{edits} (i.e., exact-match) or \emph{very few} (e.g., 1-5) edits. \omcc{Reads \omciv{that} satisfy this requirement are guaranteed to} align to \omciv{at least one location in}  the reference genome without requiring the costly read alignment process\omciv{. This \omcviii{filter} type is} particularly effective for \omcc{read sets with a large number of exactly-matching reads (e.g., ~80\% in human short read sets~\cite{10002015global,uk10k2015uk10k,nag2019gencache})}.
\omcc{While \omcviii{both} filter \omcviii{types} reduce computation overhead, they still require a large number of random memory accesses for each read,} similar to the baseline \omcc{read mapper}.  In a typical read set of several gigabytes, \omcc{\omcviii{read} filters} incur several random accesses \omciv{per read} 1)~to the \omcc{reference} index \omcc{for seeding}, and \omcviii{potentially,} 2)~to the reference genome to compare the read with \omcc{the subsequence of reference genome at \omciv{each} \omciv{potential matching} location}. 

\subsection{SSD Organization}
\label{sec:background-SSD}

\begin{figure}[h]
         \centering
         \includegraphics[width=\columnwidth]{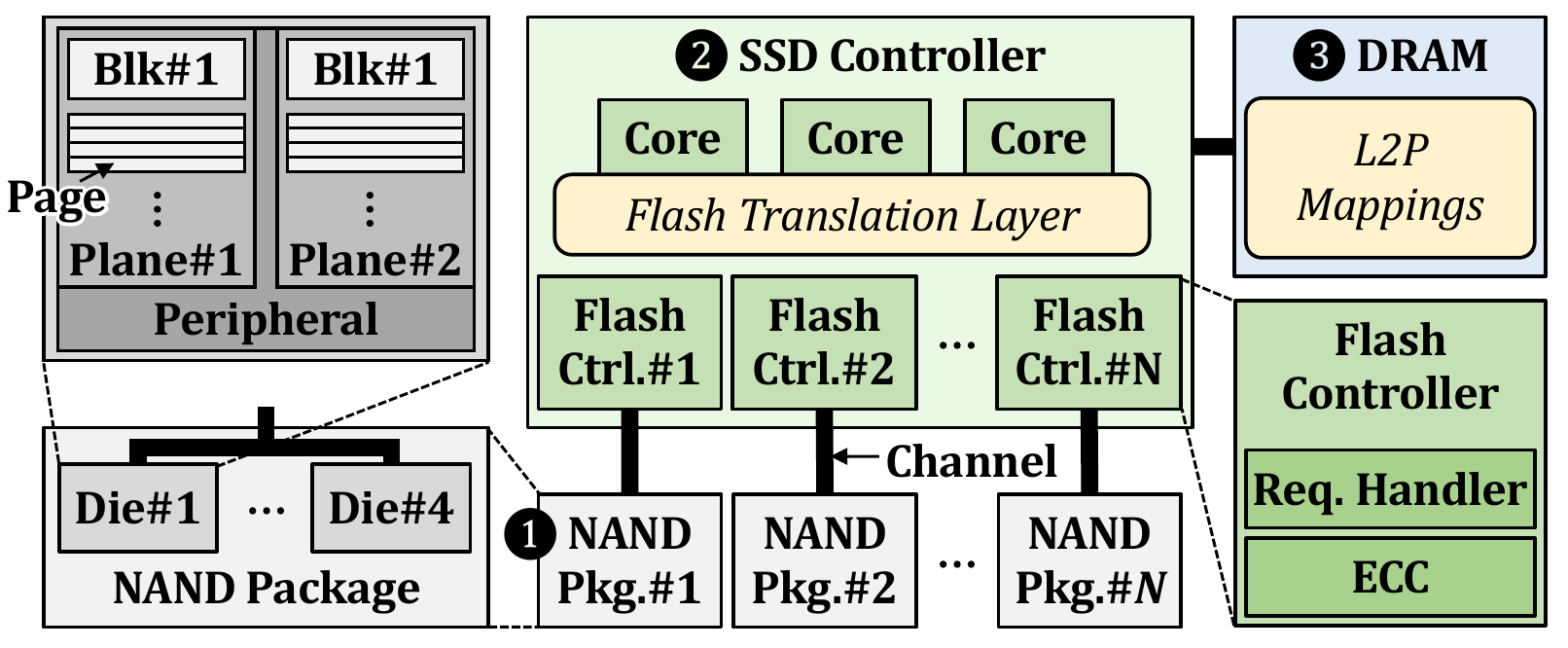}
         \caption{Organizational overview of \omc{a} modern SSD.}
         \label{fig:ssd}
\end{figure}

\fig{\ref{fig:ssd}} depicts the internal organization of a modern NAND flash-based solid-state drive (SSD) that consists of three main components: 
\circled{1}~NAND flash packages, \circled{2} an SSD controller, and \circled{3} DRAM.\linebreak
\head{\jsr{NAND Flash Memory}} 
A NAND package comprises multiple \emph{dies} \hm{(also} called \emph{chip\omcc{s}}) that share the package's \omciv{I/O pins}. 
One or more packages \omciv{share command/data busses (called \emph{channels})} to connect to the SSD controller.
Dies \omcvii{sharing the same channel} can operate independently of each other, \omcvii{but only one die can communicate with the SSD controller (e.g., for data transfer) at a time via the shared channel.}
A die has multiple (e.g., 2 or 4) \emph{planes}. \omcc{Each plane} contains thousands of \emph{blocks}.
A block includes hundreds \omcc{to} thousands of \emph{pages}, each of which is 4--16 KiB in size.
NAND flash memory performs read/write operations at page granularity but erase operations at block granularity.
Planes in the same die share \hm{the} peripheral circuitry used to \omcc{access} pages; \omciv{as such, they} can concurrently operate only \jsiv{when accessing pages (or blocks) at the same offset},
which \hm{are} called \emph{multi-plane} operations.

\head{\jsr{SSD Controller}}
An SSD controller has two main components: 
1)~multiple cores to run SSD firmware, commonly called \hm{the} \emph{flash translation layer (FTL)}, and 2) per-channel hardware flash controllers for request handling and error-correcting codes (ECC) for underlying NAND flash chips.
The FTL is responsible for communication with the host system, internal I/O scheduling, and various SSD management tasks required for hiding \hm{the} unique characteristics of NAND flash memory from the host system.
For example, a page of NAND flash memory needs to \hm{first} be erased before \hm{it is} programm\hm{ed},\footnote{\omc{This is called \omcvi{the }\emph{erase-before-write} property.}} so the FTL always performs \emph{out-of-place} updates by writing \omcvii{the} new data of a \emph{logical} page to a new \emph{physical} page that \omcc{was} erased previously. 
\omcvii{To this end,} the FTL maintain\omcvii{s} logical-to-physical (L2P) address mappings for reads and perform\omcvii{s} garbage collection to reclaim new physical pages for writes.

\head{\jsr{Internal DRAM}} A modern SSD employs large low-power DRAM (e.g., 4GB LPDDR4 \omcvii{DRAM} for a 4TB SSD~\cite{samsung860pro}) to store metadata for SSD management tasks.
Most of the DRAM capacity is used to store the L2P mappings for address translation. 
It is common practice to maintain the L2P mappings at 4KiB granularity to provide high random \omciv{access} I/O performance~\cite{park-nvmsa-2018, kim-dac-2017}, so in a 32-bit architecture, the memory overhead for the L2P mappings is \omcviii{approximately} 0.1\% of the SSD capacity (4 bytes per 4KiB data).

\head{\jsr{SSD I/O Bandwidth}}\revmark{C2}
\revp{To mitigate the \omc{large} performance gap between main memory and \omc{the} storage system, SSD manufacturers \omc{ increase the external bandwidth of SSDs by} employing advanced I/O interfaces between the host system and SSDs.
    For example, \omc{while} \omciv{older} SATA3 SSDs provide around 500MB/s sequential-read bandwidth~\cite{inteldcs4500, samsung860pro}, 
    state-of-the-art PCIe-Gen4 SSDs can provide significantly higher sequential-read bandwidth, \jsr{up} to \jsr{8} GB/s (e.g., 7 GB/s in Samsung PM1735~\cite{samsungPM1735}).}

    \revp{\omcvii{A modern SSD's} \emph{internal} bandwidth (i.e., I/O bandwidth between NAND flash chips and SSD controller) is \jsr{usually} higher than \omc{its} external bandwidth (i.e., I/O bandwidth between the host and the SSD). For example, a recent enterprise SSD controller~\cite{anandcontroller} supports 6,550MB/s external bandwidth and \omc{19.2GB/s  internal bandwidth (16 channels\omcviii{, each} with a bandwidth of 1.2 GB/s)}. Over-provisioning the internal bandwidth is  reasonable  since 1) \jsr{a modern} SSD needs to perform various internal management tasks (e.g., garbage collection~\cite{tavakkol2018flin,kim2020evanesco,cai2017error,park-dac-2019, park-dac-2016} and wear-leveling~\cite{chang2007efficient,cai2017error}), and 2) a higher channel count \omcv{reduces} contention between requests \omciv{by interleaving data between the channels}.}

\section{Motivation\omc{al Studies}}
\label{sec:motivation}

\omciv{W}e perform experimental studies to understand the potential of efficient in-storage \omcc{accelerators} for improving the performance of \omcc{genome sequence analysis} applications.

\subsection{Methodology} 

\omcc{\head{Read Mappers}} We evaluate five \omciv{read mapping} systems, each of which adopts different optimization techniques to accelerate read mapping:
\omc{1)~}\base uses Minimap2~\cite{li2018minimap2}, a state-of-the-art software tool for read mapping\omcvii{.} 
\omc{2)~}\swf extends Minimap2 to \emph{filter out exact\omc{ly}-matching reads} \omc{ (i.e., reads that  exactly \omcvii{match} \omciv{subsequences in} one or more locations in the reference genome) using simple single-instruction-multiple-data (SIMD) operations, without requiring costly ASM operations}\omcvii{.}
\omc{3)~}\isf uses an ideal In-Storage Filter \omc{(\emph{ISF})} that can \emph{concurrently} filter out exact-matching reads inside the SSD while \omcvii{the} host CPU performs read mapping for non-filtered reads\omcvii{.}
\omc{4)~}\acc uses a state-of-the-art hardware accelerator for short read mapping, GenCache~\cite{nag2019gencache}\omcvii{.}
\omc{5)~}\isfacc use\omcvii{s} an ideal in-storage filter \omc{(ISF)} that can \emph{concurrently} filter out exact\omc{ly}-matching reads inside the SSD \omc{while a hardware accelerator (ACC)} performs read mapping for non-filtered reads. 

\omcc{\head{System Configuration}} To assess the impact of the storage subsystem on end-to-end application performance, we evaluate each \omciv{of the five} systems with four different configurations:
1)~a low-end SSD (\ssdl)~\cite{inteldcs4500} with a SATA3 interface\omciv{~\cite{SATA}}, 
2)~a mid-end SSD (\ssdm)~\cite{samsung980pro} using a PCIe Gen3 M.2 interface\omciv{~\cite{PCIE}}, 
3)~a high-end SSD (\ssdh)~\cite{samsungPM1735} with a PCIe Gen4 interface\omciv{~\cite{PCIE4}}, and
4)~a system where \emph{all} of the processed data is pre-loaded to DRAM \omc{with no performance cost for pre-loading}  (\dram), as the \emph{idealized} case where storage I/O overheads are completely eliminated (we do not evaluate \dram for \isf and \isfacc since using in-storage processing is contradictory to pre-loading \omc{all} the data to main memory).
\revp{We assume\revmark{A1\\ part 1} 8 channels for \ssdl and 16 channels for \ssdm and \ssdh, where the \omciv{maximum} bandwidth per channel is 1.2~GB/s. \omciv{The maximum internal bandwidth is calculated by 1.2~GB/s $\times$ channel count.} 
The external bandwidth of \ssdl, \ssdm, and \ssdh for sequential read\omc{s} is 500~MB/s, 3.5~GB/s, and 7~GB/s, respectively. 
\omcvii{Hence, t}he internal bandwidth of \ssdl, \ssdm, \omcvii{and} \ssdh~\omcvii{for sequential reads} is 19.2$\times$, 5.48$\times$, \omcvii{and} 2.74$\times$ \omciv{that of its external} bandwidth, respectively.}

We evaluate \base and \swf by running Minimap2 on a high-end server (AMD EPYC 7742 CPU~\cite{amdepyc} with \omc{1TB} DDR4 DRAM). \omciv{We simulate} the performance of the other \omciv{three} systems using our simulation environment that faithfully models system components including DRAM and storage devices (see \sect{\ref{sec:methodology}}).
We map all reads of a short read dataset against the human reference genome, where 80\% of the reads have one or more \rev{exactly-}matching subsequences in the reference genome~\cite{10002015global,uk10k2015uk10k}.

\omcc{\head{Key Features of an Ideal In-Storage Filter}} \revp{We assume\revmark{A1\\ part 2 and\\B1} two key features for \isf and \isfacc. First, I/O overheads \omc{due to limited external SSD bandwidth} are \emph{completely} eliminated for filtered reads. Second, the system provides high in-storage filtering performance such that  
the filtering process can concurrently run in \omc{the} SSD, and the latency of this filtering process is \emph{fully hidden} by the read mapping of unfiltered reads in \omciv{the} host CPU (\isf) or hardware accelerator (\isfacc).
We assume that the accelerator or the CPU streams through the input reads in batches and analyzes a batch \emph{\omciv{concurrently}} with reading the next batch. \omciv{Thus}, the execution time of \textsf{Ideal-ISF} (+ACC) can be modeled as follows:
\begin{equation}
\label{eq:ideal-isf}
T_\text{\textsf{Ideal-ISF}} = T_\text{I/O-Ref} + \max\left\lbrace T_\text{I/O-Unfiltered}, T_\text{RM-Unfiltered}\right\rbrace,
\end{equation}

\noindent
where $T_\text{I/O-Ref}$, $T_\text{I/O-Unfiltered}$, and $T_\text{RM-Unfiltered}$ are the latency \omciv{of} \omcc{reading} the reference genome \omcc{from the SSD}  \omciv{into main memory}, the latency of \omcc{reading} the unfiltered \omcc{genomic} reads from the SSD, and the latency of read mapping of the unfiltered reads, respectively. 
\omciv{For a given input size,} $T_\text{RM-Unfiltered}$ varies depending on the computation unit used for read mapping (i.e., the host CPU or \omciv{accelerator}), while the I/O-latency values \omc{only depend on the SSD configuration}.}

\subsection{Results \& Analysis}
\begin{figure}[t]
    \centering
    \includegraphics[width=.95\linewidth]{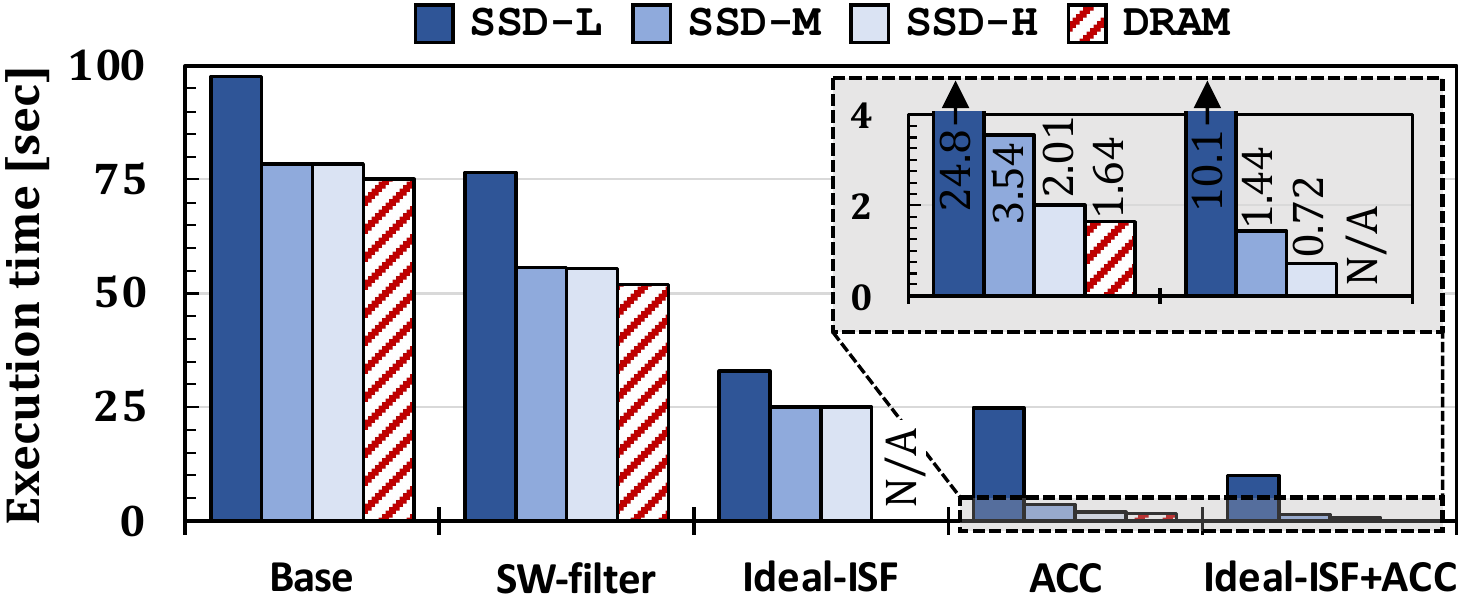}
    \vspace{-1em}
    \caption{Execution time of read mapping \omciv{with four} different \omciv{storage} configurations.}
    \vspace{-1.5em}
    \label{fig:motivation1}
\end{figure}

\fig{\ref{fig:motivation1}} shows the execution time of  read mapping in the five \omciv{evaluated} systems, \omciv{each} with \omciv{four} different storage \omciv{sub}system configurations. \omc{W}e make four key observations.

\head{Observation 1} The ideal in-storage filter  provides significant performance improvement\omcvii{s} over other systems.
\omcv{\isf  significantly outperforms \base and \swf (\omcvii{by} 3.12$\times$ and 2.21$\times$, respectively), and 
\isfacc provides a large speedup (2.78$\times$)
\omcv{over} \acc, when they all use \ssdh.}
\omcvi{These large improvements are} due to two key benefits \omcvi{provided by the ideal in-storage filter}: \omcc{1)} mitigati\omcvi{on of} data movement from the storage devices and \omcc{2)} remov\omcvi{al of} the burden of filtering out 80\% of the \omciv{input read set} from the rest of the system\omcc{, including processors and main memory}. 
\revmark{A1\\ Part 5}  \revp{To distinguish the effects of these two benefits, we analyze an ideal \emph{outside-storage} filter (\iof) \omcc{that provides only the second benefit;} \omciv{this filter} concurrently runs with the read mapper and fully overlaps the filtering process with the read mapping process of unfiltered \omcvii{reads}. The execution time of \iof(+ACC) can be formulated as follows: 
\begin{equation} 
\label{eq:ideal-osf}
T_\text{\iof} = T_\text{I/O-Ref} + \max\left\lbrace T_\text{I/O-All-Reads},  T_\text{RM-Unfiltered}\right\rbrace,
\end{equation}

\noindent
where $T_\text{I/O-All-Reads}$ is the latency for reading all \omcc{genomic} reads from the SSD \omciv{into main memory}.
\omciv{Using} \ssdh, \iof leads to an execution time of 1.15 seconds, which is 60\% slower than the \isfacc. This is because  \omcc{$T_\text{I/O-All-Reads}$} is significantly larger than \omciv{both} $T_\text{RM-Unfiltered}$ and \omcc{$T_\text{I/O-Unfiltered}$ (in Equation~\eqref{eq:ideal-isf})}.
}

\omcc{The remaining observations dive deeper into the effects of \omcvii{the} I/O bottleneck on  each read mapping \omciv{system}.}

\head{Observation 2} In \base and \swf, using high-end SSDs significantly improves read mapping performance over low-end SSDs, effectively \omciv{reducing} the storage \omciv{performance} bottleneck that exists in low-end SSDs. For example, using \ssdh instead of \ssdl reduces the execution time of \base and \swf by 24\% and 38\%, respectively, showing comparable performance to \dram, where all the data is \omcvii{pre-loaded to} main memory (i.e., \emph{no} I/O accesses). \omcc{This is because\omcvii{,} by using \ssdm and \ssdh,  the performance bottleneck of the application shift\omcvii{s} to parts of the system other than I/O \omciv{(e.g., CPU or main memory)}.}
This observation shows that I/O  has a significant impact on application performance but \omciv{this impact} can be \omcc{alleviated} at the cost of expensive storage devices and interfaces. 
Note that, while \ssdm and \ssdh provide an order-of-magnitude higher bandwidth for sequential reads
compared to \ssdl, 
\omcvii{i}t is challenging to scale a storage system's capacity using the high-end SSDs due to their significantly-higher prices and the \omcc{relatively smaller number} of the PCIe \omciv{slots} in a server.\footnote{\revp{The cost of the total storage system depends on both the price of each SSD and the available interconnection slots in the systems. High-bandwidth interconnects such as PCIe  take up very large  space in the system. \omciv{\omcv{As a result,} there are fewer PCIe slots than SATA slots in a system}. For example, building a 16-TB storage system with a single \omc{PCIe} SSD (Micron 9300 PRO~\cite{micros9300pro}) costs more than 3,000 USD, while \omc{the cost is}  less than 1,600 USD if we use \omc{four} 4-TB SATA SSDs (WD BLUE~\cite{wdblue}).}}

\head{Observation 3} \omciv{ Even though \swf outperforms \base, its filtering process is slow.
Potentially, \swf could provide significant performance benefits over \base due to two reasons;
1) as explained, 80\% of reads in the dataset exactly match  the reference genome, so only 20\% of the reads need to undergo the costly ASM \omcvii{computaion}; 
2) exact-match filtering requires only simple computation, i.e., SIMD XOR operations used by \swf.
However, even with \dram, \swf's speedup over \base is \omciv{only} 41\%.
The limited speedup is mainly due to large \omciv{number} of random memory accesses \revmark{A1\\Part 4}\revp{concurrently issued from all threads} to the reference index (\omcvii{explained in} Section~\ref{sec:background-readmapping}).
This observation highlights the potential of in-storage filtering. 
Even though both \swf and \isf filter out the same fraction of reads, the filtering process outside the SSD must compete with the read mapping process for the resources in the system (\omcv{e.g.,} the limited \omcv{main memory} bandwidth).
In contrast, \omcv{f}iltering of reads \omcvii{inside the SSD (where \omcvii{the reads} originally reside)} can remove the burden of filtering from the rest of \omcv{the} system.}

\head{Observation 4} \omcv{With a hardware accelerator (\acc)}, using the state-of-the-art SSD (\ssdh) does \emph{not} fully \omc{alleviate} the storage bottleneck, showing \jsr{23\%} longer execution time compared to when all the data is \omcvii{pre-loaded to} main memory (\dram). \omcc{While using \ssdm and \ssdh in \base and \swf \omciv{ shifts the bottleneck away from} I/O, \acc  turns I/O into a bottleneck again.}  
\omciv{This is because \acc greatly reduces the computational bottleneck, which increases the relative effect of the storage subsystem on \omcv{the} end-to-end execution time.
\omcvii{The} \acc and \isfacc} results clearly show that data movement between the storage devices and \omciv{the hardware} accelerator, which has not been properly considered in prior read mapping accelerators\omciv{~\cite{nag2019gencache, cali2020genasm, fujiki2018genax, turakhia2018darwin, madhavan2014race, cheng2018bitmapper2,houtgast2018hardware,houtgast2017efficient, goyal2017ultra,chen2016spark,chen2014accelerating,chen2021high,huangfu2018radar,khatamifard2021genvom}}, can significantly bottleneck the potential benefits of the accelerator.

\omcc{\head{Comparison to Other Near-Data \omcvii{Processing} Systems}} Even though read mapping applications could also benefit from other near-data processing (NDP) approaches such as processing-in-main memory (PIM)~\cite{kim2018grim, laguna2020seed, khatamifard2021genvom, kaplan2018rassa} or \omciv{processing}-in-caches~\cite{nag2019gencache}, in-storage processing  can \emph{fundamentally} address the data movement problem by filtering large, low-reuse data \emph{where the data initially resides}. 
As an extreme example, even if an \emph{ideal} accelerator achieved a \emph{zero} execution time for read mapping by addressing all of the computation and main memory overheads, there would still exist the need to bring the data from  storage to the accelerator.
In our motivational study, even \ssdh takes at least 1.55 seconds to read the entire dataset, which is 2.15$\times$ slower than the execution time that \isfacc provides (0.72 seconds).
Thus, even though solutions such as processing-in-memory can improve read mapping execution times, they still need to pay the cost of data movement from the storage system to the main memory\omciv{~\cite{singh2021fpga,turakhia2018darwin,cali2020genasm,kim2018grim,huangfu2018radar,khatamifard2021genvom,gupta2019rapid,li2021pim,angizi2019aligns,zokaee2018aligner}}. Therefore, \omciv{an} in-storage \omciv{filter} can be further integrated with any read mapping accelerator, \omc{including PIM accelerators,} to alleviate their data movement overhead.

\vspace{-.5em}
\subsection{Our Goal}
Based on our observations, we \rev{conclude} that an efficient in-storage filter can be \omcc{a} key enabler for read mappers to achieve high performance in both \omcc{\omciv{conventional} software\omcv{-based}  (\omciv{e.g., \base and \swf)} and new hardware-accelerated \omciv{(e.g., \acc)}} genomics systems.
\omcvii{In particular, i}n-storage filtering \omcvii{enables} the system to take full advantage of the \omcvii{high} computation capability of \omcvii{hardware} accelerators by \omcvii{fundamentally} addressing the data movement bottleneck.
\linebreak
\omciv{\textbf{Our goal} is to design an in-storage filter for genome sequence analysis in a cost-effective manner.}

We have \textbf{three key objectives} in designing \omciv{our} new system.
\jsr{First, t}he system should provide high in-storage filtering performance to overlap the filtering  with the read mapping of unfiltered data (as \isf does in our motivation study).
\jsr{Second,} it should support \omcv{reads with \omcvi{1)} different properties (e.g., lengths and error rates) and \omcvi{2)} different degrees of genetic variation in the compared \omcvi{genomes}}.
\jsr{Third,} it should \emph{not} require significant additional hardware \omciv{overhead}, e.g., complicated logic circuits or large SRAM/DRAM memory.

\section{GenStore}
\label{sec:mechanism}

\omcc{We propose \emph{\proposal}, the first in-storage processing system \jk{tailored} for genome sequence \jk{analysis. 
\proposal} greatly reduces both data movement and computational overhead\hm{s} \rev{of genome sequence analysis} by exploiting low-cost and accurate in-storage filters}.
GenStore supports \omciv{reads with different properties (lengths and error rates\omcv{)} and different degrees of genetic variation in the \omcvii{compared genomes}.}
\revmark{C1 and\\ F1}\revp{\omc{We primarily design \proposal as an in-storage accelerator, which is an \emph{extension} of the existing SSD controller and flash translation layer (FTL). \proposal is designed to be integrated into the system such that\omcvii{,} when the accelerator is not in use, the entire storage device is available to all other applications\jk{, just like} in a general-purpose system today.
}}

\omc{\subsection{Overview}}

\begin{figure}[h]
    \vspace{-1em}
    \centering
    \includegraphics[width=\linewidth]{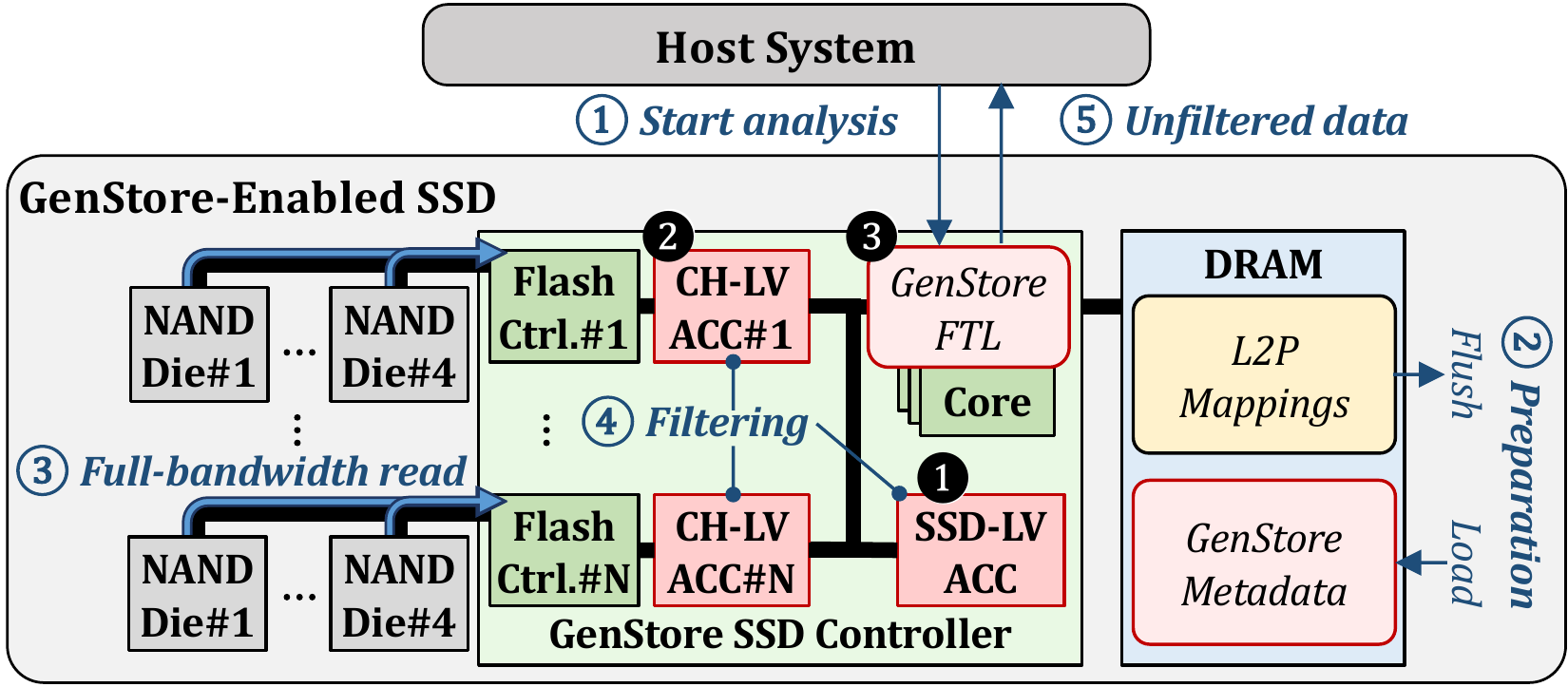}
    \vspace{-2em}
    \caption{Overview of \proposal.}
    \label{fig:genstore_overview}
    \vspace{-1em}
\end{figure}

\omcc{The key idea of \proposal is to exploit low-cost \emph{in-storage} accelerators to accurately {\emph{filter out}} the reads that do not require the expensive alignment step in read mapping and \jk{thus} significantly reduce unnecessary data movement from the storage system to main memory and processors}. 
\fig{\ref{fig:genstore_overview}} shows the overall architecture of GenStore and how it \hm{interacts} with the host system. GenStore employs two types of hardware accelerators: \circled{1}~a single \emph{SSD-level} accelerator and \circled{2}~\emph{channel-level} accelerators, each of which is dedicated to a channel. The \proposal-FTL (\circled{3}) communicates with the host system and manages the metadata and data flow over the \omcv{SSD} hardware components (i.e., NAND flash chips\jk{, \omcvii{internal} DRAM, and \omcvii{in-storage} accelerators}).

\jk{Once} the host system \omcv{indicates that the SSD should start analysis as required by} a read mapping application \omcvi{(\wcirc{1} in \fig{\ref{fig:genstore_overview}})},  \proposal prepares \hm{for operation} as an accelerator \omcv{(\wcirc{2})}. It flushes the conventional FTL metadata necessary to operate as a regular SSD (e.g., L2P mappings~\cite{kim2020evanesco}), while loading the \proposal metadata necessary for each use case \rev{(Section~\ref{sec:ftl} provides more details on \proposal FTL).}
After finishing the preparation, GenStore starts the filtering process.
\jsr{It}  keeps concurrently reading the data to process from \emph{all} \jk{NAND flash} chips \omcv{(\wcirc{3})} via multi-plane operations (i.e., it exploits the SSD's full internal bandwidth\omc{, which} is much higher than the I/O bandwidth between the SSD and host system~\cite{koo2017summarizer, mailthody2019deepstore}),  while  filtering out reads \omcv{(\wcirc{4})} that do not have to undergo \omciv{further analysis (\omcv{\omcvii{e.g.}, ASM computation})}.
Doing so is possible due to multiple channel-level accelerators that provide \jk{computational} throughput matching the SSD's internal bandwidth even for the most complicated computation required for filtering. 
The host system \omcv{performs} further computation as soon as GenStore sends unfiltered \omciv{reads} \omcv{(\wcirc{5})}, which removes GenStore's filtering process almost completely from the critical path of the application.

\rev{As explained in Section~\ref{sec:background-SSD}, most of the internal DRAM is occupied by the regular L2P mapping. Therefore, flushing the regular L2P mapping data into NAND flash memory enables \proposal} to exploit most of the GB-scale DRAM (e.g., 4GB DRAM in a 4TB SSD~\cite{samsung860pro}) \rev{for its operations}, which significantly reduces the overhead of additional internal DRAM that might \jk{otherwise} be required to store the \proposal metadata necessary for the filtering process. We carefully design \jk{the GenStore filtering algorithms} to only \emph{sequentially} access the underlying NAND flash \jk{chips, so GenStore} requires only a small amount of metadata to access the stored data. \omcc{Therefore, \proposal can use most of the internal DRAM space for \jk{such metadata.}} We envision that all \proposal metadata are built  \emph{offline} by the host or \omc{some} other system (e.g., \omc{by the sequencing machine} when the read set or reference genome are initially stored to the SSD). \omciv{Constructing \proposal metadata} is a \emph{one-time} \omc{preprocessing step} that can be performed independently of the read mapping process, while the result \omc{of the preprocessing step} can be used multiple times for different genomics applications.

There exist two main challenges in designing GenStore as an efficient in-storage filter for read mapping.
First, the behavior and data-access patterns in read mapping significantly vary depending on the read \omcv{properties} (\omccc{length and error rate) and genetic variation \omciv{between \omcvii{the} compared genomes}}.
Second, hardware resources (e.g., CPU and DRAM) are quite limited even in modern \omcvii{high-end} SSDs.
We address these challenges via thorough hardware/software co-design tailored for \omciv{filtering
1) exactly-matching reads, i.e., reads that exactly match subsequences of the reference genome  (Section~\ref{sec:SRF}), and 2) \omcv{most of the} non-matching reads, i.e., reads that would not align to any subsequence of the reference genome (Section~\ref{sec:LRF}).}

\vspace{-.3em}
\subsection{\proposal-\omccc{EM for Exactly-Matching Reads}}
\label{sec:SRF}

\subsubsection{Approach Overview}
\label{sec:sr-overview}

GenStore-EM accelerates  read mapping by using an efficient in-storage filter for reads that have at least one exact match in the reference genome. 
\omccc{Due to the low error rates of short reads, \jk{combined with low genetic variation between \omcvii{the} compared genomes}, a large fraction of short reads map \emph{exactly} to the reference genome~\cite{10002015global,uk10k2015uk10k,nag2019gencache}. For example, on average ~80\% of human short reads map \emph{exactly} to the human reference genome~\cite{10002015global,uk10k2015uk10k,nag2019gencache}.} 
Since exact-match detection is computationally cheaper than ASM, concurrently filtering exact-matching reads inside the SSD can significantly improve the runtime of read mapping \jk{(}as we demonstrate in  Section~\ref{sec:motivation}\jk{)}.
\omciv{\omcvii{Note that,} \proposal-EM is not applicable to long reads due to their \omcviii{greater} length. For example, for a \omcv{10K base pair-long} human read\omcv{,} even with zero sequencing \omcv{error \omcviii{rate}},  the \omcvii{probability} of the read exactly matching a subsequence in the reference genome is very low (e.g., $<3.6\times10^{-6}$) due to natural genetic variation.\footnote{\revp{\omc{A typical human genome contains genetic variations at $\sim$4.1 to 5~million~\cite{10002015global} out of a total of $\sim$3.2~billion base pairs~\cite{schneider2017evaluation}. Thus, each 10K-bps read contains, on average, $\sim$12.5 to 15.3~\omcv{base pairs that are different from} the reference genome.}}}
}

\head{Key Challenges} 
The key challenge in designing \proposal-EM is \omcvii{the} large number of random accesses to large data structures inside the SSD. As explained in Section~\ref{sec:background}, identifying exact matches for read mapping requires a number of random accesses \emph{\omcc{for each k-mer in a read}} to two large data structures: 1) a large k-mer index, to find \omciv{potential} \omcc{matching} locations  \omcc{of each k-mer in the reference genome}, and 2) the  reference genome, to find candidate matching sequences \omcc{at the candidate matching locations in the reference genome}.
\omcc{Handling random accesses to large data structures is challenging  for \omcvii{NAND flash-based} SSDs for two reasons.}
First, NAND flash memory  exhibits poor performance for  random \jk{access} reads. 
Second, \omcc{we cannot use in-SSD DRAM to store the \jk{large} data structures that are randomly accessed, since even}
in high-end SSDs, the size \omcvii{of internal DRAM} is \omcc{relatively small} (e.g., 4 GB~\cite{samsung860pro}) compared to the size of the data structures that \proposal-EM needs to handle \omcc{(e.g., 7 GB for \omcvii{the} human reference genome and its index~\cite{li2018minimap2})}.

\newcommand\readset{SRTable\xspace}
\newcommand\seedidx{\omcc{SKIndex}\xspace}

\head{Key Idea} 
The key idea in \proposal-EM is to \emph{sequentialize} most data accesses via a carefully designed metadata structure and data layout. \omcc{To do so, we \jk{design} a new \emph{sorted,} \emph{read-sized} k-mer index structure. This index enables \jk{\emph{sequential}} scanning \omccc{of} the read set and the new index, with only one index lookup per read while performing filtering.}

\begin{figure}[h]
        \centering
        \vspace{-.5em}
        \includegraphics[width=\linewidth]{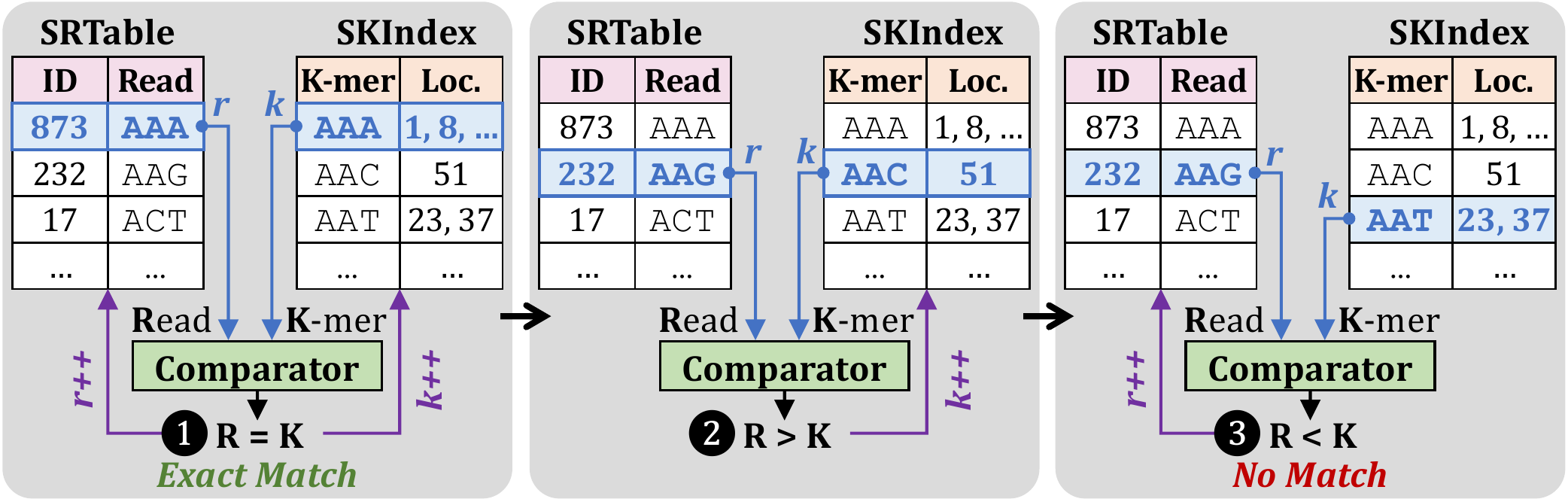}
        \vspace{-2em}
        \caption{Overview of the key idea of \proposal-EM.}
        \vspace{-1em}
        \label{fig:sr_idea}
\end{figure}

Figure~\ref{fig:sr_idea} shows \omcc{the key idea of \proposal-EM} with a simpl\omcc{ified} example in which each short read consists of \omcix{three} base pairs (bps).\footnote{\omcc{In a realistic scenario, the read \omcviii{length} is much larger (e.g., 150 bps).}}
Suppose \hm{that we have} two data structures: \rev{1)}~a \omcc{sorted read table (\emph{\readset})}, each entry of which stores a read and its unique ID, and \rev{2)}~a \omcc{sorted k-mer index (\emph{\seedidx})}, which contains \emph{all unique read-sized} k-mers of the reference genome, \omcc{along with each k-mer's corresponding locations in the reference genome.}
\omcx{Each} data structure \omcx{is} \emph{sorted} by read/k-mer in alphabetical order. 
With \jk{these two} data structures, it is possible to identify each read's exactly-matching locations \omcc{in the reference genome} by \emph{\omcix{streaming}} \jk{both} reads and k-mers \omcix{through a simple comparator}.
\omcix{We use two pointers, $r$ and $k$, which point to the current entries we are examining in \readset and \seedidx, respectively. 
We \emph{sequentially increment} the two pointers in three different ways based on the comparison result of the current read and k-mer.}
\omcix{First, when the current read and k-mer are identical (\circled{1} in Figure~\ref{fig:sr_idea}),}
we record \omcix{the read} as an \omcix{exactly-matching read} and \omcix{increment $r$ and $k$}.
\omcix{Second, if the read is alphabetically larger than the k-mer (\circled{2}), we conclude that the k-mer does not match any \omcix{read} in \readset and increment $k$ \omcx{(so that we can examine if the next k-mer matches the read)}.}
\omcix{Third, if} the k-mer is alphabetically larger than the read (\circled{3}), \omcix{we conclude that the read does not match any k-mer in \seedidx.}
We record the read as \emph{not} an exact match
and \omcix{increment $r$ \omcx{(so that we can examine the next read)}}.

\jk{GenStore\omcvi{-EM}'s two} data structures and filtering algorithm enable an exact-match filter highly suitable for in-storage processing.
\omcc{First, }since \jk{these two} data structures are \jk{only} sequentially accessed, the filtering process can be done in a \emph{stream\omcc{ing}} manner, leveraging the high sequential read bandwidth of NAND flash memory. 
\omcc{Second, we can easily perform} exact-match detection of a read and a read-sized k-mer with simple comparator logic and \omcc{fully pipeline the} filtering process with sequential \jk{access to} the data structures.

A \emph{read-sized} k-mer index increases the total amount of accessed data for read mapping. \revmark{E4}\revp{The reason is that a read-length value of k (e.g., k=150) significantly increases the number of unique k-mers, \omcvii{compared to} \jk{\omcvii{the k values} commonly used in conventional read mappers} (e.g., k=15)}.
\omcv{For example, the size of an index structure for all unique k-mers in the human reference genome is 21 GB when k=15, while the size increases to 126 GB when k=150.} 
\jk{However, our \omcvii{proposal} (\omcvii{i.e.,}~a large yet sequentially-accessed SKIndex) is feasible and desirable} \omc{for in-storage processing}
due to the large capacity and high internal bandwidth of modern NAND flash-based SSDs.

\subsubsection{Design of \proposal-EM} 
\label{sec:sr-design}

\begin{figure}[h]
        \vspace{-.5em}
        \centering
        \includegraphics[width=\linewidth]{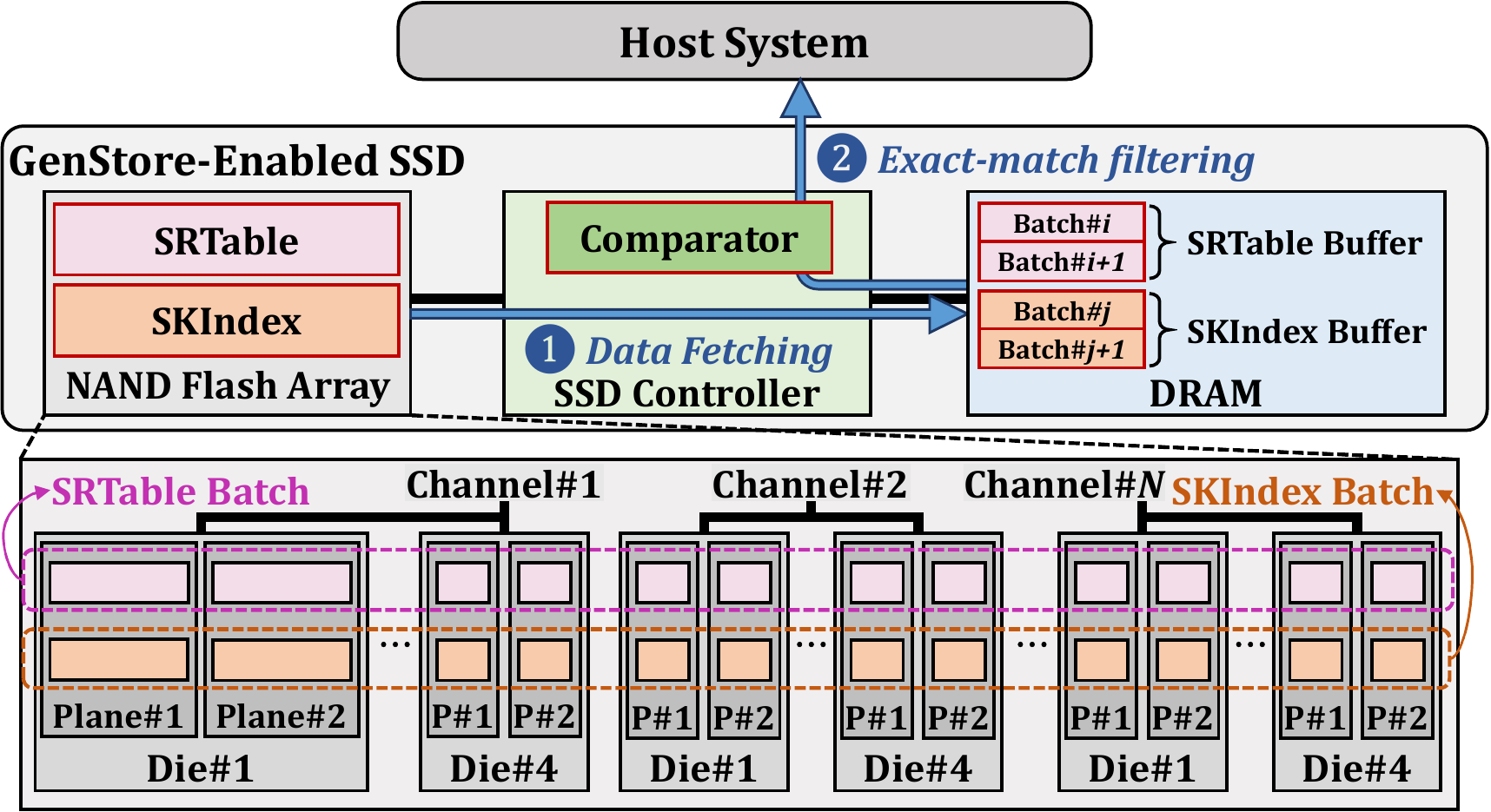}
        \vspace{-2em}
        \caption{Overview of \proposal-EM.}
        \label{fig:sr_overview}
        \vspace{-.5em}
\end{figure}

Figure~\ref{fig:sr_overview} illustrates the overall \omcc{operational flow} of \proposal-EM, which consists of two steps: \textbf{Step~1.}~data fetching and \textbf{Step~2.}~exact-match filtering.  
\proposal-EM uses \jk{the} two data structures explained in Section~\ref{sec:sr-overview}: 1) a sorted read table (\readset) for storing the read set and 2) a sorted k-mer index (\seedidx) for storing \jk{read-sized} k-mers from the reference genome.
Step~1 reads the two data structures from NAND flash chips to the SSD's internal DRAM in a \jk{batched manner} (\circled{1} in Figure~\ref{fig:sr_overview}).
Step~2 performs exact-match filtering within each read batch, using simple comparator logic \omciv{in the SSD-level accelerator} (\circled{2}).
Steps 1 and 2 are performed in a pipelined manner. \omcc{\omciv{During filtering,} \proposal-EM sends the unfiltered reads to the host system \omciv{for} full read mapping. This enables the \omciv{concurrent} filtering of reads in \omcv{the} SSD and read mapping of the unfiltered reads in the host system.}

\head{Data Structures}
We carefully design \readset and \seedidx to minimize performance and storage overheads of \proposal-EM,
by extending the two data structures described in Figure~\ref{fig:sr_idea} in two aspects. 
First, both \readset and \seedidx contain a \emph{strong} hash value (e.g., SHA-1~\cite{dang2015secure} or MD5~\cite{rivest1992rfc1321}) of each read and \omcc{read-sized} k-mer, respectively, which is used as both the sorting criterion of the data structures and a \emph{fingerprint} of each read and \omcc{k-mer} \jk{that is used by the comparator logic}.
Second, \omcc{\omcvii{unlike} the data structures described in Figure~\ref{fig:sr_idea}}, \seedidx no longer contains the k-mers of the reference genome \jk{but only the fingerprints of the k-mers}.
Using strong hash values enables \proposal-EM to determine exact matches between \jk{a read and a k-mer by comparing only} their fingerprints, \omcc{which provides two benefits.}  \omcc{First, it} reduces the storage overhead of \seedidx by obviating the need to store the raw \omcc{read-sized} k-mers.\footnote{We design \readset to \omcc{store} the raw reads \omcc{so that we can} transfer \emph{unfiltered} reads to the host \omcc{for full read mapping} after \omcc{we detect them as non-exactly-matching reads}.} 
For \omcv{the human reference genome, the size of the optimized \seedidx \omcvi{ (}which stores fingerprints instead of read-sized k-mers) is \rev{32}~GB when the read size is 150 bps, which is 3.9$\times$ smaller than the size of the unoptimized \seedidx.} 
Second, \omcc{using fingerprints} significantly reduces the performance overhead of exact-match detection by avoiding comparisons of reads and k-mers \jk{that are hundreds of bytes in size}.

Note that such exact-match detection does \emph{not} affect the accuracy of read mapping due to the extremely low collision rate of strong hash functions. 
\revmark{C3}\revp{Even in an \jk{extremely  rare} case of a hash collision, \omc{the} impact \omc{of the collision} on \proposals’s accuracy will be negligible~\cite{arxivGS}, since the DNA information loss due to the falsely filtered read will highly likely be compensated by \omc{\jsr{other} reads generated from the neighboring locations in the DNA\jsr{, which almost \omciv{\emph{fully}} overlap with the falsely filtered read but have totally different hash values}.} This is because it is common practice to sequence each DNA fragment several times (i.e., with high coverage) to improve the accuracy of downstream genetic analyses~\cite{sims2014sequencing,quail2012tale,levy2016advancements,hu2021next}.
}

\omcc{We  envision that all \proposal data structures are built  \emph{offline} by the host or some other system (e.g., \jk{by the sequencing machine} when the read set or reference genome are initially \omcvii{written} to the SSD)}. \revmark{CQ2}\revp{\omcc{This} preprocessing overhead can be \jk{hidden} by two essential initial steps of the genome \jk{sequence} analysis pipeline: 1)~sequencing and 2) 
basecalling.
\omcv{For example, in the \omcv{current highest-throughput} Illumina \omcv{Nova}Seq \omcv{6000} sequencer~\cite{illumina}, sequencing and basecalling \omc{work in a pipelined manner and generate genomic read data at a} limited \omc{throughput of} \omcv{18.9} MB/s~\cite{illumina}.} 
\omcv{We analyze the throughput of \proposals's preprocessing step (i.e., generating hash values and sorting reads) and observe that even a personal laptop~\cite{lenovot740p} can provide 174 MB/s of preprocessing throughput.}
Therefore, \proposals’s preprocessing can be done in a pipeline\omc{d} manner with sequencing/basecalling, without decreasing \omc{the overall throughput of these steps}. 
This low preprocessing overhead can be \omc{further} amortized since the preprocessed data can be reused multiple times \omcc{in different read mapping experiments}.
}

\head{Step 1. Data Fetching}
\proposal-EM \jk{reads \readset and \seedidx in batches}, while exploiting the \emph{full} internal bandwidth of the SSD. \omcc{In Figure~\ref{fig:sr_overview}, we refer to each batch of \readset as a\omcvi{n} \emph{\readset Batch} and each batch of \seedidx as a\omcvi{n} \emph{\seedidx batch}}. The batch size is equal to the size of data that can be read in parallel by a multi-plane read operation for each chip (i.e., \emph{Number of Planes in the SSD}~$\times$~\emph{Page Size}), 
which enables 100\% utilization of the NAND flash chips \jk{while} reading a batch. 
As shown in Figure~\ref{fig:sr_overview} (bottom), \proposal-EM stores the \readset and \seedidx to NAND flash chips in an \emph{\jk{interleaved}} manner so that each of the data structures can be \emph{sequentially}, \emph{evenly} distributed across all the NAND flash chips.

\proposal-EM \omcc{exploits} the SSD's full internal bandwidth using double buffering.
As shown in Figure~\ref{fig:sr_overview}, \proposal-EM employs two \omcc{sets of} batch buffers in the internal DRAM: \omcc{\emph{\readset Buffer}} (for \readset) and \omcc{\emph{\seedidx Buffer}} (for \seedidx), each of which can store two batches \omcc{of \jk{the respective} data structure}. 
\omcc{After Step 1 finishes fetching $Batch\#i$, it \omcvii{proceeds} to fetching $Batch\#i+1$, while Step 2 starts working on $Batch\#i$. If the two steps work with the same throughput, we only need to buffer two batches for each data structure. For example, to enable double-buffering} in an 8-channel SSD (with four 2-plane dies per channel and 16-KiB pages), the batch buffers require 8MB DRAM space in total.

\head{Step 2. Exact-Match Filtering}
Step 2 scans \omcc{through} each batch of \readset and \seedidx \omcc{ stored in \readset Buffer and \seedidx Buffer}, \omcvii{respectively,} comparing the fingerprints (strong hash values) with a simple hardware comparator.
When $FP(r_i)>FP(k_j)$ (where $FP(x)$ is the fingerprint of $x$, and $r_i$ and $k_j$ are the $i$-th read and $j$-th k-mer in the current batch, respectively), Step 2 scans \seedidx while increasing $j$ until it finds $k_j$ such that $FP(r_i)\leq{}FP(k_j)$.
If $FP(r_i)<FP(\omcviii{k}_j)$, it is guaranteed that no exact match exists \omcvii{for read~$r_i$}, so \proposal-EM sends \omcvii{$r_i$} to the host for the read mapping process.
When $FP(r_i)=FP(\omcviii{k}_j)$, i.e., the two fingerprints are identical, \omcc{and \proposal-EM marks the read as an exactly matching read.}

\omcv{Due to the simple computation in Step 2, the execution time of \proposal-EM is bottlenecked by Step~1 (Data Fetching). As explained, Step~1 only streams the data structures in batches from the underlying NAND flash chips to the internal DRAM, leveraging the SSD's full internal bandwidth. 
Therefore, the performance of \proposal-ER can be easily scaled \omcvi{up} by increasing the SSD's internal parallelism (e.g., by deploying more channels or using low-latency NAND flash memory~\omcvi{\cite{cheong-isscc-2018, park-nvmsa-2018, park-asplos-2021}}).}

\subsection{\omciv{\proposal-NM for Non-Matching Reads}}
\label{sec:LRF}

\subsubsection{Approach Overview}
\label{sec:lr-overview}

    \proposal-NM filters \omciv{most of the} \emph{\underline{n}on-\underline{m}atching} reads, i.e., reads that \omciv{would not} align to any subsequence in the reference genome.
    \omcv{This is motivated by the fact that in read mapping, a \omcvi{large} fraction of reads might \emph{not} align to the reference genome due to} 1) the high sequencing error rate (in long reads) and/or 2) high genetic variation between \omcvii{the} compared genomes (in both short and long reads).
    \omcv{To \omcvi{illustrate} this, we analyze four \omcvi{read} mapping use cases, one with read sets with high sequencing error \omcvii{rates}, and three with high genetic variation in the compared \omcvi{genomes}. 
    Use cases with high genetic variation include 1) \omcvii{samples from rapidly-evolving species} (such as SARS-CoV-2~\omcvi{\cite{bhoyar2021high}}) that have high genetic variation compared to the reference genome,} 2)~samples with no known reference genomes, and
    3)~mapping \omcvi{a read set} against the human reference genome to filter out \omcvi{\emph{human contamination}, i.e.,~reads that have been sequenced from human-origin contaminant DNA in a non-human sample.\footnote{\omcvi{Contamination of non-human samples with human DNA is commonly observed~\cite{breitwieser2019human,human2012structure} and corrected for~\cite{danko2021global,knight2018best,human2012structure}}.}}
    \omcv{For each use case \omcvi{(except for the \omcvii{\emph{Contamination}} use case)}, we analyze \omcvi{two different combinations of \omcvi{the input} read set and \omcvi{the} reference}.} 
    Table~\ref{table:lr-profile} \omcv{summarize\omcvi{s} the result of this analysis by showing \omcvi{input}} read sets, \omcv{the propert\omcvi{ies of each read set (e.g., \omcvii{read length and dataset size})}}, reference genomes, and the percentage of reads in each read set that align to subsequences in the reference genome.  
    We observe that a large fraction of reads \omciv{\omcvi{(31.7\%--99.6\%)} within a read set} does \emph{not align to \omcvii{any} subsequence} in the reference genome. 
    \omcv{\omcvi{Quickly f}iltering this large fraction of non-aligning reads in the SSD can reduce the \omcvi{large} data movement from the storage and \omcvi{expensive} ASM computation for reads that would not align.}

\begin{table}[h]
\centering
\caption{\omcv{Fraction of aligning reads in various read mapping use cases with short and long reads.}}
\vspace{-.5em}
\resizebox{1.0\columnwidth}{!}{%
\begin{tabular}{c|c|c|c|c}
\toprule
\multirow{2}{*}{\textbf{\omcv{Use case}}} & \multirow{2}{*}{\textbf{Input read \omcvi{set} (S}hort/\textbf{L}ong\textbf{)}}  & \textbf{Size} & \multirow{2}{*}{\textbf{Reference}} & \omcv{\textbf{Align}} \\
& & \textbf{[GB]} & & \textbf{[\%]}\\
\midrule
\midrule
\multirow{2}{*}{Sequencing errors} & \omciv{ERR3988483 (L)~\cite{sayers2021database}} & 54 & \multirow{2}{*}{hg38~\cite{schneider2017evaluation}} & \omcv{{\bf 47.\omcvi{4}}}\\
& \omciv{HG002\_ONT\_20200204 (L)~\cite{zook2016extensive}} & 371 & & \omcv{{\bf 69.\omcvi{3}}}\\
\midrule
Rapidly evolving & SRR5413248 (L)~\cite{sayers2021database} & 1.69& NZ\_NJEX02~\cite{clark2016genbank} & \omcv{{\bf 60\omcvi{.0}}}\\
samples & \omciv{SRR12423642 (S)~\cite{sayers2021database}} & 0.466 & NC\_045512.2~\cite{wu2020new} & \omcv{{\bf 23.\omcvi{1}}}\\
\midrule
\multirow{2}{*}{No reference} & SRR6767727 (L)~\cite{sayers2021database} & 12.4& \multirow{2}{*}{NZ\_NJEX02~\cite{clark2016genbank}} & \omcv{{\bf 0.35 }}\\
& SRR9953689 (L)~\cite{sayers2021database} & 15.9&  & \omcv{{\bf 37\omcvi{.0} }} \\
\midrule
Contamination & SRR9953689 (L)~\cite{sayers2021database} & 15.9& hg38~\cite{schneider2017evaluation} & \omcv{{\bf  1\omcvi{.0} }} \\
\bottomrule   
\end{tabular}}
\vspace{-1.5em}
\label{table:lr-profile}
\end{table}

\omciv{To avoid expensive ASM \omcv{computation} for reads that would not be aligned, state-of-the-art read mappers commonly employ a step called \emph{chaining} (described in Section~\ref{sec:background-readmapping}), which calculates each read's similarity score (called \emph{chaining score}) to the reference genome and filters out reads with a low score.
\omciv{\proposal-NM uses this basic idea of chaining to build an in-storage filter.}}

    \head{Key Challenges} Calculating a chaining score in the SSD is challenging because finding the best chaining score requires performing an expensive dynamic programming (DP) algorithm on every \omciv{potential matching location (i.e., seed)} within a read (explained in Section~\ref{sec:background-readmapping}).
    Normally this operation has $\mathcal{O}(N^2)$ time and space complexity, where $N$ is the number of seeds.
    \omcvii{Even though}
    \omciv{\omcvii{a} chaining score} can be approximated accurately \omcv{in} $\mathcal{O}(hN)$ time and space by considering only $h < 50$ seeds at a time~\cite{li2018minimap2}\omcvii{, w}e observe that some long reads can have up to thousands of seeds \omciv{due to their length}. Therefore, designing \omcv{a} chaining accelerator to \omciv{perform an expensive \omcvii{DP} algorithm on \omcvii{these} large \omcvii{reads} ($N > 1000$) can incur significant performance or area \omcv{overheads}.}

\begin{figure}[b]
        \centering
        \vspace{-1em}
        \includegraphics[width=0.75\linewidth]{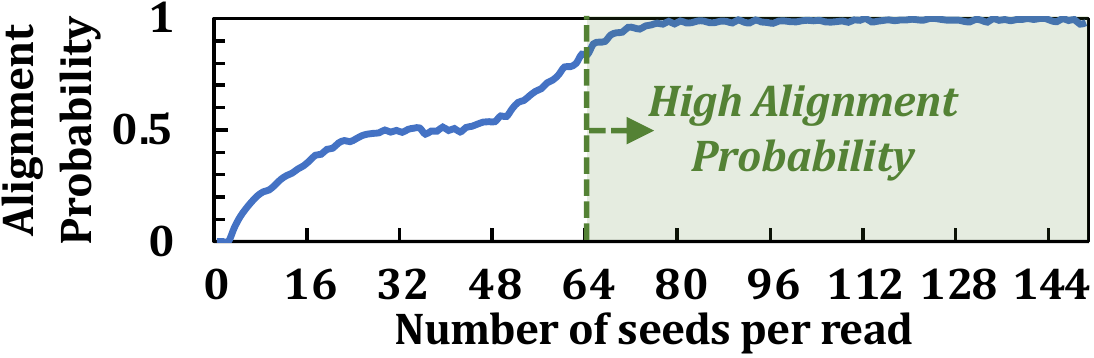}
        \vspace{-1em}
        \caption{Alignment probability as a function of the number of seeds per read in a long read mapping use case.}
        \label{fig:lr_align}
\end{figure}

    \head{Key Ideas} 
    \omcv{T}\omciv{o avoid such expensive chaining, \omcv{\proposal-NM} selectively perform\omcv{s} \omcv{a fast version of} chaining  \emph{only} on reads with a small number of seeds} and send\omcv{s} other reads to the host system for full read mapping (including \omcv{complete} chaining). This idea is based on \omciv{our} key observation that 1) a read with a large number of seeds most likely aligns to the reference \omciv{genome and does not require in-storage filtering, and 2) selective chaining \omcvi{can filter} many non-aligning long reads, without requiring costly hardware resources in the SSD}. \fig{\ref{fig:lr_align}} shows the \omcvi{alignment} probability of a \omciv{read in a \omcv{long read dataset (SRR5413248 \cite{sichtig2019fda} in Table~\ref{table:lr-profile})} \omcvi{to} subsequences} in the reference genome~\omcvii{(NZ\_NJEX02~\cite{clark2016genbank})}, \omcviii{as a function of the} \omcv{number of} seeds \omcv{per read \omcviii{($N$)}}.
    \omciv{The average read length in this dataset is 10K base pairs and the number of seeds goes up to several thousands for some reads.} 
    We observe that reads with \omcv{a sufficiently large number of seeds} are \omcv{very} likely to align \omciv{to subsequences in the reference genome} \omcv{(e.g., at least 85\% of reads with \omcvi{$N\geq64$} seeds align). Such reads }can be directly sent to the CPU for \omciv{full read mapping} (bypassing the \omciv{in-storage \omcviii{chaining-based}} filter). 
    
We extend this analysis to read datasets from various organisms commonly used in genomics studies: \textit{E. coli}~\cite{clark2016genbank}, yeast~\cite{engel2014reference}, thale cress~\cite{berardini2015arabidopsis}, fruit fly~\cite{larkin2020}, mouse~\cite{church2009lineage}, and human~\cite{schneider2017evaluation}. 
We observe that when a read has $N = 64$, $128$, or $256$ seeds, it aligns with average probabilities of $88.87\%$, $91.32\%$, and $93.84\%$, respectively. 
Based on these observations, we \omcvi{design \proposal-NM to \emph{selectively} perform chaining only on} reads with \omcvii{fewer} than $N$ seeds\omcvi{, while sending \omcvii{reads with \omcviii{at least} $N$ seeds} to the host system.}\footnote{\omcvi{\proposal-NM's design can be tuned based on different values of $N$.}} 
\omcv{This selective chaining} significantly reduces the chaining execution time and \omcv{additional hardware} area \omcv{cost,} while filtering most reads \omciv{that would not align to the reference genome}.

\newcommand\kindex{KmerIndex\xspace}
\subsubsection{Design of \proposal-NM}
\label{sec:lr-design}

\begin{figure}[h]
        \vspace{-.5em} 
        \centering
        \includegraphics[width=\linewidth]{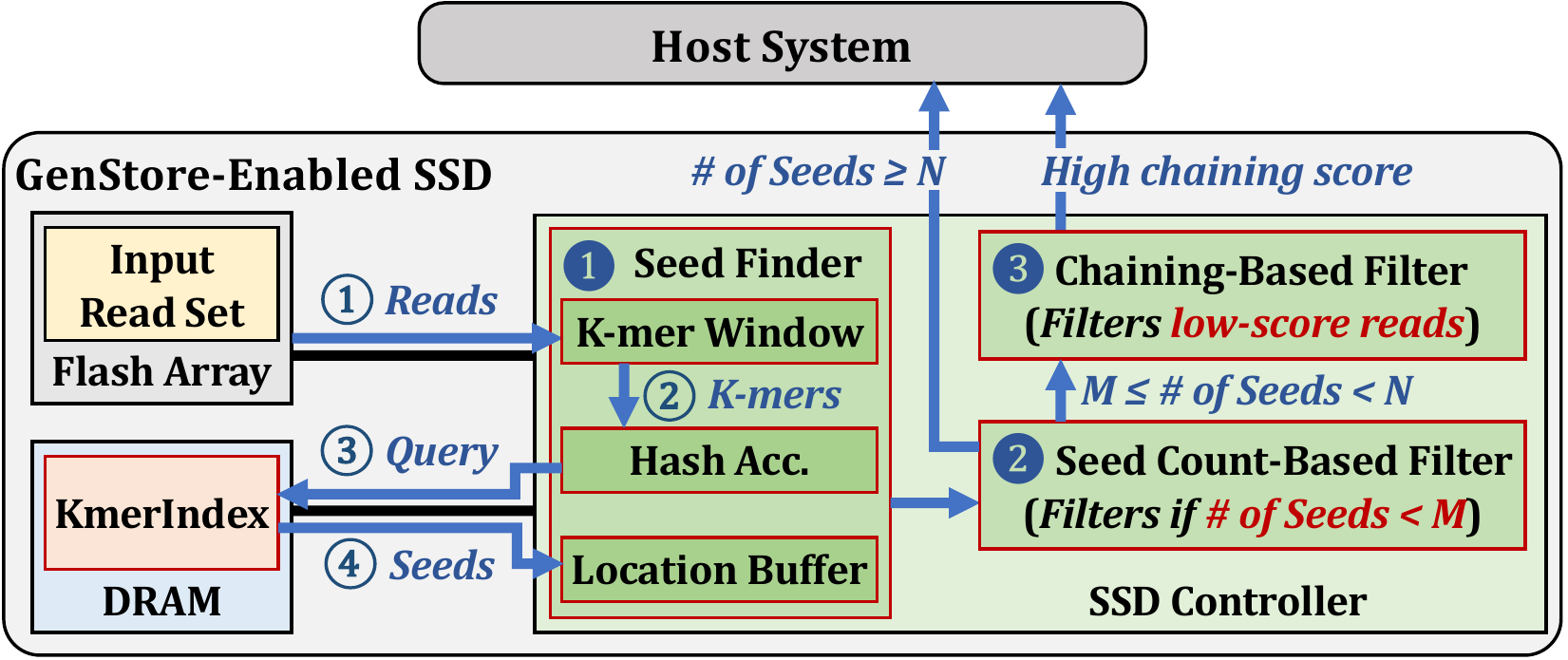}
        \vspace{-2em}
        \caption{Overview of \proposal-NM.}
        \vspace{-.5em}
        \label{fig:lr_overview}
\end{figure}

    Figure~\ref{fig:lr_overview} shows the overview of \proposal-NM that filters out most of the non-matching reads in three steps. 
     In Step 1, \omciv{\proposal-NM} reads the input read set from the flash chips, generates minimizer \omciv{k-mers} for each read (as explained in Section~\ref{sec:background-readmapping}), and looks up each minimizer in a \emph{K-mer Index} (\kindex) to find \omciv{the potential matching locations, i.e., seeds  \omcv{(\circled{1} in \fig{\ref{fig:lr_overview}})}}.  In Step 2, \proposal-NM \omciv{counts the number of seeds \omcv{in each} read \omcv{to decide} if the read needs to go through \omcv{chaining (\circled{2})}.} 
    \omcvi{To further improve \omcviii{overall} performance, Step 2 also filters out reads with \emph{too few} seeds \omcviii{(i.e., $< M$)}, which would not align to the reference and thus would be filtered anyway by the baseline read mapper~\cite{li2018minimap2}.}
    In Step 3, \proposal-NM filters reads based on their chaining scores using a \omcv{fast and} efficient chaining accelerator \omcv{(\circled{3})}.  If a read has at least one chain with a score above a specified threshold, it \omciv{is sent} to the CPU for mapping. Otherwise, the read \omciv{is} filtered. \omciv{All three} steps run in a pipelined manner.

    \head{Data Structures} We carefully design \omcv{the} \kindex to reduce the \omcv{SSD's internal} DRAM capacity required for storing it. \kindex, similar to the index used by the baseline read mapping tool (i.e., \omcvii{M}inimap2~\cite{li2018minimap2}), is a hierarchical hash table.
    To reduce the memory capacity required for storing the \kindex, we \omcv{make} three \omcv{modifications}: 1)~not stor\omc{e} the reference genome since it is not needed by our approach, 2)~not stor\omc{e} seeds with \omcv{\emph{many}} matching locations (\rev{e.g., \omcv{more than} 495 locations in the baseline read mapper~\cite{li2018minimap2}})\footnote{\revp{Each matching k-mer can map to one or more locations in the reference genome.}}
    \revmark{E5} \omcv{since} \omc{read mappers usually ignore such} seeds in the chaining process~\cite{li2018minimap2}, \omcv{and} 3)~increas\omcv{e} the number of \omciv{\omcv{hash table}} buckets so that each bucket holds one minimizer. \omciv{Increasing the number of buckets} increases the false positive rate in the \omciv{filter}, which leads to \omciv{finding extra seeds and performing} extra chaining operations (without loss of accuracy). We make this trade-off to reduce \omcv{\omcvi{the} \kindex's} size and the required memory space. \omciv{These optimization}s \omcvii{reduce} the size of the index for the human reference genome~\cite{li2018minimap2} from 5.8 GB to 2.9 GB, \omciv{which enables \proposal-NM to \omcv{easily} store} the \kindex in the \omcv{limited} DRAM \omcv{space} inside the SSD.\revmark{E7}
    
    \head{Step 1: Seed \omciv{Finding}} 
     \omcv{This} step find\omcv{s} the \omciv{potential} matching locations of a read \omciv{(i.e., seeds)} in the \kindex.
    \omciv{~\proposal-NM} \omcv{first} reads the input read set from all \omcvi{f}lash chips \omcv{(\wcirc{1} in Figure~\ref{fig:lr_overview})}. 
    \omcvi{\proposal-NM} exploits the \emph{full} internal bandwidth of the SSD by reading the read set in parallel \omcv{via} a multi-plane read operation for each chip \omcv{(same as \proposal-EM, Section~\ref{sec:sr-design})}. In a shifting buffer (\omcv{called the} \emph{K-mer Window}) \omciv{in the channel-level accelerator}, \omciv{\proposal-NM} stores the \omciv{$w$} most recently read k-mers of each read. 
     For each sliding window of $w$ k-mers\omcvi{,}\footnote{\omciv{The default value for $w$ in \omcvii{M}inimap2~\cite{li2018minimap2} is 10, but \proposals's design can be tuned for different values.}} \proposal-NM \omcvi{finds the minimizer of the window (described in \sect{\ref{sec:background-readmapping}})} using a 64-bit Integer Mix hash function (hash64)\omcv{~\cite{wang2007integer}} \omciv{in the SSD-level accelerator}
     \omcv{(\wcirc{2})}. 
     \omciv{\proposal\omcvi{-NM} queries} each minimizer in the \kindex until it reads $N$ seeds (\omciv{e.g.},  $N$ = 64) or it reaches the end of the read \omcv{(\wcirc{3})}. Each index query takes up to two memory accesses (to visit the two levels of the hash table). \omciv{\proposal-NM} stores the \omciv{seeds} in the \emph{Location Buffer} \omciv{in the channel-level accelerator} \omcv{(\wcirc{4})}, and moves to Step 2.

    \head{Step 2: Seed Count-Based Filtering} \omcv{This step} compare\omcv{s} the number of seeds \omciv{in the Location Buffer} with a lower bound $M$ and an upper bound $N$.
    If a read has \omcvii{fewer} than $M$ matching seeds, it is filtered since it will not meet the minimum chaining score requirement \omciv{of the baseline read mapper}\omcv{~\cite{li2018minimap2}}.\footnote{\omcv{We use $M$=3, as \omcvi{in}~\cite{li2018minimap2}, but \proposal-NM's design trivially supports different values of $M$.}} 
    If the read has more than $N$ seeds, it means that \omcvi{the read} will
    \omcv{very} likely \omcvi{align to the reference (e.g., reads with \omcviii{at least} 64 seeds in \fig{\ref{fig:lr_align}})}
    and \omcvi{thus} require the full read mapping process. \omciv{\proposal-NM} sends such reads to the \omciv{host system for the full read mapping process}. This way, the read mapping process of the unfiltered reads can run concurrently with \proposal's \omciv{filtering} operations. For any other read, \omcv{\proposal-NM} performs chaining in the SSD in Step 3.
    
    \head{Step 3: Chaining-Based Filtering} \omcv{This} step performs chaining \omcv{(see \sect{\ref{sec:background-readmapping}), a step commonly used in state-of-the-art read mappers~\cite{li2018minimap2,dobin2012,li2016minimap}}} to filter out reads with low chaining scores and send reads with high chaining scores to the CPU to undergo the full read mapping process. 
    \omcv{The key difference in \proposal-NM's chaining process compared to existing read mappers is that \proposal-NM \emph{selectively} performs chaining only \omcvi{on} reads with lower seed counts than a threshold $N$. 
    Such selective chaining is based on our two key observations; 1) a long read with many \omcvi{seeds} most likely aligns to the reference genome and \omcvi{thus} does not require in-storage filtering, and 2)~selective chaining can filter many non-aligning long reads, without requiring costly hardware resources in the SSD (\omcvi{as shown in} Section~\ref{sec:lr-overview}).
    
We design \proposal-NM's chaining unit based on the chaining algorithm used in Minimap2~\cite{li2018minimap2}, a state-of-the-art baseline read mapper.   }
A chain is computed from a sequence of seeds $S_1,\ldots, S_N$\footnote{\omcvii{Sorted based on their locations in the reference genome.}} \omciv{of lengths $w_1,\ldots, w_N$, respectively. For a given seed $S_i$, $x_i$ ($y_i$) denotes the ending position of the seed's matching location in the reference genome (the read).}
The chaining score acts as an approximation of an alignment score, \omciv{increasing linearly with the number of matching base pairs between the read and the reference, and decreasing with the size of gaps between the seeds. The chaining score is defined as follows (based on ~\cite{li2018minimap2}):}
\begin{equation}
\label{eq:chain1}
f(i) = \max\left\lbrace\mathop{\max}_{i>j\geq 1}\left\lbrace f(j)+\alpha(j,i)-\beta(j,i)\right\rbrace, w_i\right\rbrace,
\end{equation}
\noindent where $f(i)$ is the best chaining score which can be computed with the seeds $S_1,\ldots, S_i$. Given a chain whose last seed is $S_j$, $\alpha(j,i)$ (also called the \emph{match score}) is the number of new base pairs added to the chain after adding $S_i$. $\beta(j,i)$ (also called the \emph{gap penalty}) subtracts from the chain score based on \omcv{the distance} between $S_i$ and $S_j$ \omcv{in the reference genome}.\footnote{\omcv{See Minimap2~\cite{li2018minimap2} for more details regarding $\alpha(j,i)$ and $\beta(j,i)$.}}

\begin{figure}[h]
        \centering
        \includegraphics[width=.9\linewidth]{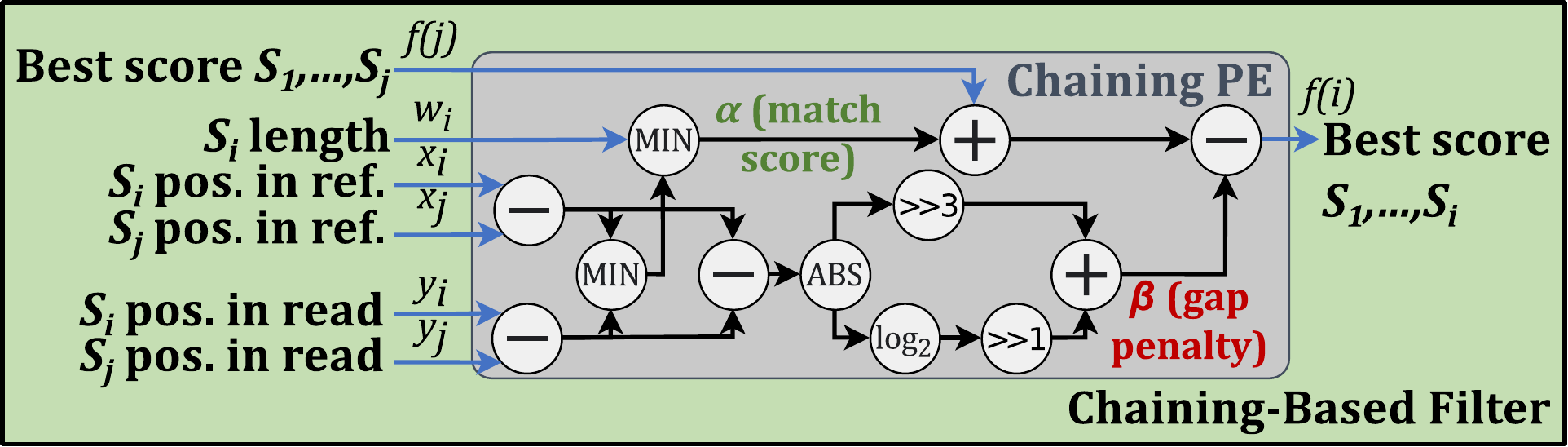}
        \vspace{-.5em}
        \caption{\omcv{Chaining \omcvi{processing element (PE)} in \proposal-NM}.}
        \vspace{-1em}
        \label{fig:chaining_unit}
\end{figure}

\omcv{\proposal-NM's Chaining-Based Filter (in Figure~\ref{fig:lr_overview}) consists of 1)~a~\emph{Chaining Buffer} to store $x_i$, $y_i$, $w_i$, and $f(i)$ values and 2)~a~\emph{Chaining Processing Element (PE)} based on Equation~\eqref{eq:chain1}. } Figure~\ref{fig:chaining_unit} shows the \omcv{design of the Chaining PE} \omciv{in the channel-level accelerator}. We reduce the latency and size of this unit by approximating multiplication with shift operat\omciv{ions}. We ensure that \omcv{our hardware} optimizations always over-estimate the chaining score so that we do not \omciv{filter out} any potential read \omciv{mappings}. \proposal-NM does not affect the accuracy of the read mapper because the filter performs the same computations as the baseline chaining filter for reads with seeds $< N$ inside SSD. 
The chaining \omcv{PE} needs to be executed  $N \times M$ \omciv{times} per read where $N$ is the number of seeds per read and $M$ is the DP-iterations of each seed. Due to the \omcv{limited} number of seeds that go through the chaining step, we limit the number of chaining units that are needed to match the full internal bandwidth of SSD.

Similar to \proposal-EM, the steps of \proposal-NM \omciv{are} pipelined. The performance of \proposal-NM can be easily scaled \omcv{\omcvi{up} by increasing the SSD's internal parallelism (e.g., by deploying more channels or using low-latency NAND flash memory)~\omcvi{\cite{cheong-isscc-2018, park-nvmsa-2018, park-asplos-2021}}}.

\subsection{\proposal Flash Translation Layer (FTL)}\label{sec:ftl}

\proposal requires simple changes to the existing FTL code\omciv{.}

\head{\proposal FTL Metadata} 
\proposal metadata includes the mapping information of the data structures necessary for read mapping acceleration, which enables access to the data structures without the L2P mapping table of the regular FTL. 
In accelerator mode, where \proposal operates as an accelerator, \proposal also keeps \omciv{in internal DRAM} other metadata structures of the regular FTL (e.g., the page status table and block read counts~\cite{park-asplos-2021,cai2017error}) which need to be updated \omciv{during} the filtering process.
We carefully design GenStore to only \emph{sequentially} access the underlying NAND flash chips while operating as an accelerator, so it requires only a small amount of metadata to access the stored data. 

\head{Data Placement} 
\proposal needs to properly \omciv{place its} data structures to enable the full utilization of \omciv{the} internal SSD bandwidth in accelerator mode. 
When \jsiv{each} data \jsiv{structure} is initially written \jsiv{to the SSD}, \proposal sequentially and evenly distribute\jsiv{s} it across NAND flash chips. 
We design \proposal to store \jsiv{every} data structure using multiple sets of NAND flash blocks \jsiv{such that} each block set consists of NAND flash blocks with the same block offset across the planes in a die\jsiv{.
For example,} a die’s block set \jsiv{$k$} includes all NAND flash blocks whose offset is \jsiv{$k$} in the die.
\jsiv{Such data placement} enables \proposal \jsiv{ not only} to always perform \omciv{\emph{multi-plane}} read operations \omciv{during} the filtering process, \jsiv{but also to significantly} reduce the size of \proposal metadata. 
For example, \jsiv{suppose that a \proposal-enabled SSD consists of} 128 2-plane NAND flash dies\jsiv{, and the block size is 12 MB (i.e., the size of each block set} would be 24 MB). 
In such a case, \proposal can specify the physical location of a 30-GB data structure by maintaining only the list of 1,250 (30 GB/24 MB) physical block addresses. 
This way, \proposal significantly reduces the size of the necessary mapping information from 300 MB (\omciv{with conventional} 4-KiB page mapping) to \omciv{only} 5 KB\omciv{ (1,250$\times$4 bytes), and} the saved internal DRAM space can be used for loading data structures in the accelerator mode.

\head{SSD Management Tasks} 
In accelerator mode, \proposal stops working as a regular SSD\omciv{.
It only reads} data structures \omciv{to perform filtering, and does not \jsiv{write} any} \jsiv{new data. 
Therefore, \proposal does not require any} write-related SSD-management tasks such as garbage collection~\cite{tavakkol2018flin,kim2020evanesco,cai2017error,park-dac-2016, park-dac-2019} and wear-leveling~\cite{chang2007efficient,cai2017error,cai-insidessd-2018}.

The other tasks necessary for ensuring data reliability, such as refreshing data to avoid \jsiv{uncorrectable errors due to }\omciv{data} retention and read disturb~\omciv{\cite{cai-hpca-2017, luo2018improving,luo-hpca-2018,cai2015read,cai2013error, cai2012flash, cai2017error, ha2015integrated, cai-insidessd-2018,luo2015warm}}, can be done before or after the filtering process\omciv{,} for two reasons. 
First, \proposal significantly limits the amount of data whose retention age would exceed the manufacturer-specified threshold for reliable operation (e.g., 1 year~\cite{micron3dnandflyer}) during the filtering process, since
\proposals’s filtering process takes a short time.\footnote{\omciv{E.g.,} less than 7 minutes for a 1TiB dataset even with a low-end SSD \omciv{(\ssdl)}\jsiv{~\cite{samsung860pro}}.}
\proposal simply refreshes such data~\cite{cai2013error,luo2018improving, cai2017error,cai2012flash} before starting the filtering process. 
Second, \proposal-FTL can easily avoid read disturbance errors for data with high read counts~\cite{cai2015read, ha2015integrated} since \proposal sequentially reads NAND flash blocks only \emph{once} during filtering.
After the filtering process, to prevent read disturb errors in \omciv{future} filtering processes \omciv{or regular accesses}, \proposal refreshes pages that store a data structure if the structure’s read count exceeds a certain threshold.

\section{Evaluation Methodology}
\label{sec:methodology}

{
\head{Evaluated Systems}
We show the benefits of \proposal when it is integrated with the state-of-the-art software and hardware read mappers.
\vspace{-.2em}
 To this end, we evaluate the following systems:}
 \squishlist
     \item {\textbf{\base:} Minimap2~\cite{li2018minimap2} \omciv{is} a state-of-the-art software read mapper baseline for \jsr{both} short and long reads\omciv{.} GenCache~\cite{nag2019gencache} and Darwin~\cite{turakhia2018darwin} \omciv{are} state-of-the-art hardware read mappers for short and long reads, respectively.}
     \item {\textbf{\gsos:} \revmark{CQ4\\Part 1}
     \base integrated with an implementation of the \proposal filter without in-storage support \omciv{(Ext stands for \emph{external} to storage)}.
     \gsos concurrently filters reads while using \base to perform read mapping for unfiltered reads. 
     The goal of evaluating \gsos is to decouple the effects of \omcvii{\proposals's} two major benefits: 1) alleviating I/O bottlenecks via efficient in-storage processing and 2) reducing the workload of the read mapper by filtering reads with simple operations. \omciv{\gsos obtains the second benefit but not the first.} %
     For software read mappers, we \omciv{evaluate} a pure software implementation of \proposal ~\omciv{that} concurrently run\omcv{s} with \base.\footnote{\revp{We do not evaluate a separate \gsos configuration for long read software mapper since \base (Minimap2~\cite{li2018minimap2}) already incorporates the chaining filter used in \proposal-NM. \proposal-NM implements part of this chaining filter at low cost (enabled by the key observations in Section~4.3) to fit within the constraints of in-storage processing.}} For hardware read mappers, we \omciv{evaluate} a hardware implementation of \proposal \emph{outside} the SSD.}
     \item  {\textbf{\gs:} \base integrated with the {hardware} \proposal filter\omciv{ing accelerators, \proposal-EM and \proposal-NM, }as described in Sections~4.2 and 4.3\omciv{. GS} concurrently filters reads \emph{inside} the SSD while {using} \base to perform read mapping for unfiltered reads.}
 \squishend
 \vspace{-.2em}
\omciv{The source code of \proposal and scripts and datasets can be freely downloaded from \url{https://github.com/CMU-SAFARI/GenStore.}}

\head{Area and Power} 
We \hm{implement}  \rev{\proposal's logic components} in Verilog HDL. We synthesize our designs  using \hm{the} Synopsys Design Compiler~\cite{synopsysdc} \hm{with a} 65nm process {technology node} to estimate latency, area and power consumption.
We use \hm{the} SSD power values \hm{of the} Samsung 3D NAND flash-based SSD~\cite{samsung860pro}, and DRAM power values  based on DDR4 model~\cite{ddr4sheet, ghose2019demystifying}.

\head{Performance Model\omciv{ing}}  
{We evaluate \textbf{hardware} configurations using two state-of-the-art simulators} to analyze the performance of \proposal.
We model DRAM timing with \hm{the} DDR4 \rev{interface~\cite{ghose2019demystifying,ghose2018your}} in Ramulator~\cite{kim2016ramulator, ramulatorsource}, a widely-used, cycle-accurate DRAM simulator.
We model SSD performance using MQSim~\cite{tavakkol2018mqsim}\omcv{, a} widely-used \omcv{simulator for} modern SSDs. 
\revmark{A3}\revp{We {model} the end-to-end throughput of \proposal based on the throughput of each \hm{\proposal pipeline} stage: accessing NAND flash chips, accessing internal DRAM, accelerator \omciv{computation}, and transferring unfiltered data to the host.} \hm{We estimate the performance of GenCache and Darwin accelerators based on the data reported in the original works~\cite{nag2019gencache,turakhia2018darwin}.}

\omciv{\head{Real System Results}} {We use real systems to evaluate \omciv{all} \textbf{software} configurations.} For a given reference genome, separate \omciv{reference} indexes are generated for each sequencing technology using \omciv{the software read mapper's default settings for each technology~\cite{li2018minimap2}}.
All other \omciv{read mapping} parameters are kept at their default values. We perform all experiments on an AMD$^\text{\textregistered}$ EPYC$^\text{\textregistered}$ 7742 CPU with {1TB} DDR4 DRAM (available in user space). We measure power in these systems using AMD$^\text{\textregistered}$ \textmu{}Prof~\cite{microprof}.
We carefully evaluate GenStore’s benefits over the baseline in a conservative manner \hm{by {using optimized configurations} for all baselines}. Our baselines fetch data from the SSD by sequentially reading the data in batch\hm{es}~\cite{li2018minimap2}, \hm{while providing} sufficiently large DRAM \hm{capacity} to contain all data that gets reused \emph{during} {read mapping}. \rev{We \omcv{use} the number of threads that leads to each software configuration's best performance (i.e., execution time):  128 threads for \base and \revp{16 threads for \gsos}  in our experimental setup}.\footnote{\revp{Performance of \gsos saturates after 16 threads since it becomes bottlenecked by the external SSD bandwidth.}}

\rev{
\head{SSD Configurations} We analyze the benefits of \proposal \omciv{ on} three SSD configurations: 
1)~a low-end SSD (\ssdl)~\cite{inteldcs4500} with a SATA3 interface~\cite{SATA}, 
2)~a mid-end SSD (\ssdm)~\cite{samsung980pro} using a PCIe Gen3 M.2 interface~\cite{PCIE}, and
3)~a high-end SSD (\ssdh)~\cite{samsungPM1735} with a PCIe Gen4 interface~\cite{PCIE4}. 
}

\head{Datasets} 
\label{sec:datasets}
For short read experiments, we use the \texttt{hg38} human reference genome~\cite{schneider2017evaluation}. {In Section~\ref{sec:motivation},} \omciv{w}e use real short reads\linebreak
(\texttt{SRR2052419}~\cite{zook2016extensive}, 19.6 GB). 
\rev{{In Section~\ref{sec:results}}, to flexibly \omciv{and controllably} analyze the effect of read sets with different features,} we simulate \rev{read sets with various sizes (up to 440 GB) and different fractions of exactly-matching reads (75\% and 85\%)
} 
using \hm{the} Mason 2 \rev{genomic read simulator~\cite{holtgrewe2010mason}}. 
We generate reads with different exact-match rates by introducing sequence mutations \hm{(uniformly randomly drawn} from the gold\hm{-}standard mutation list for the \hm{human} sample \texttt{NA12878}~\cite{zook2016extensive}\hm{) to the reference genome}.
For the long read experiments, we use the reference genome and read set from  Table~\ref{table:lr-profile} in Section~\ref{sec:lr-overview}. \rev{To flexibly analyze the effect of data size, we generate larger read sets by concatenating the original read sets several times.}

\section{Evaluation}
\label{sec:results}

\subsection{Area \atb{and Power} Analysis}
\label{sec:area}

\omciv{We propose \proposal as an in-storage processing system that supports both accurate short reads and error-prone long reads with different \omcv{levels of} genetic variation \omcvi{between} \omcv{compared \omcvi{genomes}}}.
Table~\ref{tab:area-power} shows the area and power \omcc{consumption} of each logic unit used in \proposal. In the first column, the table shows \omcvii{the} different logic units used in \proposal along with the width of each entry. \omcvii{The s}econd column shows the number of instances of each unit \omciv{in an 8-channel SSD}. We find the number of instances based on the frequency  of and the required throughput from each unit such that \proposal can process data concurrently read from all planes in all SSD channels. 
\omcv{\proposals's hardware units work at different clock frequencies, with the lowest frequency being 300 MHz (for each channel-level Chaining PE). 
While the hardware units can be designed to operate at \omcvi{higher} frequency, their throughput is sufficient when operating at lower frequencies since the end-to-end execution time of \proposal is bottlenecked by \omcvi{NAND} flash reads.} 
The third and fourth columns show the area and power of one instance of each unit.

\begin{table}[h]
        \vspace{-.5em}
        \centering
        \caption{%
        Area and power breakdown of \proposal's logic}
        \vspace{-.5em}
        \resizebox{1.0\columnwidth}{!}{%
        \begin{tabular}{c|c|c|c}
        \toprule
        \textbf{Logic unit} & \textbf{\# \omcvi{of \omcviii{instances}}} & \textbf{Area [mm$^2$]} & \textbf{Power [mW]}\\
        \midrule
        \midrule
        Comparator (64-bit) & \omciv{1 per SSD} &  0.0007 & 0.14\\
        $K$-mer Window ($10\times19\text{-bit}$) & 2 per  channel & 0.0018 & 0.27\\
        \omcviii{Hash} Accelerator (64-bit) & \omciv{2 per SSD} & 0.008 & 1.8\\
        Location Buffer \rev{($64\times64\text{-bit}$)} & 1 per  channel & \rev{0.00725} & \rev{0.37375}\\
        Chaining Buffer ($50\times(16\text{-bit}+64\text{-bit})$) & 1 per  channel & 0.008 & 0.95\\
        Chaining PE & 1 per  channel & 0.004& 0.98\\
        Control & \omciv{1 per SSD} & 0.0002 & 0.11\\
        \midrule
        \textit{Total for an 8-channel SSD} & - & 0.20 & 26.6\\
        \bottomrule
        \end{tabular}}
        \vspace{-.5em}
        \label{tab:area-power}
\end{table}

\omcv{T}he \omciv{total hardware} area \omciv{needed for \proposal} is very small (0.20~mm$^2$ at 65~nm and 0.02~mm$^2$ at 14~nm\hm{,} only 0.006\% of a 14nm Intel Processor~\cite{wikichipcascade}).\footnote{Since lower technology nodes are not available publicly, we scale the area consumption down to {lower} process technolog\omciv{y nodes} using the methodology  in~\cite{stillmaker2017Scaling}.} 
The area overhead of GenStore hardware (0.06~mm$^2$ at 32~nm) is less than 9.5\% of the three 28nm ARM Cortex R4
processors~\omciv{\cite{cortexr4}  \omcv{in a SATA} SSD controller~\cite{samsung860pro}}.

\revp{The area and power\revmark{B2} of most logic units (except for the comparator, the hash64 accelerator, and the control unit) increase linearly with the channel count. Each instance of the comparator and hash64 accelerator supports multiple channels (up to 12 and 4 channels, respectively). Therefore, these units scale by $\lceil\frac{\#channels}{12}\rceil$ or $\lceil\frac{\#channels}{4}\rceil$, respectively. The area and power of \proposal's control unit remains the same across different channel counts.}

\subsection{GenStore-EM Analysis}

We analyze the benefits of \proposal-EM  \rev{for a 22-GB short read set} \revmark{E6}\revp{\omcvii{where} 80\% of  reads exactly match \omcvii{some subsequence in} the reference genome \omciv{(see Section~\ref{sec:methodology} for input generation methodology)},} on a system with three different SSD configurations: \ssdl, \ssdm, \omcv{and} \ssdh.

\begin{figure}[!tbh]
\centering
\vspace{-.5em}
 \includegraphics[width=\linewidth]{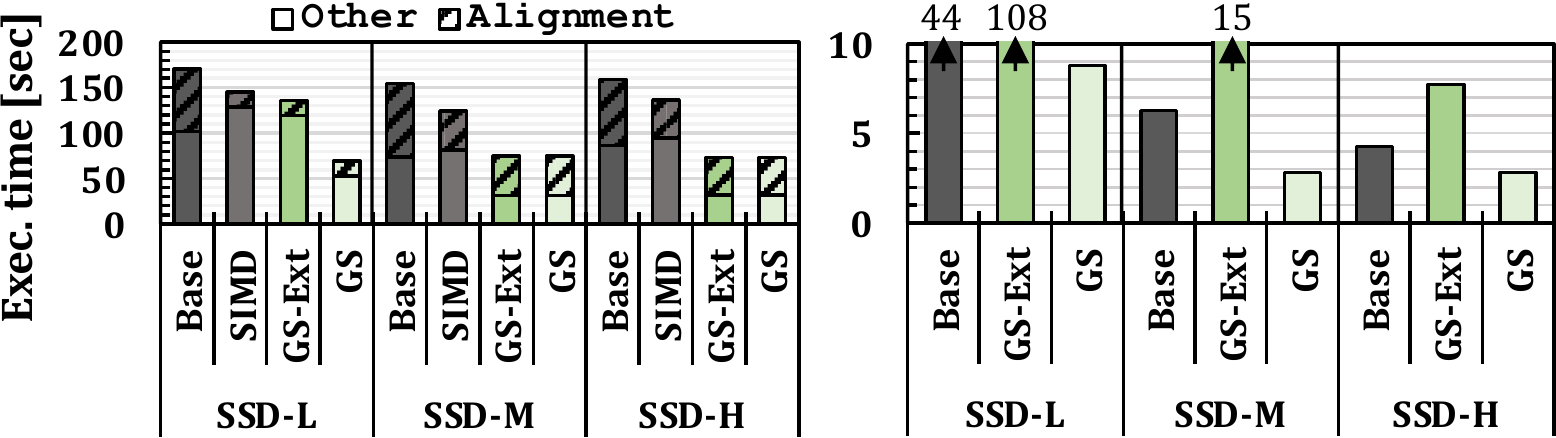}
\begin{subfigure}{.6\linewidth}
  \centering
  \vspace{-.5em}
  \caption{\revp{Software read mapper.\vspace{-1em}}}
  \label{fig:GS-SR-main-SW}
\end{subfigure}\hfill%
\begin{subfigure}{.4\linewidth}
  \centering
  \vspace{-.5em}
  \caption{\rev{Hardware read mapper.\vspace{-1em}}}
  \label{fig:GS-SR-main-HW}
\end{subfigure}%
\caption{\rev{\omciv{\proposal-EM} performance under different SSDs.}}
\vspace{-.5em}
\label{fig:GS-SR-main}
\end{figure}

\head{\rev{Integration with a Software Read Mapper}}
Figure~\ref{fig:GS-SR-main-SW} shows the execution time of \jsr{four read mapper configurations: 
1) } \base (Minimap2~\cite{li2018minimap2}),
2) \simd, an extension of \base with \hm{a} SIMD implementation of a baseline exactly-matching read filter \omciv{using 128-bit SIMD instructions}, 
\rev{3) \gsos}, and
4) \gs. 
\revp{We divide the execution time between \texttt{Alignment} (chaining and alignment's contribution to end-to-end execution time) and \texttt{Other} (file access, seeding, and exact match filtering's contribution to end-to-end execution time).\revmark{B3\\Part 1}}

We make four observations \jsr{from} Figure~\ref{fig:GS-SR-main-SW}. 
\rev{First, for all SSD types\omciv{,} \gs significantly outperforms \base and \simd by $2.07$-$2.45\times$ and $1.66$-$2.09\times$, respectively\hm{, by alleviating} the cost of moving data from the SSD to the host CPU and remov\hm{ing} the burden of finding exactly-matching reads from the rest of the system (DRAM and the processor).
\revp{Second, \revmark{CQ4\\Part3}\gsos provides significant performance improvements over \omciv{both} \base and \simd in \ssdm and \ssdh. However, \gsos provides limited benefits over \simd in \ssdl since its performance is \jsr{bottlenecked} by the low external I/O bandwidth.}
Third, \hm{e}ven though \simd also filters out the same \hm{number} }\rev{of reads and reduces the alignment time similar\omciv{ly to} \gs and \gsos, it\hm{s} average performance benefit over \base ($1.19\times$) \omciv{is} quite limited compared to those of \gs ($2.23\times$) and \gsos ($1.83\times$). This is because %
1)~both \gs and \gsos reduce the number of memory accesses per read and convert the random memory accesses to more efficient streaming accesses, and 2)~\gs addresses the I/O bottlenecks due to }\rev{limited external SSD bandwidth.
Based on our observations, we draw two conclusions.
First, \proposal-EM leads to \omciv{large} performance improvements \omciv{in short read genomic analysis} by efficiently filtering large amounts of data in storage.
\revp{\omciv{Second, \revmark{CQ4\\Part4}even without in-storage processing support, the key idea of \proposal-EM can significantly improve the performance of the state-of-the-art software read mapper especially when using high-bandwidth SSDs (e.g., \ssdm and \ssdh).}}
}

\rev{
\head{Integration with a Hardware Read Mapper} 
Figure~\ref{fig:GS-SR-main-HW} shows the execution time of \rev{three hardware read mapper configurations:
1)} \base (GenCache~\cite{nag2019gencache}),
\rev{2) \gsos, \omcvii{and}}
\jsr{3)} \gs.\footnote{\omciv{We do not show the execution breakdown of mapping in Figure~\ref{fig:GS-SR-main-HW}} since we cannot obtain the execution time breakdown for the hardware read mapper from \cite{nag2019gencache}.}
\revp{Integrating \proposal with \revmark{D2}existing accelerators requires no architectural changes to \proposal and the accelerators because \proposal operates directly on the original data that is stored in the SSD, before any accelerator-specific operation starts.
The only factor that \proposal needs to consider is to generate \omciv{its} outputs (i.e., the information of unfiltered reads) in the format that the accelerator needs as its inputs. The host system is responsible for orchestrating the execution and data flows between \proposal and the accelerator.

}

We make \omciv{two} observations based on Figure ~\ref{fig:GS-SR-main-HW}.
First, \gs significantly outperforms \base by $3.32\times$, $2.55\times$, and $1.52\times$ on systems with \ssdl, \ssdm, and \ssdh, respectively. 
Second, \gsos performs significantly slower than \texttt{Base} ($2.28$-$1.91\times$) on all systems since \proposal-EM requires accessing the large SSIndex data structure (Section~\ref{sec:SRF}), while \gsos suffers from limited SSD external bandwidth.
We conclude that \omciv{\omcvii{the} in-storage processing approach in \proposal effectively addresses the I/O bottleneck, which becomes even more significant with hardware accelerators that address the computation bottleneck.}
}

\begin{figure}[t]
\centering
\includegraphics[width=\linewidth]{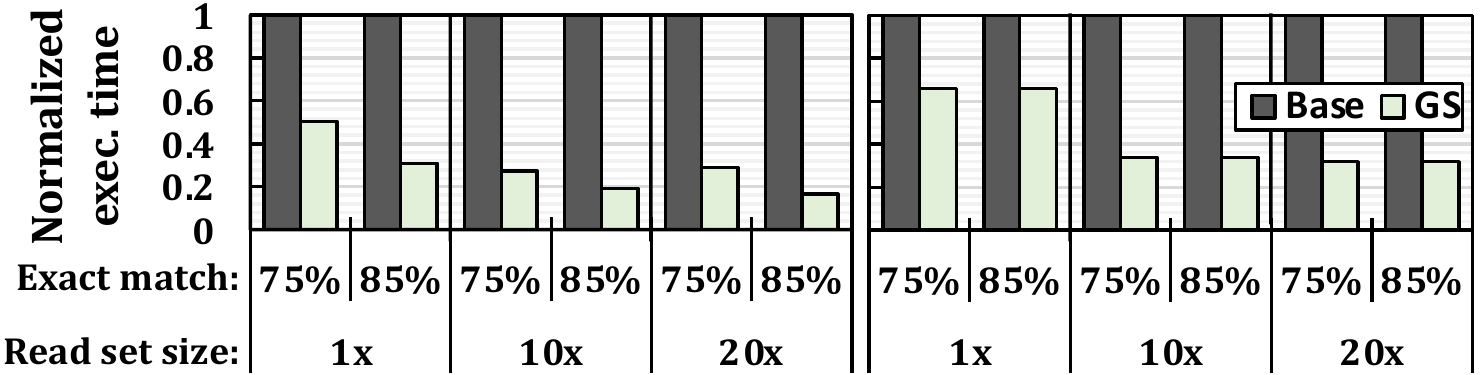}
\begin{subfigure}{.6\linewidth}
  \centering
  \vspace{-.5em}
  \caption{\revp{Software read mapper.\vspace{-1em}}}
  \label{fig:GS-SR-sens-SW}
\end{subfigure}\hfill%
\begin{subfigure}{.4\linewidth}
  \centering
  \vspace{-.5em}
  \caption{\revp{Hardware read mapper.\vspace{-1em}}}
  \label{fig:GS-SR-sens-HW}
\end{subfigure}%
\caption{\revp{\proposal-EM performance versus input size and exact match rate.}}
\vspace{-1em}
\label{fig:GS-SR-sens}
\end{figure}

\revp{\head{Effect of Read Set Features on Performance} \omciv{We study the benefits of \proposal-EM  depending on the characteristics of input read sets. 
To this end, we evaluate \proposal-EM while changing two key \omcv{characteristics}:
1) the input read set \revmark{G1\\Part1}size and 2) exactly-matching read rate. 
These two factors affect the data movement savings ($DM\_Saving$) of GenStore as governed by} Equation~\eqref{eq:dm-saving}:
\begin{equation}
\label{eq:dm-saving}
DM\_Saving = \frac{{Size_{Ref}+Size_{\omciv{ReadSet}}}}{Size_{Ref}+{Size_{\omciv{ReadSet}} \times (1 - Ratio_{Filter})}},
\end{equation}
where $Size_{Ref}$ is the size of the reference genome and its index (e.g., 7 GB for humans~\omciv{\cite{li2018minimap2}}),  $Size_{\omciv{ReadSet}}$ is the size of the read set, and $Ratio_{Filter}$ in \proposal-EM is the exact\omcvii{ly}-match\omcvii{ing read} rate of the read set.

Figure~\ref{fig:GS-SR-sens-SW} shows the execution time of \base and \gs for the \omciv{baseline} software read mapper (Minimap2~\cite{li2018minimap2}) for input read sets with different sizes (1x, 10x, and 20x larger than \omciv{the size of our default} 22-GB short read set) \omciv{\jsr{and} different} exactly-matching read rates (75\% and 85\%) on a system with \ssdh. We make two observations.
First, \gs's performance benefit grows \omcv{as} input size \omcv{increases} (from $2.62\times$ to $4.75\times$) due to \omciv{larger} data movement savings.  Since $Size_{Ref}$ is constant, $DM\_Saving$ and the performance benefits of \proposal-EM increase with larger $Size_{\omciv{ReadSet}}$ (see Equation~\eqref{eq:dm-saving}).
Second, \gs's performance benefit increases with higher exact\jsr{ly}-match\jsr{ing read} rates (from $3.46\times$ to $6.05\times$ for the largest read set) because with \omciv{a larger exactly-matching read rate (i.e., $Ratio_{Filter}$ in Equation~\eqref{eq:dm-saving}), data movement saving ($DM\_Saving$)} increases, and the time spent on mapping the unfiltered reads decreases. Mapping the unfiltered reads is the key contributor to the end-to-end execution \omciv{time} of \gs since it \omciv{has larger} execution time compared to concurrently running \proposal-EM operations in the SSD. \omciv{We conclude that the benefits of \proposal-EM\omcv{, when integrated with the software read mapper,} increases with larger read sets and with \omcv{larger} exactly-matching read rates.}

Figure~\ref{fig:GS-SR-sens-HW} shows the execution time of \base and \gs for \omciv{a baseline} hardware read mapper  (GenCache~\cite{nag2019gencache}). We make two observations.
First, \gs's performance benefit increases with larger input sizes (from $1.52\times$ to $3.13\times$) due to its \omciv{larger data movement reduction}.  
Second,  \gs's performance benefit does \emph{not} increase with higher exact\jsr{ly}-match\jsr{ing read} rates. \jsr{This is because,} with the hardware read mapper, the end-to-end performance of \gs is dominated by the execution time of \proposal-EM's filter operations inside \omciv{the} SSD, which only depends on the input \omcviii{read set} size and not on \omciv{the} exact-match rate. \omciv{We conclude that the benefits of \proposal-EM, \omcv{when integrated} with the hardware read mapper, increases with larger \omcviii{input read set} sizes.}
}

\subsection{GenStore-NM Analysis}

\begin{figure}[h]
\centering
\vspace{-.5em}
\includegraphics[width=\linewidth]{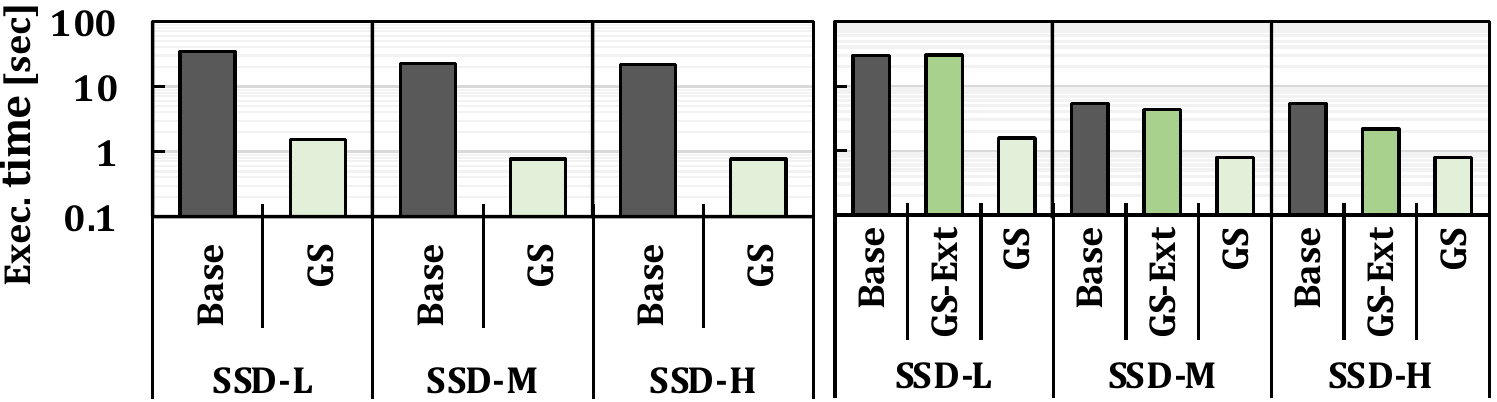}
\begin{subfigure}{.57\linewidth}
  \centering
  \vspace{-.5em}
  \caption{\rev{Software read mapper.\vspace{-1em}}}
  \label{fig:GS-LR-main-SW}
\end{subfigure}\hfill%
\begin{subfigure}{.43\linewidth}
  \centering
  \vspace{-.5em}
  \caption{\rev{Hardware read mapper.\vspace{-1em}}}
  \label{fig:GS-LR-main-HW}
\end{subfigure}%
\caption{\rev{\omciv{\proposal-NM} performance  under different SSDs.}}
 \vspace{-1em}
\label{fig:GS-LR-main}
\end{figure}

\rev{We  analyze the benefits of \proposal-NM  \rev{for a 12.4-GB read set (\omcvi{the first \emph{No reference} use case} in Table~\ref{table:lr-profile})} with 99.65\% of reads \emph{not} aligning on a system with three different SSD configurations: \ssdl, \ssdm, and \ssdh.

\head{Integration with a Software Read Mapper}
Figure~\ref{fig:GS-LR-main-SW} shows the execution time of \jsr{two read mapper configurations: 
1) }\base (Minimap2~\cite{li2018minimap2}), which already incorporates the chaining filter, and 
\jsr{2)} \gs.\footnote{\revp{Unlike \fig{\ref{fig:GS-SR-main-SW}}, we do not divide the execution time\omcv{s of} \texttt{Alignment} and \texttt{Other} in \fig{\ref{fig:GS-LR-main-SW}} since the execution time of \gs is dominated by the filter operations of \proposal-NM. This is because, in this case, a large fraction \omcvi{(e.g., 99.65\% in our evaluated dataset)} of reads get filtered in the SSD and the execution time of mapping the unfiltered reads is very small.}}\revmark{B3\\Part2}
  We observe that \gs  outperforms \base by $22.4\times$, $29.0\times$, and $27.9\times$ on systems with \ssdl, \ssdm, and \ssdh, respectively.\footnote{In this case, \gs provides larger benefits for systems with \ssdm and \ssdh since these SSDs have larger internal bandwidth compared to \ssdl.} The reason is that \gs alleviates the cost of data movement from SSD to \omciv{the processor} and removes the burden of \omciv{mapping a large fraction of} reads from the rest of the system (DRAM and \omciv{the processor}).
}

\rev{
\head{Integration with a Hardware Read Mapper} 
Figure~\ref{fig:GS-LR-main-HW} shows the execution time of \rev{three hardware read mapper configurations:
1)} \base (Darwin~\cite{turakhia2018darwin}), 
\rev{2) \gsos, and}
\jsr{3)} \gs. 
We make two observations based on Figure ~\ref{fig:GS-LR-main-HW}.
First, \gs significantly outperforms \base by $19.2\times$, $6.86\times$, and $6.85\times$ on systems with \ssdl, \ssdm, and \ssdh, respectively. The reason is that \gs alleviates the cost of data movement from SSD to the accelerator and removes the burden of \omciv{read mapping for the filtered reads} from the rest of the system (DRAM and the accelerator).
Second, \gsos does not provide significant performance benefits compared to \base in systems with \ssdl and \ssdm since the execution time of \gsos is dominated by the I/O overhead of bringing reads from SSD to the accelerator. \gsos performs $2.50\times$ faster than \base in systems with \ssdh since the I/O bottlenecks are partially alleviated with \ssdh. However, the benefits of \gsos are limited compared to \gs since \gs significantly alleviates the I/O overhead of bringing reads from SSD to the accelerator with all three SSD configurations.
}

\revp{\head{Effect of Read Set Features on Performance} \omciv{We study how the benefits of \proposal-NM vary depending on the characteristics of input read sets. 
To this end, we evaluate \proposal-NM while changing two key \omcv{characteristics}: 1) the input read set size\revmark{G1\\Part2} and 2) alignment rates (\omcv{\omcvii{the} fraction} of reads \omcv{in} the read set that align to the reference genome). 
We use input sets with different sizes (1$\times$, 10$\times$, and 20$\times$ larger than \omciv{the size of our default 12.5GB}  read set) \omciv{with} different alignment rates ($0.3\%$ and $37\%$, corresponding to \omcvi{the first and second \emph{No reference} use cases} in Table~1), on a system with \ssdh.} 
Figure~\ref{fig:GS-LR-sens} shows the execution time of \base and \gs for the \omciv{baseline} software read mapper~\cite{li2018minimap2} (Figure~\ref{fig:GS-LR-sens-SW}) and the \omciv{baseline} hardware read mapper~\cite{turakhia2018darwin} (Figure~\ref{fig:GS-LR-sens-HW})\omcvi{.}

\begin{figure}[h]
\centering
\vspace{-0.5em}
\includegraphics[width=\linewidth]{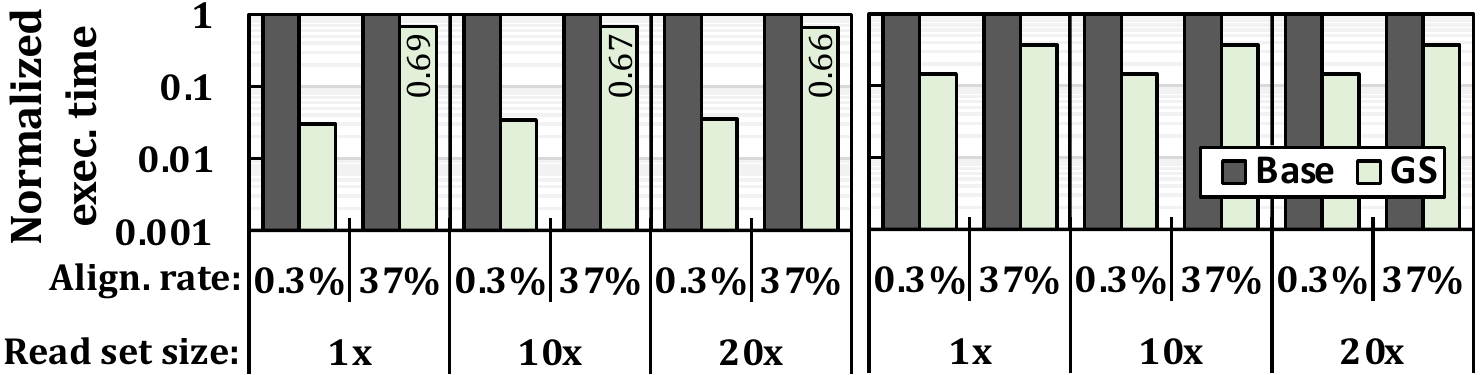}
\begin{subfigure}{.6\linewidth}
  \centering
  \vspace{-1em}
  \caption{\revp{Software read mapper.\vspace{-1em}}}
  \label{fig:GS-LR-sens-SW}
\end{subfigure}\hfill%
\begin{subfigure}{.4\linewidth}
  \centering
  \vspace{-1em}
  \caption{\revp{Hardware read mapper.\vspace{-1em}}}
  \label{fig:GS-LR-sens-HW}
\end{subfigure}%
\vspace{-.5em}
\caption{\revp{\proposal-NM performance versus input size and \omciv{read alignment} rate.}}
\vspace{-1em}
\label{fig:GS-LR-sens}
\end{figure}

\omciv{W}e make two \omcv{main} observations \omcvi{from \fig{\ref{fig:GS-LR-sens}}}.
First, for both hardware and software read mappers, \gs's \omcv{performance benefits vary little as} the input size \omcv{changes.}
The reason is that $Size_{Ref}$ is very small (14.6~MB) for this experiment. Therefore, \omciv{data movement saving ($DM\_Saving$ in Equation~\eqref{eq:dm-saving})} \omcv{mainly} correlates with \omciv{alignment rate (i.e., $1 - Ratio_{Filter}$)}. 
Second, for both software and hardware read mappers, \gs's performance benefit increase\omcv{s as} the ratio of non-aligning reads \omcv{increases (i.e., as alignment rate reduces)}, because with higher values of $Ratio_{Filter}$, both 1) $DM\_Saving$  increases and 2) the time spent on mapping unfiltered reads decreases. 
We conclude the \proposal-NM provides high performance benefits to both software and hardware read mappers with different input sizes and its benefits increase with lower \omciv{alignment} rates.
}

\subsection{Energy \omciv{Analysis}}

To \omcv{demonstrate the}  energy \omcv{benefits} of different \proposal modes, we \omciv{obtain} the energy of the host processor, the host-side DRAM, the DRAM inside the SSD, the communication between the SSD and the host, active and idle energy of the SSD, and the energy of the logic units used in \proposal (Section~\ref{sec:area}). \omcv{We calculate the energy of each component based on its idle and dynamic power consumption and its execution time.}
We observe that by filtering out large amounts of data, \proposal reduces the end-to-end energy consumption of read mapping in all of our evaluations \omciv{compared to \omcv{\base (Minimap2~\cite{li2018minimap2})}}.
We measure the energy consumption for experiments on all SSD configurations. \omciv{By filtering exact matches \omcv{in a short read set with 80\% exactly-matching read rate (see Section~\ref{sec:methodology})}}, \proposal-EM reduces the energy consumption by \omcv{on average (up to)} %
$3.92\times$ ($3.97\times$) \omcvi{across all storage configurations}. 
\omciv{By filtering non-matching reads in a long long read set with \omcvi{a} read alignment rate of 0.35\% (\omcvi{the first \emph{No reference} use case} in Table~\ref{table:lr-profile})}, \proposal-NM reduces the energy consumption by  \omcv{on average (up to)} $27.17\times$ ($29.25\times$)~\omcvi{across all storage configurations}.

\section{Related Work} 
\label{sec_related}

To our knowledge, \proposal is the \emph{first} in-storage \rev{system \omciv{designed for} accelerating genome sequence analysis.} \omcv{\proposal works with both} short and long reads, \omciv{\omcv{and} different \omcv{\omcvii{degrees} of} genetic variation \omcv{between \omcvii{the} compared genomes}}.

\rev{\head{Accelerating Read Mapping}}
There exists a large \omcv{body} of \omcv{work} on accelerating read mapping, which tend to follow two \omcv{general} directions: \omciv{1) non-filtering and 2) filtering approaches}~\cite{alser2020accelerating}. \omciv{\omcviii{Neither} of these two approaches are designed for processing \omcv{in storage}.}
\omciv{\omcv{N}on-filtering approaches} accelerate one or more \omciv{non-filtering} steps of read mapping \omcv{(e.g., seeding and approximate string matching)} using hardware accelerators. 
\omciv{Examples of} these \omcv{accelerators} include processing-in-memory architectures~\cite{huangfu2018radar,khatamifard2021genvom, cali2020genasm, gupta2019rapid,li2021pim,angizi2019aligns,zokaee2018aligner},
\omcv{ASICs}~\cite{turakhia2018darwin, fujiki2018genax, madhavan2014race},
GPUs~\cite{cheng2018bitmapper2,houtgast2018hardware,houtgast2017efficient, zeni2020logan,ahmed2019gasal2,nishimura2017accelerating,de2016cudalign,liu2015gswabe,liu2013cudasw++,liu2009cudasw++,liu2010cudasw++,wilton2015arioc},
\rev{and FPGAs~\cite{goyal2017ultra,chen2016spark,chen2014accelerating,chen2021high,fujiki2020seedex, banerjee2018asap,fei2018fpgasw,waidyasooriya2015hardware,chen2015novel,rucci2018swifold,haghi2021fpga,li2021pipebsw,ham2020genesis,ham2021accelerating,wu2019fpga}}.
\omciv{\omcv{F}iltering approaches} accelerate \omcv{the} \omciv{pre-alignment filtering} step of read mapping. \omcv{These works provide} highly-parallel read filtering heuristics that quickly eliminate dissimilar sequences before invoking computationally-expensive  alignment algorithms. 
\omciv{Examples of} these \omcv{accelerators} include
\omcvii{processing-near-memory architectures~\cite{nag2019gencache,kim20111,singh2021fpga,hameed2021alpha},}
\omcv{FPGAs}~\cite{alser2017gatekeeper, alser2017magnet, alser2019shouji, alser2020sneakysnake,guo2019hardware},
GPUs~\cite{alser2020sneakysnake, bingol2021gatekeeper,guo2019hardware},
or traditional CPU\omcv{-based acceleration}~\cite{xin2015shifted,xin2013accelerating,alser2020sneakysnake}.

In contrast to \proposal, prior works \omcv{on read mapping acceleration} suffer from two key issues.
First, they all \emph{still} need to access the \omciv{genomic} data that is initially \rev{stored} in \rev{the} storage \rev{device}, which \rev{incurs time and energy costs due to} data movement \rev{overhead} from the storage \rev{device} to the main memory and later to the CPU cores
\omc{or accelerators}.
Second, most \omcv{prior} works support processing \emph{only} one type of sequencing 
reads~\cite{turakhia2018darwin,fujiki2018genax,alser2017gatekeeper,bingol2021gatekeeper,khatamifard2021genvom,liu2015gswabe,angizi2019aligns,zokaee2018aligner,cheng2018bitmapper2,houtgast2017efficient,houtgast2018hardware,alser2019shouji,alser2017magnet,ham2020genesis,wilton2015arioc,li2021pipebsw,banerjee2018asap,waidyasooriya2015hardware,chen2021high,liu2013cudasw++,guo2019hardware,xin2015shifted,xin2013accelerating},
which limits their applicability.
\revmark{E1}\revp{Compared to existing filtering approaches, our work 1) is \omc{the} first to identify the I/O inefficiencies that are not addressed by existing techniques and 2) introduces new filtering techniques that enable efficient in-storage processing for both short and long reads.}

\rev{\head{In-Storage Processing Systems}}
Prior works \omcc{explore} in-storage processing \omcc{in the form of application-specific accelerators}~\cite{mailthody2019deepstore,pei2019registor,jun2018grafboost, do2013query, seshadri2014willow,kim2016storage, riedel2001active,riedel1998active}, general-purpose processing inside \omcx{ storage devices}~\cite{gu2016biscuit, kang2013enabling, wang2019project,acharya1998active,keeton1998case,riedel1998active,riedel2001active}, SSD\omciv{s} closely integrated with FPGAs~\cite{jun2015bluedbm, jun2016bluedbm, torabzadehkashi2019catalina, lee2020smartssd, ajdari2019cidr, koo2017summarizer}\omcv{,} or SSD\omciv{s} closely integrated with GPU\omciv{s}~\cite{cho2013xsd}. 
Several works propose techniques for general pattern matching in storage~\cite{jeong2019react, jun2016storage} \emph{without} \omcv{a} specific focus on read mapping \emph{nor} support for different sequencing \omciv{read} types \omciv{and data structures}.
\omcc{None of these works perform} in-storage \omcc{filtering for read mapping to accelerate} \rev{genome sequence analysis}\omciv{.}

\section{Discussion}
\label{sec:discussion}

\omc{
\revmark{CQ3 and B5}As sequencing technologies develop in the future, we expect that \proposal will still play an important role in genomic sequence analysis.
We witness three major trends in sequencing technologies. First, the length and the accuracy of reads are expected to \omcv{increase~\cite{levy2016advancements,hu2021next,alser2020accelerating,morrison2020nanopore,nanopore2020}}. Second, short reads will continue to be widely used due to their \omcv{\emph{very}} high accuracy and low cost~\cite{quail2008large,pacbio2021,quail2012tale}\omcv{.} Third, DNA sequencing machines are increasingly adopting data processing capabilities to perform \omcv{\emph{both}} sequencing and genomic analysis within the same machine~\omcv{\cite{wang2021nanopore,dunn2021,loka2019reliable,zhang2021real,ardui2018single,kovaka2020targeted}}. 

Even with increases in long read accuracy (\emph{trend 1}),  \proposal-NM \omcvi{would} continue to filter large numbers of reads that do \omcvi{\emph{not}} align to \omcv{a} reference genome due to genetic differences between the \omcvi{compared genomes}~\cite{danko2021global,afshinnekoo2015geospatial,hsu2016urban,sherman2019assembly,li2021building,miga2021need}. 
\omcvi{Given that accurate short reads are essential for many processes in genome analysis (\emph{trend 2}), e.g.,}~for polishing~\cite{zhang2020comprehensive,firtina2020apollo,miga2020telomere} and validating~\cite{logsdon2021structure} long read sequences\omcvi{,} \proposal-EM can \omcv{continue to} accelerate short read mapping by filtering \omcv{out} exactly-matching reads.
With the push to provide DNA sequencing and preliminary genetic analysis on more portable integrated devices with limited compute resources and available DRAM capacity~\omcv{\cite{wang2021nanopore,alser2020accelerating,dunn2021,peresini2021nanopore}} (\emph{trend 3}), there will be a growing demand for \omcvi{data movement-minimizing} systems like \proposal to \omcvi{quickly} filter out a large fraction of reads at low cost and \omcv{high efficiency}~\cite{singh2021fpga,dunn2021,bhoyar2021high,ahmed2021pan,kovaka2020targeted,pomerantz2018real}.

With \omcvi{higher availability} of portable genome sequenc\omcvi{er\omcvii{s}}, genomic analyses will be performed more routinely~\cite{sunagawa2015structure,uk10k2015uk10k,danko2021global,alser2020accelerating,bloom2021massively,morrison2020nanopore,pomerantz2018real,clark2019diagnosis,farnaes2018rapid,sweeney2021rapid,alkan2009personalized,flores2013p4,ginsburg2009genomic,chin2011cancer,Ashley2016} \omcv{and will need to provide much faster results}~\cite{alser2020accelerating,singh2021fpga}. 
\omcv{I}n-storage processing is crucial for meeting the huge demands \omcv{for faster analyses} that scale well to large numbers of \omcvi{genomic samples}. 
Examples of these analyses include 
gene detection~\cite{lax2014longitudinal}, alignment of reads from rapidly mutating organisms such as SARS-CoV-2~\cite{bhoyar2021high,bloom2021massively}, and studies of as-of-yet undiscovered microbes~\cite{danko2021global,afshinnekoo2015geospatial,hsu2016urban, sunagawa2015structure}. 
In these analyses, \proposal-NM can effectively filter out $>$99.7\%~\cite{lax2014longitudinal,hsu2016urban}, $\sim$36.1\%~\cite{bhoyar2021high}, and $\sim$47.3\%~\cite{danko2021global,afshinnekoo2015geospatial,hsu2016urban} of the reads, respectively, \omcv{th\omcvi{ereby} greatly improving the end-to-end throughput \omcvi{and energy efficiency} of genome sequence analysis}.
We hope that our proposed techniques provide a foundation for future efforts \omcv{to} accelerat\omcv{e} genome analysis.
}

\section{Conclusion}
\label{sec:conclusion}

We propose \proposal, a \omcv{new} in-storage \rev{processing system for genome sequence analysis}. \omcv{\proposal} can be integrated with \omcv{both} hardware and software read mappers to improve the end-to-end performance of \omcv{both} short \omcc{and} long read mapping.
We address the challenges of in-storage processing for genomic \omciv{read} filter\omciv{ing via}   
\omcc{new} hardware/software co-design\omciv{ed techniques and} develop \omciv{new} in-storage filtering accelerators for both \omciv{short and long} reads. 
\damlaa{\omciv{Our} evaluat\omciv{ions show that}}
\proposal provides \omcc{large performance and energy} 
improvements when integrated \omcc{in}to the state-of-the-art software and hardware \omciv{read mappers}. 
\omcc{We hope that \proposal inspires \omciv{other in-storage \omcv{processing and storage system design} ideas for genome sequence analysis. \omcv{A promising avenue for future work is to} enable \omcv{the} integration \omcv{of \proposal-like accelerators} into sequencing devices for extreme scalability and efficiency.}}

\section*{Acknowledgments}

We thank the anonymous reviewers of MICRO 2021 and ASPLOS 2022 for feedback. We thank the SAFARI group members for feedback and the stimulating intellectual environment. We acknowledge the generous gifts \omcviii{and support} provided by our industrial partners: Google, Huawei, Intel, Microsoft, VMware\omcv{, and the Semiconductor Research Corporation. This research was partially supported by the ETH Future Computing Laboratory. \omcvii{\omcviii{Jisung Park}
was in part supported by the National Research Foundation (NRF) of Korea (NRF-2020R1A6A3A03040573).}}

\balance
\begingroup
\let\clearpage\relax
\bibliographystyle{unsrtlim}
\bibliography{sample-base}

\begin{thebibliography}{100}

\bibitem{clark2019diagnosis}
Michelle~M Clark, Amber Hildreth, Sergey Batalov, Yan Ding, Shimul Chowdhury,
  et~al.
\newblock {Diagnosis of Genetic Diseases in Seriously Ill Children by Rapid
  Whole-genome Sequencing and Automated Phenotyping and Interpretation}.
\newblock {\em Science Translational Medicine}, 2019.

\bibitem{farnaes2018rapid}
Lauge Farnaes, Amber Hildreth, Nathaly~M Sweeney, Michelle~M Clark, Shimul
  Chowdhury, et~al.
\newblock {Rapid Whole-genome Sequencing Decreases Infant Morbidity and Cost of
  Hospitalization}.
\newblock {\em NPJ Genomic Medicine}, 2018.

\bibitem{sweeney2021rapid}
Nathaly~M Sweeney, Shareef~A Nahas, Shimul Chowdhury, Sergey Batalov, Michelle
  Clark, et~al.
\newblock {Rapid Whole Genome Sequencing Impacts Care and Resource Utilization
  in Infants with Congenital Heart Disease}.
\newblock {\em NPJ Genomic Medicine}, 2021.

\bibitem{alkan2009personalized}
Can Alkan, Jeffrey~M Kidd, Tomas Marques-Bonet, Gozde Aksay, Francesca
  Antonacci, et~al.
\newblock {Personalized Copy Number and Segmental Duplication Maps Using
  Next-Generation Sequencing}.
\newblock {\em Nature Genetics}, 2009.

\bibitem{flores2013p4}
Mauricio Flores, Gustavo Glusman, Kristin Brogaard, Nathan~D Price, and Leroy
  Hood.
\newblock {P4 Medicine: How Systems Medicine Will Transform the Healthcare
  Sector and Society}.
\newblock {\em Personalized Medicine}, 2013.

\bibitem{ginsburg2009genomic}
Geoffrey~S Ginsburg and Huntington~F Willard.
\newblock {Genomic and Personalized Medicine: Foundations and Applications}.
\newblock {\em Translational Research}, 2009.

\bibitem{chin2011cancer}
Lynda Chin, Jannik~N Andersen, and P~Andrew Futreal.
\newblock {Cancer Genomics: From Discovery Science to Personalized Medicine}.
\newblock {\em Nature Medicine}, 2011.

\bibitem{Ashley2016}
Euan~A Ashley.
\newblock {Towards Precision Medicine}.
\newblock {\em Nature Reviews Genetics}, 2016.

\bibitem{bloom2021massively}
Joshua~S Bloom, Laila Sathe, Chetan Munugala, Eric~M Jones, Molly Gasperini,
  et~al.
\newblock {Massively Scaled-up Testing for SARS-CoV-2 RNA via Next-generation
  Sequencing of Pooled and Barcoded Nasal and Saliva Samples}.
\newblock {\em Nature Biomedical Engineering}, 2021.

\bibitem{yelagandula2021multiplexed}
Ramesh Yelagandula, Aleksandr Bykov, Alexander Vogt, Robert Heinen, Ezgi
  {\"O}zkan, et~al.
\newblock {Multiplexed Detection of SARS-CoV-2 and Other Respiratory Infections
  in High Throughput by SARSeq}.
\newblock {\em Nature Communications}, 2021.

\bibitem{le2013selected}
Vien Thi~Minh Le and Binh~An Diep.
\newblock {Selected Insights from Application of Whole Genome Sequencing for
  Outbreak Investigations}.
\newblock {\em Current Opinion in Critical Care}, 2013.

\bibitem{nikolayevskyy2016whole}
Vlad Nikolayevskyy, Katharina Kranzer, Stefan Niemann, and Francis Drobniewski.
\newblock {Whole Genome Sequencing of Mycobacterium Tuberculosis for Detection
  of Recent Transmission and Tracing Outbreaks: A Systematic Review}.
\newblock {\em Tuberculosis}, 2016.

\bibitem{qiu2015whole}
Shaofu Qiu, Peng Li, Hongbo Liu, Yong Wang, Nan Liu, et~al.
\newblock {Whole-genome Sequencing for Tracing the Transmission Link between
  Two ARD Outbreaks Caused by A Novel HAdV Serotype 7 Variant, China}.
\newblock {\em Scientific Reports}, 2015.

\bibitem{gilchrist2015whole}
Carol~A Gilchrist, Stephen~D Turner, Margaret~F Riley, William~A Petri, and
  Erik~L Hewlett.
\newblock {Whole-genome Sequencing in Outbreak Analysis}.
\newblock {\em Clinical Microbiology Reviews}, 2015.

\bibitem{hoban2016finding}
Sean Hoban, Joanna~L Kelley, Katie~E Lotterhos, Michael~F Antolin, Gideon
  Bradburd, et~al.
\newblock {Finding the Genomic Basis of Local Adaptation: Pitfalls, Practical
  Solutions, and Future Directions}.
\newblock {\em The American Naturalist}, 2016.

\bibitem{romiguier2014comparative}
J~Romiguier, Philippe Gayral, Marion Ballenghien, Arnaud Bernard, Vincent
  Cahais, et~al.
\newblock {Comparative Population Genomics in Animals Uncovers the Determinants
  of Genetic Diversity}.
\newblock {\em Nature}, 2014.

\bibitem{ellegren2016determinants}
Hans Ellegren and Nicolas Galtier.
\newblock {Determinants of Genetic Diversity}.
\newblock {\em Nature Reviews Genetics}, 2016.

\bibitem{prohaska2019human}
Ana Prohaska, Fernando Racimo, Andrew~J Schork, Martin Sikora, Aaron~J Stern,
  et~al.
\newblock {Human Disease Variation in the Light of Population Genomics}.
\newblock {\em Cell}, 2019.

\bibitem{ellegren2014genome}
Hans Ellegren.
\newblock {Genome Sequencing and Population Genomics in Non-Model Organisms}.
\newblock {\em Trends in Ecology \& Evolution}, 2014.

\bibitem{Prado-Martinez2013}
Javier Prado-Martinez, Peter~H. Sudmant, Jeffrey~M. Kidd, Heng Li, Joanna~L.
  Kelley, et~al.
\newblock {Great Ape Genetic Diversity and Population History}.
\newblock {\em Nature}, 2013.

\bibitem{Prohaska2019}
Ana Prohaska, Fernando Racimo, Andrew~J Schork, Martin Sikora, Aaron~J Stern,
  et~al.
\newblock {Human Disease Variation in the Light of Population Genomics}.
\newblock {\em Cell}, 2019.

\bibitem{reuter2015high}
Jason~A Reuter, Damek~V Spacek, and Michael~P Snyder.
\newblock {High-Throughput Sequencing Technologies}.
\newblock {\em Molecular Cell}, 2015.

\bibitem{van2014ten}
Erwin~L van Dijk, H{\'e}l{\`e}ne Auger, Yan Jaszczyszyn, and Claude Thermes.
\newblock {Ten Years of Next-Generation Sequencing Technology}.
\newblock {\em {Trends in Genetics}}, 2014.

\bibitem{glenn2011field}
Travis~C Glenn.
\newblock {Field Guide to Next-Generation DNA Sequencers}.
\newblock {\em Molecular Ecology Resources}, 2011.

\bibitem{goodwin2016coming}
Sara Goodwin, John~D McPherson, and W~Richard McCombie.
\newblock {Coming of Age: Ten Years of Next-generation Sequencing
  Technologies}.
\newblock {\em Nature Reviews Genetics}, 2016.

\bibitem{quail2012tale}
Michael~A Quail, Miriam Smith, Paul Coupland, Thomas~D Otto, Simon~R Harris,
  et~al.
\newblock {A Tale of Three Next Generation Sequencing Platforms: Comparison of
  Ion Torrent, Pacific Biosciences and Illumina MiSeq Sequencers}.
\newblock {\em BMC Genomics}, 2012.

\bibitem{kchouk2017generations}
Mehdi Kchouk, Jean-Francois Gibrat, and Mourad Elloumi.
\newblock {Generations of Sequencing Technologies: from First to Next
  Generation}.
\newblock {\em Biology and Medicine}, 2017.

\bibitem{pfeiffer2018systematic}
Franziska Pfeiffer, Carsten Gr{\"o}ber, Michael Blank, Kristian H{\"a}ndler,
  Marc Beyer, et~al.
\newblock {Systematic Evaluation of Error Rates and Causes in Short Samples in
  Next-generation Sequencing}.
\newblock {\em Scientific Reports}, 2018.

\bibitem{amarasinghe2020opportunities}
Shanika~L Amarasinghe, Shian Su, Xueyi Dong, Luke Zappia, Matthew~E Ritchie,
  and Quentin Gouil.
\newblock {Opportunities and Challenges in Long-read Sequencing Data Analysis}.
\newblock {\em Genome Biology}, 2020.

\bibitem{cali2017nanopore}
Damla Senol~Cali, Jeremie~S Kim, Saugata Ghose, Can Alkan, and Onur Mutlu.
\newblock {Nanopore Sequencing Technology and Tools for Genome Assembly:
  Computational Analysis of the Current State, Bottlenecks and Future
  Directions}.
\newblock {\em Briefings in Bioinformatics}, 2018.

\bibitem{ardui2018single}
Simon Ardui, Adam Ameur, Joris~R Vermeesch, and Matthew~S Hestand.
\newblock {Single Molecule Real-Time (SMRT) Sequencing Comes of Age:
  Applications and Utilities for Medical Diagnostics}.
\newblock {\em Nucleic Acids Research}, 2018.

\bibitem{weirather2017comprehensive}
Jason~L Weirather, Mariateresa de~Cesare, Yunhao Wang, Paolo Piazza, Vittorio
  Sebastiano, et~al.
\newblock {Comprehensive Comparison of Pacific Biosciences and Oxford Nanopore
  Technologies and Their Applications to Transcriptome Analysis}.
\newblock {\em F1000Research}, 2017.

\bibitem{van2018third}
Erwin~L van Dijk, Yan Jaszczyszyn, Delphine Naquin, and Claude Thermes.
\newblock {The Third Revolution in Sequencing Technology}.
\newblock {\em Trends in Genetics}, 2018.

\bibitem{wang2021nanopore}
Yunhao Wang, Yue Zhao, Audrey Bollas, Yuru Wang, and Kin~Fai Au.
\newblock {Nanopore Sequencing Technology, Bioinformatics and Applications}.
\newblock {\em Nature Biotechnology}, 2021.

\bibitem{alser2020accelerating}
Mohammed Alser, Z{\"u}lal Bing{\"o}l, Damla~Senol Cali, Jeremie Kim, Saugata
  Ghose, et~al.
\newblock {Accelerating Genome Analysis: A Primer on An Ongoing Journey}.
\newblock {\em IEEE Micro}, 2020.

\bibitem{alser2020technology}
Mohammed Alser, Jeremy Rotman, Kodi Taraszka, Huwenbo Shi, Pelin~Icer Baykal,
  et~al.
\newblock {Technology Dictates Algorithms: Recent Developments in Read
  Alignment}.
\newblock {\em Genome Biology}, 2021.

\bibitem{xin2013accelerating}
Hongyi Xin, Donghyuk Lee, Farhad Hormozdiari, Samihan Yedkar, Onur Mutlu, and
  Can Alkan.
\newblock {Accelerating Read Mapping with FastHASH}.
\newblock {\em BMC Genomics}, 2013.

\bibitem{xin2015shifted}
Hongyi Xin, John Greth, John Emmons, Gennady Pekhimenko, Carl Kingsford, et~al.
\newblock {Shifted Hamming Distance: A Fast and Accurate SIMD-friendly Filter
  to Accelerate Alignment Verification in Read Mapping}.
\newblock {\em Bioinformatics}, 2015.

\bibitem{turakhia2018darwin}
Yatish Turakhia, Gill Bejerano, and William~J Dally.
\newblock {Darwin: A Genomics Co-processor Provides up to 15,000 x Acceleration
  on Long Read Assembly}.
\newblock In {\em ASPLOS}, 2018.

\bibitem{cali2020genasm}
Damla Senol~Cali, Gupreet Kalsi, Zulal Bing{\"o}l, Lavanya Subramanian, Can
  Firtina, et~al.
\newblock {GenASM: A High-Performance, Low-Power Approximate String Matching
  Acceleration Framework for Genome Sequence Analysis}.
\newblock In {\em MICRO}, 2020.

\bibitem{alser2020sneakysnake}
Mohammed Alser, Taha Shahroodi, Juan G{\'o}mez-Luna, Can Alkan, and Onur Mutlu.
\newblock {SneakySnake: A Fast and Accurate Universal Genome Pre-alignment
  Filter for CPUs, GPUs and FPGAs}.
\newblock {\em Bioinformatics}, 2020.

\bibitem{alser2017gatekeeper}
Mohammed Alser, Hasan Hassan, Hongyi Xin, O{\u{g}}uz Ergin, Onur Mutlu, and Can
  Alkan.
\newblock {GateKeeper: A New Hardware Architecture for Accelerating
  Pre-alignment in DNA Short Read Mapping}.
\newblock {\em Bioinformatics}, 2017.

\bibitem{nag2019gencache}
Anirban Nag, CN~Ramachandra, Rajeev Balasubramonian, Ryan Stutsman, Edouard
  Giacomin, et~al.
\newblock {GenCache: Leveraging In-cache Operators for Efficient Sequence
  Alignment}.
\newblock In {\em MICRO}, 2019.

\bibitem{kim2019airlift}
Jeremie~S Kim, Can Firtina, Meryem~Banu Cavlak, Damla~Senol Cali, Mohammed
  Alser, et~al.
\newblock {AirLift: A Fast and Comprehensive Technique for Remapping Alignments
  between Reference Genomes}.
\newblock {\em arXiv}, 2019.

\bibitem{kim2018grim}
Jeremie~S Kim, Damla~Senol Cali, Hongyi Xin, Donghyuk Lee, Saugata Ghose,
  et~al.
\newblock {GRIM-Filter: Fast Seed Location Filtering in DNA Read Mapping Using
  Processing-in-memory Technologies}.
\newblock {\em BMC Genomics}, 2018.

\bibitem{firtina2020apollo}
Can Firtina, Jeremie~S Kim, Mohammed Alser, Damla Senol~Cali, A~Ercument Cicek,
  et~al.
\newblock {Apollo: A Sequencing-technology-independent, Scalable and Accurate
  Assembly Polishing Algorithm}.
\newblock {\em Bioinformatics}, 2020.

\bibitem{vsovsic2017edlib}
Martin {\v{S}}o{\v{s}}i{\'c} and Mile {\v{S}}iki{\'c}.
\newblock {Edlib: A C/C++ Library for Fast, Exact Sequence Alignment Using Edit
  Distance}.
\newblock {\em Bioinformatics}, 2017.

\bibitem{alser2017magnet}
Mohammed Alser, Onur Mutlu, and Can Alkan.
\newblock {MAGNET: Understanding and Improving the Accuracy of Genome
  Pre-alignment Filtering}.
\newblock {\em arXiv}, 2017.

\bibitem{alser2019shouji}
Mohammed Alser, Hasan Hassan, Akash Kumar, Onur Mutlu, and Can Alkan.
\newblock {Shouji: A Fast and Efficient Pre-alignment Filter for Sequence
  Alignment}.
\newblock {\em Bioinformatics}, 2019.

\bibitem{needleman1970general}
Saul~B Needleman and Christian~D Wunsch.
\newblock {A General Method Applicable to the Search for Similarities in the
  Amino Acid Sequence of Two Proteins}.
\newblock {\em Journal of Molecular Biology}, 1970.

\bibitem{smith1981identification}
Temple~F Smith, Michael~S Waterman, et~al.
\newblock {Identification of Common Molecular Subsequences}.
\newblock {\em Journal of Molecular Biology}, 1981.

\bibitem{gotoh1982improved}
Osamu Gotoh.
\newblock {An Improved Algorithm for Matching Biological Sequences}.
\newblock {\em Journal of Molecular Biology}, 1982.

\bibitem{levy2016advancements}
Shawn~E Levy and Richard~M Myers.
\newblock {Advancements in Next-generation Sequencing}.
\newblock {\em Annual Review of Genomics and Human Genetics}, 2016.

\bibitem{hu2021next}
Taishan Hu, Nilesh Chitnis, Dimitri Monos, and Anh Dinh.
\newblock {Next-Generation Sequencing Technologies: An Overview}.
\newblock {\em Human Immunology}, 2021.

\bibitem{10002015global}
1000 Genomes~Project Consortium et~al.
\newblock {A Global Reference for Human Genetic Variation}.
\newblock {\em Nature}, 2015.

\bibitem{zhang2000greedy}
Zheng Zhang, Scott Schwartz, Lukas Wagner, and Webb Miller.
\newblock {A Greedy Algorithm for Aligning DNA Sequences}.
\newblock {\em Journal of Computational Biology}, 2000.

\bibitem{slater2005automated}
Guy St~C Slater and Ewan Birney.
\newblock {Automated Generation of Heuristics for Biological Sequence
  Comparison}.
\newblock {\em BMC Bioinformatics}, 2005.

\bibitem{li2018minimap2}
Heng Li.
\newblock {Minimap2: Pairwise Alignment for Nucleotide Sequences}.
\newblock {\em Bioinformatics}, 2018.

\bibitem{myers1999fast}
Gene Myers.
\newblock {A Fast Bit-vector Algorithm for Approximate String Matching Based on
  Dynamic Programming}.
\newblock {\em JACM}, 1999.

\bibitem{marco2021fast}
Santiago Marco-Sola, Juan~Carlos Moure, Miquel Moreto, and Antonio Espinosa.
\newblock {Fast Gap-affine Pairwise Alignment Using the Wavefront Algorithm}.
\newblock {\em Bioinformatics}, 2021.

\bibitem{huangfu2018radar}
Wenqin Huangfu, Shuangchen Li, Xing Hu, and Yuan Xie.
\newblock {RADAR: A 3D-ReRAM based DNA Alignment Accelerator Architecture}.
\newblock In {\em DAC}, 2018.

\bibitem{fujiki2018genax}
Daichi Fujiki, Arun Subramaniyan, Tianjun Zhang, Yu~Zeng, Reetuparna Das,
  et~al.
\newblock {Genax: A Genome Sequencing Accelerator}.
\newblock In {\em ISCA}, 2018.

\bibitem{fujiki2020seedex}
Daichi Fujiki, Shunhao Wu, Nathan Ozog, Kush Goliya, David Blaauw, et~al.
\newblock {SeedEx: A Genome Sequencing Accelerator for Optimal Alignments in
  Subminimal Space}.
\newblock In {\em MICRO}, 2020.

\bibitem{banerjee2018asap}
Subho~Sankar Banerjee, Mohamed El-Hadedy, Jong~Bin Lim, Zbigniew~T Kalbarczyk,
  Deming Chen, et~al.
\newblock {ASAP: Accelerated Short-read Alignment on Programmable Hardware}.
\newblock {\em IEEE Transactions on Computers}, 2019.

\bibitem{khatamifard2021genvom}
S~Karen Khatamifard, Zamshed Chowdhury, Nakul Pande, Meisam Razaviyayn, Chris~H
  Kim, and Ulya~R Karpuzcu.
\newblock {GeNVoM: Read Mapping Near Non-Volatile Memory}.
\newblock {\em TCBB}, 2021.

\bibitem{gupta2019rapid}
Saransh Gupta, Mohsen Imani, Behnam Khaleghi, Venkatesh Kumar, and Tajana
  Rosing.
\newblock {RAPID: A ReRAM Processing In-memory Architecture for DNA Sequence
  Alignment}.
\newblock In {\em ISLPED}, 2019.

\bibitem{li2021pim}
Xue-Qi Li, Guang-Ming Tan, and Ning-Hui Sun.
\newblock {PIM-Align: A Processing-in-Memory Architecture for FM-Index Search
  Algorithm}.
\newblock {\em Journal of Computer Science and Technology}, 2021.

\bibitem{angizi2019aligns}
Shaahin Angizi, Jiao Sun, Wei Zhang, and Deliang Fan.
\newblock {Aligns: A Processing-in-memory Accelerator for DNA Short Read
  Alignment Leveraging SOT-MRAM}.
\newblock In {\em DAC}, 2019.

\bibitem{zokaee2018aligner}
Farzaneh Zokaee, Hamid~R Zarandi, and Lei Jiang.
\newblock {Aligner: A Process-in-memory Architecture for Short Read Alignment
  in ReRAMs}.
\newblock {\em IEEE Computer Architecture Letters}, 2018.

\bibitem{madhavan2014race}
Advait Madhavan, Timothy Sherwood, and Dmitri Strukov.
\newblock {Race Logic: A Hardware Acceleration for Dynamic Programming
  Algorithms}.
\newblock {\em ACM SIGARCH Computer Architecture News}, 2014.

\bibitem{cheng2018bitmapper2}
Haoyu Cheng, Yong Zhang, and Yun Xu.
\newblock {Bitmapper2: A GPU-accelerated All-mapper Based on The Sparse Q-gram
  Index}.
\newblock {\em TCBB}, 2018.

\bibitem{houtgast2018hardware}
Ernst~Joachim Houtgast, Vlad-Mihai Sima, Koen Bertels, and Zaid Al-Ars.
\newblock {Hardware Acceleration of BWA-MEM Genomic Short Read Mapping for
  Longer Read Lengths}.
\newblock {\em Computational Biology and Chemistry}, 2018.

\bibitem{houtgast2017efficient}
Ernst~Joachim Houtgast, VladMihai Sima, Koen Bertels, and Zaid AlArs.
\newblock {An Efficient GPU-accelerated Implementation of Genomic Short Read
  Mapping with BWA-MEM}.
\newblock {\em ACM SIGARCH Computer Architecture News}, 2017.

\bibitem{goyal2017ultra}
Amit Goyal, Hyuk~Jung Kwon, Kichan Lee, Reena Garg, Seon~Young Yun, et~al.
\newblock {Ultra-fast Next Generation Human Genome Sequencing Data Processing
  Using DRAGENTM Bio-IT Processor for Precision Medicine}.
\newblock {\em Open Journal of Genetics}, 2017.

\bibitem{chen2016spark}
Yu-Ting Chen, Jason Cong, Zhenman Fang, Jie Lei, and Peng Wei.
\newblock {When Spark Meets FPGAs: A Case Study for Next-Generation DNA
  Sequencing Acceleration}.
\newblock In {\em USENIX HotCloud}, 2016.

\bibitem{chen2014accelerating}
Peng Chen, Chao Wang, Xi~Li, and Xuehai Zhou.
\newblock {Accelerating the Next Generation Long Read Mapping with the
  FPGA-based System}.
\newblock {\em TCBB}, 2014.

\bibitem{chen2021high}
Yen-Lung Chen, Bo-Yi Chang, Chia-Hsiang Yang, and Tzi-Dar Chiueh.
\newblock {A High-Throughput FPGA Accelerator for Short-Read Mapping of the
  Whole Human Genome}.
\newblock {\em IEEE TPDS}, 2021.

\bibitem{zeni2020logan}
Alberto Zeni, Giulia Guidi, Marquita Ellis, Nan Ding, Marco~D Santambrogio,
  et~al.
\newblock {Logan: High-performance GPU-based X-drop Long-read Alignment}.
\newblock In {\em IPDPS}, 2020.

\bibitem{ahmed2019gasal2}
Nauman Ahmed, Jonathan L{\'e}vy, Shanshan Ren, Hamid Mushtaq, Koen Bertels, and
  Zaid Al-Ars.
\newblock {GASAL2: A GPU Accelerated Sequence Alignment Library for
  High-Throughput NGS Data}.
\newblock {\em BMC Bioinformatics}, 2019.

\bibitem{nishimura2017accelerating}
Takahiro Nishimura, Jacir~L Bordim, Yasuaki Ito, and Koji Nakano.
\newblock {Accelerating the Smith-waterman Algorithm Using Bitwise Parallel
  Bulk Computation Technique on GPU}.
\newblock In {\em IPDPSW)}, 2017.

\bibitem{de2016cudalign}
Edans~Flavius de~Oliveira~Sandes, Guillermo Miranda, Xavier Martorell, Eduard
  Ayguade, George Teodoro, and Alba Cristina~Magalhaes Melo.
\newblock {CUDAlign 4.0: Incremental Speculative Traceback for Exact
  Chromosome-wide Alignment in GPU Clusters}.
\newblock {\em IEEE TPDS}, 2016.

\bibitem{liu2015gswabe}
Yongchao Liu and Bertil Schmidt.
\newblock {GSWABE: Faster GPU-accelerated Sequence Alignment with Optimal
  Alignment Retrieval for Short DNA Sequences}.
\newblock {\em Concurrency and Computation: Practice and Experience}, 2015.

\bibitem{liu2013cudasw++}
Yongchao Liu, Adrianto Wirawan, and Bertil Schmidt.
\newblock {CUDASW++ 3.0: Accelerating Smith-Waterman Protein Database Search by
  Coupling CPU and GPU SIMD Instructions}.
\newblock {\em BMC Bioinformatics}, 2013.

\bibitem{wilton2015arioc}
Richard Wilton, Tamas Budavari, Ben Langmead, Sarah~J Wheelan, Steven~L
  Salzberg, and Alexander~S Szalay.
\newblock {Arioc: High-throughput Read Alignment with GPU-accelerated
  Exploration of The Seed-and-extend Search Space}.
\newblock {\em PeerJ}, 2015.

\bibitem{fei2018fpgasw}
Xia Fei, Zou Dan, Lu~Lina, Man Xin, and Zhang Chunlei.
\newblock {FPGASW: Accelerating Large-scale Smith--Waterman Sequence Alignment
  Application with Backtracking on FPGA Linear Systolic Array}.
\newblock {\em Interdisciplinary Sciences: Computational Life Sciences}, 2018.

\bibitem{waidyasooriya2015hardware}
Hasitha~Muthumala Waidyasooriya and Masanori Hariyama.
\newblock {Hardware-acceleration of Short-read Alignment Based on the
  Burrows-wheeler Transform}.
\newblock {\em TPDS}, 2015.

\bibitem{chen2015novel}
Yu-Ting Chen, Jason Cong, Jie Lei, and Peng Wei.
\newblock {A Novel High-throughput Acceleration Engine for Read Alignment}.
\newblock In {\em FCCM}, 2015.

\bibitem{rucci2018swifold}
Enzo Rucci, Carlos Garcia, Guillermo Botella, Armando De~Giusti, Marcelo
  Naiouf, and Manuel Prieto-Matias.
\newblock {SWIFOLD: Smith-Waterman Implementation on FPGA with OpenCL for Long
  DNA Sequences}.
\newblock {\em BMC Systems Biology}, 2018.

\bibitem{haghi2021fpga}
Abbas Haghi, Santiago Marco-Sola, Lluc Alvarez, Dionysios Diamantopoulos,
  Christoph Hagleitner, and Miquel Moreto.
\newblock {An FPGA Accelerator of the Wavefront Algorithm for Genomics Pairwise
  Alignment}.
\newblock In {\em FPL}, 2021.

\bibitem{li2021pipebsw}
Luyi Li, Jun Lin, and Zhongfeng Wang.
\newblock {PipeBSW: A Two-Stage Pipeline Structure for Banded Smith-Waterman
  Algorithm on FPGA}.
\newblock In {\em ISVLSI}, 2021.

\bibitem{ham2020genesis}
{Ham, Tae Jun and Bruns-Smith, David and Sweeney, Brendan and Lee, Yejin and
  Seo, Seong Hoon and Song, U Gyeong and Oh, Young H and Asanovic, Krste and
  Lee, Jae W and Wills, Lisa Wu}.
\newblock {Genesis: A Hardware Acceleration Framework for Genomic Data
  Analysis}.
\newblock In {\em ISCA}, 2020.

\bibitem{ham2021accelerating}
Tae~Jun Ham, Yejin Lee, Seong~Hoon Seo, U~Gyeong Song, Jae~W Lee, et~al.
\newblock {Accelerating Genomic Data Analytics With Composable Hardware
  Acceleration Framework}.
\newblock {\em IEEE Micro}, 2021.

\bibitem{wu2019fpga}
Lisa Wu, David Bruns-Smith, Frank~A Nothaft, Qijing Huang, Sagar Karandikar,
  et~al.
\newblock {FPGA Accelerated Indel Realignment in the Cloud}.
\newblock In {\em HPCA}, 2019.

\bibitem{singh2021fpga}
Gagandeep Singh, Mohammed Alser, Damla~Senol Cali, Dionysios Diamantopoulos,
  Juan G{\'o}mez-Luna, et~al.
\newblock {FPGA-based Near-Memory Acceleration of Modern Data-Intensive
  Applications}.
\newblock {\em IEEE Micro}, 2021.

\bibitem{kim20111}
Jung-Sik Kim, Chi~Sung Oh, Hocheol Lee, Donghyuk Lee, Hyong-Ryol Hwang, et~al.
\newblock {A 1.2 v 12.8 gb/s 2gb Mobile Wide-i/o Dram with 4$\times$ 128 i/os
  Using TSV-based Stacking}.
\newblock In {\em ISSCC}, 2011.

\bibitem{bingol2021gatekeeper}
Z{\"u}lal Bing{\"o}l, Mohammed Alser, Onur Mutlu, Ozcan Ozturk, and Can Alkan.
\newblock Gatekeeper-gpu: Fast and accurate pre-alignment filtering in short
  read mapping.
\newblock {\em IPDPSW}, 2021.

\bibitem{hameed2021alpha}
Fazal Hameed, Asif~Ali Khan, and Jeronimo Castrillon.
\newblock {ALPHA: A Novel Algorithm-Hardware Co-design for Accelerating DNA
  Seed Location Filtering}.
\newblock {\em ITETC}, 2021.

\bibitem{guo2019hardware}
Licheng Guo, Jason Lau, Zhenyuan Ruan, Peng Wei, and Jason Cong.
\newblock {Hardware Acceleration of Long Read Pairwise Overlapping in Genome
  Sequencing: A Race between FPGA and GPU}.
\newblock In {\em FCCM}, 2019.

\bibitem{uk10k2015uk10k}
UK10K consortium et~al.
\newblock {The UK10K Project Identifies Rare Variants in Health and Disease}.
\newblock {\em Nature}, 2015.

\bibitem{bhoyar2021high}
Rahul~C Bhoyar, Abhinav Jain, Paras Sehgal, Mohit~Kumar Divakar, Disha Sharma,
  et~al.
\newblock {High Throughput Detection and Genetic Epidemiology of SARS-CoV-2
  Using COVIDSeq Next-generation Sequencing}.
\newblock {\em PloS One}, 2021.

\bibitem{kim2016ramulator}
Yoongu Kim, Weikun Yang, and Onur Mutlu.
\newblock {Ramulator: A Fast and Extensible DRAM Simulator}.
\newblock {\em CAL}, 2015.

\bibitem{tavakkol2018mqsim}
Arash Tavakkol, Juan G{\'o}mez-Luna, Mohammad Sadrosadati, Saugata Ghose, and
  Onur Mutlu.
\newblock {MQSim: A Framework for Enabling Realistic Studies of Modern
  Multi-queue SSD Devices}.
\newblock In {\em FAST}, 2018.

\bibitem{syed2009next}
Fraz Syed, Haiying Grunenwald, and Nicholas Caruccio.
\newblock {Next-generation Sequencing Library Preparation: Simultaneous
  Fragmentation and Tagging Using in Vitro Transposition}.
\newblock {\em Nature Methods}, 2009.

\bibitem{shendure2008next}
Jay Shendure and Hanlee Ji.
\newblock {Next-generation DNA Sequencing}.
\newblock {\em Nature Biotechnology}, 2008.

\bibitem{gamaarachchi2022fast}
Hasindu Gamaarachchi, Hiruna Samarakoon, Sasha~P Jenner, James~M Ferguson,
  Timothy~G Amos, et~al.
\newblock {Fast Nanopore Sequencing Data Analysis with SLOW5}.
\newblock {\em Nature Biotechnology}, 2022.

\bibitem{minion21}
Oxford~Nanopore Technologies.
\newblock {MinION Mk1B IT Requirements}.
\newblock
  \url{https://community.nanoporetech.com/requirements_documents/minion-it-reqs.pdf},
  2021.

\bibitem{leinonen2010sequence}
Rasko Leinonen, Hideaki Sugawara, Martin Shumway, and International Nucleotide
  Sequence~Database Collaboration.
\newblock {The Sequence Read Archive}.
\newblock {\em Nucleic Acids Research}, 2010.

\bibitem{kostlbacher2021pangenomics}
Stephan K{\"o}stlbacher, Astrid Collingro, Tamara Halter, Frederik Schulz,
  Sean~P Jungbluth, and Matthias Horn.
\newblock {Pangenomics Reveals Alternative Environmental Lifestyles among
  Chlamydiae}.
\newblock {\em Nature Communications}, 2021.

\bibitem{van2020emergence}
Lucy van Dorp, Mislav Acman, Damien Richard, Liam~P Shaw, Charlotte~E Ford,
  et~al.
\newblock {Emergence of Genomic Diversity and Recurrent Mutations in
  SARS-CoV-2}.
\newblock {\em Infection, Genetics and Evolution}, 2020.

\bibitem{lapierre2020metalign}
Nathan LaPierre, Mohammed Alser, Eleazar Eskin, David Koslicki, and Serghei
  Mangul.
\newblock {Metalign: Efficient Alignment-based Metagenomic Profiling Via
  Containment Min Hash}.
\newblock {\em Genome Biology}, 2020.

\bibitem{meyer2021critical}
F~Meyer, A~Fritz, Z-L Deng, D~Koslicki, A~Gurevich, et~al.
\newblock {Critical Assessment of Metagenome Interpretation-the second round of
  challenges}.
\newblock {\em bioRxiv}, 2021.

\bibitem{khayat2021hidden}
Michael~M Khayat, Sayed Mohammad~Ebrahim Sahraeian, Samantha Zarate, Andrew
  Carroll, Huixiao Hong, et~al.
\newblock {Hidden Biases in Germline Structural Variant Detection}.
\newblock {\em Genome Biology}, 2021.

\bibitem{luo2014balsa}
Ruibang Luo, Yiu-Lun Wong, Wai-Chun Law, Lap-Kei Lee, Jeanno Cheung, et~al.
\newblock {BALSA: Integrated Secondary Analysis for Whole-genome and
  Whole-exome Sequencing, Accelerated by GPU}.
\newblock {\em PeerJ}, 2014.

\bibitem{xin2016optimal}
Hongyi Xin, Sunny Nahar, Richard Zhu, John Emmons, Gennady Pekhimenko, et~al.
\newblock {Optimal Seed Solver: Optimizing Seed Selection in Read Mapping}.
\newblock {\em Bioinformatics}, 2016.

\bibitem{altschul1990basic}
Stephen~F Altschul, Warren Gish, Webb Miller, Eugene~W Myers, and David~J
  Lipman.
\newblock {Basic Local Alignment Search Tool}.
\newblock {\em Journal of Molecular Biology}, 1990.

\bibitem{wood2019improved}
Derrick~E Wood, Jennifer Lu, and Ben Langmead.
\newblock {Improved Metagenomic Analysis with Kraken 2}.
\newblock {\em Genome Biology}, 2019.

\bibitem{schleimer2003winnowing}
Saul Schleimer, Daniel~S Wilkerson, and Alex Aiken.
\newblock {Winnowing: Local Algorithms for Document Fingerprinting}.
\newblock In {\em ACM SIGMOD}, 2003.

\bibitem{roberts2004reducing}
Michael Roberts, Wayne Hayes, Brian~R Hunt, Stephen~M Mount, and James~A Yorke.
\newblock {Reducing Storage Requirements for Biological Sequence Comparison}.
\newblock {\em Bioinformatics}, 2004.

\bibitem{marccais2017improving}
Guillaume Mar{\c{c}}ais, David Pellow, Daniel Bork, Yaron Orenstein, Ron
  Shamir, and Carl Kingsford.
\newblock {Improving the Performance of Minimizers and Winnowing Schemes}.
\newblock {\em Bioinformatics}, 2017.

\bibitem{li2016minimap}
Heng Li.
\newblock {Minimap and Miniasm: Fast Mapping and De Novo Assembly for Noisy
  Long Sequences}.
\newblock {\em Bioinformatics}, 2016.

\bibitem{samsung860pro}
Samsung.
\newblock {Samsung SSD 860 PRO}.
\newblock
  \url{https://www.samsung.com/semiconductor/minisite/ssd/product/consumer/860pro/},
  2018.

\bibitem{park-nvmsa-2018}
Jisung Park, Myungsuk Kim, Sungjin Lee, and Jihong Kim.
\newblock {Improving {I/O} Performance of Large-page Flash Storage Systems
  Using Subpage-parallel Reads}.
\newblock In {\em NVMSA}, 2018.

\bibitem{kim-dac-2017}
Myungsuk Kim, Jaehoon Lee, Sungjin Lee, Jisung Park, Youngsun Song, and Jihong
  Kim.
\newblock {Improving Performance and Lifetime of Large-page {NAND} Storages
  Using Erase-free Subpage Programming}.
\newblock In {\em DAC}, 2017.

\bibitem{inteldcs4500}
Intel.
\newblock {Intel SSD DC S4500 Series}.
\newblock
  \url{https://ark.intel.com/content/www/us/en/ark/products/120521/intel-ssd-dc-s4500-series-480gb-2-5in-sata-6gbs-3d1-tlc.html},
  2017.

\bibitem{samsungPM1735}
Samsung.
\newblock {Samsung SSD PM1735}.
\newblock
  \url{https://www.samsung.com/semiconductor/ssd/enterprise-ssd/MZPLJ3T2HBJR-00007/},
  2020.

\bibitem{anandcontroller}
AnandTech.
\newblock {New Enterprise SSD Controllers}.
\newblock
  \url{https://www.anandtech.com/show/16275/new-enterprise-ssd-controllers-from-silicon-motion-phison-fadu}.

\bibitem{tavakkol2018flin}
Arash Tavakkol, Mohammad Sadrosadati, Saugata Ghose, Jeremie Kim, Yixin Luo,
  et~al.
\newblock {FLIN: Enabling Fairness and Enhancing Performance in Modern NVMe
  Solid State Drives}.
\newblock In {\em ISCA}, 2018.

\bibitem{kim2020evanesco}
Myungsuk Kim, Jisung Park, Genhee Cho, Yoona Kim, Lois Orosa, et~al.
\newblock {Evanesco: Architectural Support for Efficient Data Sanitization in
  Modern Flash-Based Storage Systems}.
\newblock In {\em ASPLOS}, 2020.

\bibitem{cai2017error}
Yu~Cai, Saugata Ghose, Erich~F Haratsch, Yixin Luo, and Onur Mutlu.
\newblock {Error Characterization, Mitigation, and Recovery in
  Flash-Memory-Based Solid-state Drives}.
\newblock In {\em Proc. IEEE}, 2017.

\bibitem{park-dac-2019}
Jisung Park, Youngdon Jung, Jonghoon Won, Minji Kang, Sungjin Lee, and Jihong
  Kim.
\newblock {RansomeBlocker: a Low-Overhead Ransomware-Proof SSD}.
\newblock In {\em DAC}, 2019.

\bibitem{park-dac-2016}
Jisung Park, Jaeyong Jeong, Sungjin Lee, Youngsun Song, and Jihong Kim.
\newblock {Improving Performance and Lifetime of {NAND} Storage Systems Using
  Relaxed Program Sequence}.
\newblock In {\em DAC}, 2016.

\bibitem{chang2007efficient}
Li-Pin Chang.
\newblock {On Efficient Wear Leveling for Large-scale Flash-memory Storage
  Systems}.
\newblock In {\em ACM SAC}, 2007.

\bibitem{SATA}
{Serial ATA International Organization}.
\newblock {SATA revision 3.0 specifications}.
\newblock \url{https://www.sata-io.org}.

\bibitem{samsung980pro}
Samsung.
\newblock {Samsung SSD 980 PRO}.
\newblock
  \url{https://www.samsung.com/semiconductor/minisite/ssd/product/consumer/980pro/},
  2020.

\bibitem{PCIE}
PCI-SIG.
\newblock {PCI Express Base Specification Revision 3.0}.
\newblock \url{https://pcisig.com/specifications}.

\bibitem{PCIE4}
PCI-SIG.
\newblock {PCI Express Base Specification Revision 4.0, Version 1.0}.
\newblock \url{https://pcisig.com/specifications}.

\bibitem{amdepyc}
AMD.
\newblock {AMD$^\text{\textregistered}$ EPYC$^\text{\textregistered}$ 7742
  CPU}.
\newblock \url{https://www.amd.com/en/products/cpu/amd-epyc-7742}.

\bibitem{micros9300pro}
Micron.
\newblock {Micron 9300 SSD}.
\newblock \url{https://www.micron.com/products/ssd/product-lines/9300}, 2019.

\bibitem{wdblue}
Western Digital.
\newblock {WD Blue SATA Internal SSD Hard Drive}.
\newblock
  \url{https://www.westerndigital.com/en-ca/products/internal-drives/wd-blue-sata-2-5-ssd#WDS400T2B0A}.

\bibitem{laguna2020seed}
Ann~Franchesca Laguna, Hasindu Gamaarachchi, Xunzhao Yin, Michael Niemier, Sri
  Parameswaran, and X~Sharon Hu.
\newblock {Seed-and-vote Based In-memory Accelerator for DNA Read Mapping}.
\newblock In {\em ICCAD}, 2020.

\bibitem{kaplan2018rassa}
Roman Kaplan, Leonid Yavits, and Ran Ginosar.
\newblock {RASSA: Resistive Prealignment Accelerator for Approximate DNA Long
  Read Mapping}.
\newblock {\em IEEE Micro}, 2018.

\bibitem{koo2017summarizer}
Gunjae Koo, Kiran~Kumar Matam, I~Te, HV~Krishna~Giri Narra, Jing Li, et~al.
\newblock {Summarizer: Trading Communication with Computing Near Storage}.
\newblock In {\em MICRO}, 2017.

\bibitem{mailthody2019deepstore}
Vikram~Sharma Mailthody, Zaid Qureshi, Weixin Liang, Ziyan Feng, Simon~Garcia
  De~Gonzalo, et~al.
\newblock {Deepstore: In-storage Acceleration for Intelligent Queries}.
\newblock In {\em MICRO}, 2019.

\bibitem{schneider2017evaluation}
Valerie~A Schneider, Tina Graves-Lindsay, Kerstin Howe, Nathan Bouk, Hsiu-Chuan
  Chen, et~al.
\newblock {Evaluation of GRCh38 and De Novo Haploid Genome Assemblies
  Demonstrates the Enduring Quality of the Reference Assembly}.
\newblock {\em Genome Research}, 2017.

\bibitem{dang2015secure}
Quynh Dang.
\newblock {Secure Hash Standard}.
\newblock \url{https://doi.org/10.6028/NIST.FIPS.180-4}, 2015.

\bibitem{rivest1992rfc1321}
Ronald Rivest.
\newblock {RFC1321: The MD5 Message-digest Algorithm}.
\newblock \url{https://datatracker.ietf.org/doc/rfc1321/}, 1992.

\bibitem{arxivGS}
Nika Mansouri~Ghiasi, Jisung Park, Harun Mustafa, Jeremie Kim, Ataberk Olgun,
  et~al.
\newblock {GenStore: A High-Performance and Energy-Efficient In-Storage
  Computing System for Genome Sequence Analysis}.
\newblock In {\em arXiv}, 2022.

\bibitem{sims2014sequencing}
David Sims, Ian Sudbery, Nicholas~E Ilott, Andreas Heger, and Chris~P Ponting.
\newblock {Sequencing Depth and Coverage: Key Considerations in Genomic
  Analyses}.
\newblock {\em Nature Reviews Genetics}, 2014.

\bibitem{illumina}
Illumina.
\newblock {NovaSeq 6000 System Specifications}.
\newblock
  \url{https://emea.illumina.com/systems/sequencing-platforms/novaseq/specifications.html},
  2020.

\bibitem{lenovot740p}
Lenovo.
\newblock {ThinkPad T470p}.
\newblock
  \url{https://www.lenovo.com/ch/en/laptops/thinkpad/t-series/ThinkPad-T470p/p/22TP2TT470P},
  2016.

\bibitem{cheong-isscc-2018}
Wooseong Cheong, Chanho Yoon, Seonghoon Woo, Kyuwook Han, Daehyun Kim, et~al.
\newblock {A Flash Memory Controller for 15\textmu{}s Ultra-Low-Latency SSD
  Using High-Speed 3D NAND Flash with 3\textmu{}s Read Time}.
\newblock In {\em ISSCC}, 2018.

\bibitem{park-asplos-2021}
Jisung Park, Myungsuk Kim, Myoungjun Chun, Lois Orosa, Jihong Kim, and Onur
  Mutlu.
\newblock {Reducing Solid-State Drive Read Latency by Optimizing Read-Retry}.
\newblock In {\em ASPLOS}, 2021.

\bibitem{breitwieser2019human}
Florian~P Breitwieser, Mihaela Pertea, Aleksey~V Zimin, and Steven~L Salzberg.
\newblock {Human Contamination in Bacterial Genomes has Created Thousands of
  Spurious Proteins}.
\newblock {\em Genome Research}, 2019.

\bibitem{human2012structure}
Human Microbiome~Project Consortium et~al.
\newblock {Structure, Function and Diversity of the Healthy Human Microbiome}.
\newblock {\em Nature}, 2012.

\bibitem{danko2021global}
David Danko, Daniela Bezdan, Evan~E Afshin, Sofia Ahsanuddin, Chandrima
  Bhattacharya, et~al.
\newblock {A Global Metagenomic Map of Urban Microbiomes and Antimicrobial
  Resistance}.
\newblock {\em Cell}, 2021.

\bibitem{knight2018best}
Rob Knight, Alison Vrbanac, Bryn~C Taylor, Alexander Aksenov, Chris Callewaert,
  et~al.
\newblock {Best Practices for Analysing Microbiomes}.
\newblock {\em Nature Reviews Microbiology}, 2018.

\bibitem{sayers2021database}
Eric~W Sayers, Jeffrey Beck, Evan~E Bolton, Devon Bourexis, James~R Brister,
  et~al.
\newblock {Database Resources of the National Center for Biotechnology
  Information}.
\newblock {\em Nucleic Acids Research}, 2021.

\bibitem{zook2016extensive}
Justin~M Zook, David Catoe, Jennifer McDaniel, Lindsay Vang, Noah Spies, et~al.
\newblock {Extensive Sequencing of Seven Human Genomes to Characterize
  Benchmark Reference Materials}.
\newblock {\em Scientific data}, 2016.

\bibitem{clark2016genbank}
Karen Clark, Ilene Karsch-Mizrachi, David~J Lipman, James Ostell, and Eric~W
  Sayers.
\newblock {GenBank}.
\newblock {\em Nucleic Acids Research}, 2016.

\bibitem{wu2020new}
Fan Wu, Su~Zhao, Bin Yu, Yan-Mei Chen, Wen Wang, et~al.
\newblock {A New Coronavirus Associated with Human Respiratory Disease in
  China}.
\newblock {\em Nature}, 2020.

\bibitem{sichtig2019fda}
Heike Sichtig, Timothy Minogue, Yi~Yan, Christopher Stefan, Adrienne Hall,
  et~al.
\newblock {FDA-ARGOS is A Database with Public Quality-controlled Reference
  Genomes for Diagnostic Use and Regulatory Science}.
\newblock {\em Nature Communications}, 2019.

\bibitem{engel2014reference}
Stacia~R Engel, Fred~S Dietrich, Dianna~G Fisk, Gail Binkley, Rama
  Balakrishnan, et~al.
\newblock {The Reference Genome Sequence of Saccharomyces Cerevisiae: Then and
  Now}.
\newblock {\em G3: Genes, Genomes, Genetics}, 2014.

\bibitem{berardini2015arabidopsis}
Tanya~Z Berardini, Leonore Reiser, Donghui Li, Yarik Mezheritsky, Robert
  Muller, et~al.
\newblock {The Arabidopsis Information Resource: Making and Mining the “Gold
  Standard” Annotated Reference Plant Genome}.
\newblock {\em Genesis}, 2015.

\bibitem{larkin2020}
Aoife Larkin, Steven~J Marygold, Giulia Antonazzo, Helen Attrill, Gilberto dos
  Santos, et~al.
\newblock {FlyBase: Updates to the Drosophila Melanogaster Knowledge Base}.
\newblock {\em Nucleic Acids Research}, 2020.

\bibitem{church2009lineage}
Deanna~M Church, Leo Goodstadt, LaDeana~W Hillier, Michael~C Zody, Steve
  Goldstein, et~al.
\newblock {Lineage-specific Biology Revealed by a Finished Genome Assembly of
  the Mouse}.
\newblock {\em PLoS Biology}, 2009.

\bibitem{wang2007integer}
Thomas Wang.
\newblock {Integer Hash Function}.
\newblock
  \url{http://web.archive.org/web/20071223173210/http://www.concentric.net/~Ttwang/tech/inthash.htm},
  2007.

\bibitem{dobin2012}
Alexander Dobin, Carrie~A. Davis, Felix Schlesinger, Jorg Drenkow, Chris
  Zaleski, et~al.
\newblock {{STAR: Ultrafast Universal RNA-seq Aligner}}.
\newblock {\em Bioinformatics}, 2012.

\bibitem{cai-insidessd-2018}
Yu~Cai, Saugata Ghose, Erich~F. Haratsch, Yixin Luo, and Onur Mutlu.
\newblock {Reliability Issues in Flash-memory-based Solid-state Drives:
  Experimental Analysis, Mitigation, Recovery}.
\newblock In {\em Inside Solid State Drives}. Springer, 2018.

\bibitem{cai-hpca-2017}
Yu~Cai, Saugata Ghose, Yixin Luo, Ken Mai, Onur Mutlu, and Erich~F. Haratsch.
\newblock {Vulnerabilities in MLC NAND Flash Memory Programming: Experimental
  Analysis, Exploits, and Mitigation Techniques}.
\newblock In {\em HPCA}, 2017.

\bibitem{luo2018improving}
Yixin Luo, Saugata Ghose, Yu~Cai, Erich~F Haratsch, and Onur Mutlu.
\newblock {Improving 3D NAND Flash Memory Lifetime by Tolerating Early
  Retention Loss and Process Variation}.
\newblock {\em ACM POMACS}, 2018.

\bibitem{luo-hpca-2018}
Yixin Luo, Saugata Ghose, Yu~Cai, Erich~F. Haratsch, and Onur Mutlu.
\newblock {HeatWatch: Improving 3D NAND Flash Memory Device Reliability by
  Exploiting Self-recovery and Temperature Awareness}.
\newblock In {\em HPCA}, 2018.

\bibitem{cai2015read}
Yu~Cai, Yixin Luo, Saugata Ghose, and Onur Mutlu.
\newblock {Read Disturb Errors in MLC NAND Flash Memory: Characterization,
  Mitigation, and Recovery}.
\newblock In {\em IEEE/IFIP DSN}, 2015.

\bibitem{cai2013error}
Yu~Cai, Gulay Yalcin, Onur Mutlu, Erich~F Haratsch, Adrian Crista, et~al.
\newblock {Error Analysis and Management for MLC NAND Flash Memory.}
\newblock {\em Intel Technology}, 2013.

\bibitem{cai2012flash}
Yu~Cai, Gulay Yalcin, Onur Mutlu, Erich~F Haratsch, Adrian Cristal, et~al.
\newblock {Flash Correct-and-refresh: Retention-aware Error Management for
  Increased Flash Memory lifetime}.
\newblock In {\em ICCD}, 2012.

\bibitem{ha2015integrated}
Keonsoo Ha, Jaeyong Jeong, and Jihong Kim.
\newblock {An Integrated Approach for Managing Read Disturbs in High-density
  NAND Flash Memory}.
\newblock {\em IEEE TCAD}, 2015.

\bibitem{luo2015warm}
Yixin Luo, Yu~Cai, Saugata Ghose, Jongmoo Choi, and Onur Mutlu.
\newblock {WARM: Improving NAND Flash Memory Lifetime with Write-Hotness Aware
  Retention Management}.
\newblock In {\em MSST}, 2015.

\bibitem{micron3dnandflyer}
Micron.
\newblock {Product Flyer: Micron 3D NAND Flash Memory}.
\newblock
  \url{https://www.micron.com/-/media/client/global/documents/products/product-flyer/3d_nand_flyer.pdf?la=en},
  2016.

\bibitem{synopsysdc}
{Synopsys, Inc.}
\newblock {Design Compiler}.
\newblock
  \url{https://www.synopsys.com/implementation-and-signoff/rtl-synthesis-test/design-compiler-graphical.html}.

\bibitem{ddr4sheet}
Micron~Technology Inc.
\newblock {4Gb: x4, x8, x16 DDR4 SDRAM Data Sheet}, 2016.

\bibitem{ghose2019demystifying}
Saugata Ghose, Tianshi Li, Nastaran Hajinazar, Damla~Senol Cali, and Onur
  Mutlu.
\newblock {Demystifying Complex Workload-DRAM Interactions: An Experimental
  Study}.
\newblock {\em ACM POMACS}, 2019.

\bibitem{ghose2018your}
Saugata Ghose, Abdullah~Giray Yaglik{\c{c}}i, Raghav Gupta, Donghyuk Lee, Kais
  Kudrolli, et~al.
\newblock {What Your DRAM Power Models Are Not Telling You: Lessons from a
  Detailed Experimental Study}.
\newblock {\em POMACS}, 2018.

\bibitem{ramulatorsource}
{Ramulator Source Code}.
\newblock \url{https://github.com/CMU-SAFARI/ramulator}.

\bibitem{microprof}
{Advanced Micro Devices}.
\newblock {AMD{\textregistered} $\mu$Prof}.
\newblock \url{https://developer.amd.com/amd-uprof/}, 2021.

\bibitem{holtgrewe2010mason}
M.~Holtgrewe.
\newblock {Mason - A Read Simulator for Second Generation Sequencing Data}.
\newblock {\em Technical Report FU Berlin}, 2010.

\bibitem{wikichipcascade}
WikiChip.
\newblock {Cascade Lake SP - Intel}.
\newblock \url{https://en.wikichip.org/wiki/intel/cores/cascade\_lake\_sp}.

\bibitem{stillmaker2017Scaling}
Aaron Stillmaker and Bevan Baas.
\newblock {Scaling Equations for the Accurate Prediction of {{CMOS}} Device
  Performance from 180 Nm to 7 Nm}.
\newblock {\em Integration}, 2017.

\bibitem{cortexr4}
ARM Holdings.
\newblock {Cortex-R4}.
\newblock
  \url{https://developer.arm.com/ip-products/processors/cortex-r/cortex-r4},
  2011.

\bibitem{liu2009cudasw++}
Yongchao Liu, Douglas~L Maskell, and Bertil Schmidt.
\newblock {CUDASW++: Optimizing Smith-Waterman Sequence Database Searches for
  CUDA-enabled Graphics Processing Units}.
\newblock {\em BMC Research Notes}, 2009.

\bibitem{liu2010cudasw++}
Yongchao Liu, Bertil Schmidt, and Douglas~L Maskell.
\newblock {CUDASW++ 2.0: Enhanced Smith-Waterman Protein Database Search on
  CUDA-enabled GPUs Based on SIMT and Virtualized SIMD Abstractions}.
\newblock {\em BMC Research Notes}, 2010.

\bibitem{pei2019registor}
Shuyi Pei, Jing Yang, and Qing Yang.
\newblock {REGISTOR: A Platform for Unstructured Data Processing inside SSD
  Storage}.
\newblock {\em ACM TOS}, 2019.

\bibitem{jun2018grafboost}
Sang-Woo Jun, Andy Wright, Sizhuo Zhang, Shuotao Xu, et~al.
\newblock {GraFBoost: Using Accelerated Flash Storage for External Graph
  Analytics}.
\newblock In {\em ISCA}, 2018.

\bibitem{do2013query}
Jaeyoung Do, Yang-Suk Kee, Jignesh~M Patel, Chanik Park, Kwanghyun Park, and
  David~J DeWitt.
\newblock {Query Processing on Smart SSDs: Opportunities and Challenges}.
\newblock In {\em ACM SIGMOD}, 2013.

\bibitem{seshadri2014willow}
Sudharsan Seshadri, Mark Gahagan, Sundaram Bhaskaran, Trevor Bunker, Arup De,
  et~al.
\newblock {Willow: A User-Programmable SSD}.
\newblock In {\em USENIX OSDI}, 2014.

\bibitem{kim2016storage}
Sungchan Kim, Hyunok Oh, Chanik Park, Sangyeun Cho, Sang-Won Lee, and Bongki
  Moon.
\newblock {In-storage Processing of Database Scans and Joins}.
\newblock {\em Information Sciences}, 2016.

\bibitem{riedel2001active}
Erik Riedel, Christos Faloutsos, Garth~A Gibson, and David Nagle.
\newblock {Active Disks for Large-Scale Data Processing}.
\newblock {\em Computer}, 2001.

\bibitem{riedel1998active}
Erik Riedel, Garth Gibson, and Christos Faloutsos.
\newblock {Active Storage for Large-Scale Data Mining and Multimedia
  Applications}.
\newblock {\em VLDB}, 1998.

\bibitem{gu2016biscuit}
Boncheol Gu, Andre~S Yoon, Duck-Ho Bae, Insoon Jo, Jinyoung Lee, et~al.
\newblock {Biscuit: A Framework for Near-data Processing of Big Data
  Workloads}.
\newblock {\em ISCA}, 2016.

\bibitem{kang2013enabling}
Yangwook Kang, Yang-suk Kee, Ethan~L Miller, and Chanik Park.
\newblock {Enabling Cost-effective Data Processing with Smart SSD}.
\newblock In {\em MSST}, 2013.

\bibitem{wang2019project}
Xiaohao Wang, Yifan Yuan, You Zhou, Chance~C Coats, and Jian Huang.
\newblock {Project Almanac: A Time-traveling Solid-state Drive}.
\newblock In {\em EuroSys}, 2019.

\bibitem{acharya1998active}
Anurag Acharya, Mustafa Uysal, and Joel Saltz.
\newblock {Active Disks: Programming Model, Algorithms and Evaluation}.
\newblock {\em ASPLOS}, 1998.

\bibitem{keeton1998case}
Kimberly Keeton, David~A Patterson, and Joseph~M Hellerstein.
\newblock {A Case for Intelligent Disks (IDISKs)}.
\newblock {\em {SIGMOD Record}}, 1998.

\bibitem{jun2015bluedbm}
Sang-Woo Jun, Ming Liu, Sungjin Lee, Jamey Hicks, John Ankcorn, et~al.
\newblock {Bluedbm: An Appliance for Big Data Analytics}.
\newblock In {\em ISCA}, 2015.

\bibitem{jun2016bluedbm}
Sang-Woo Jun, Ming Liu, Sungjin Lee, Jamey Hicks, John Ankcorn, et~al.
\newblock {Bluedbm: Distributed Flash Storage for Big Data Analytics}.
\newblock {\em ACM TOCS}, 2016.

\bibitem{torabzadehkashi2019catalina}
Mahdi Torabzadehkashi, Siavash Rezaei, Ali Heydarigorji, Hosein Bobarshad,
  Vladimir Alves, and Nader Bagherzadeh.
\newblock {Catalina: In-storage Processing Acceleration for Scalable Big Data
  Analytics}.
\newblock In {\em Euromicro PDP}, 2019.

\bibitem{lee2020smartssd}
Joo~Hwan Lee, Hui Zhang, Veronica Lagrange, Praveen Krishnamoorthy, Xiaodong
  Zhao, and Yang~Seok Ki.
\newblock {SmartSSD: FPGA Accelerated Near-Storage Data Analytics on SSD}.
\newblock {\em IEEE Computer Architecture Letters}, 2020.

\bibitem{ajdari2019cidr}
Mohammadamin Ajdari, Pyeongsu Park, Joonsung Kim, Dongup Kwon, and Jangwoo Kim.
\newblock {CIDR: A Cost-effective In-line Data Reduction System for
  Terabit-per-second Scale SSD Arrays}.
\newblock In {\em HPCA}, 2019.

\bibitem{cho2013xsd}
Benjamin~Y Cho, Won~Seob Jeong, Doohwan Oh, and Won~Woo Ro.
\newblock {Xsd: Accelerating Mapreduce by Harnessing the GPU inside an SSD}.
\newblock In {\em Near-Data Processing}, 2013.

\bibitem{jeong2019react}
Won~Seob Jeong, Changmin Lee, Keunsoo Kim, Myung~Kuk Yoon, Won Jeon, et~al.
\newblock {REACT: Scalable and High-performance Regular Expression Pattern
  Matching Accelerator for In-storage Processing}.
\newblock {\em IEEE TPDS}, 2019.

\bibitem{jun2016storage}
Sang-Woo Jun, Huy~T Nguyen, Vijay Gadepally, et~al.
\newblock {In-storage Embedded Accelerator for Sparse Pattern Processing}.
\newblock In {\em HPEC}, 2016.

\bibitem{morrison2020nanopore}
Gretchen~A Morrison, Jianmin Fu, Grace~C Lee, Nathan~P Wiederhold, Connie~F
  Ca{\~n}ete-Gibas, et~al.
\newblock {Nanopore Sequencing of the Fungal Intergenic Spacer Sequence as a
  Potential Rapid Diagnostic Assay}.
\newblock {\em Journal of Clinical Microbiology}, 2020.

\bibitem{nanopore2020}
Oxford~Nanopore Technologies.
\newblock {R10.3: the Newest Nanopore for High Accuracy Nanopore Sequencing –
  Now Available in Store}.
\newblock
  \url{https://nanoporetech.com/about-us/news/r103-newest-nanopore-high-accuracy-nanopore-sequencing-now-available-store/},
  2020.

\bibitem{quail2008large}
Michael~A Quail, Iwanka Kozarewa, Frances Smith, Aylwyn Scally, Philip~J
  Stephens, et~al.
\newblock {A Large Genome Center's Improvements to the Illumina Sequencing
  System}.
\newblock {\em Nature Methods}, 2008.

\bibitem{pacbio2021}
PacBio Sequencing.
\newblock {Pacific Biosciences Closes Acquisition of Omniome and Establishes
  San Diego Presence}.
\newblock
  \url{https://www.pacb.com/press_releases/pacific-biosciences-closes-acquisition-of-omniome-and-establishes-san-diego-presence/},
  2021.

\bibitem{dunn2021}
Tim Dunn, Harisankar Sadasivan, Jack Wadden, Kush Goliya, Kuan-Yu Chen, et~al.
\newblock Squigglefilter: An accelerator for portable virus detection.
\newblock {\em MICRO}, 2021.

\bibitem{loka2019reliable}
Tobias~P Loka, Simon~H Tausch, and Bernhard~Y Renard.
\newblock {Reliable Variant Calling during Runtime of Illumina Sequencing}.
\newblock {\em Scientific Reports}, 2019.

\bibitem{zhang2021real}
Haowen Zhang, Haoran Li, Chirag Jain, Haoyu Cheng, Kin~Fai Au, et~al.
\newblock {Real-time Mapping of Nanopore Raw Signals}.
\newblock {\em Bioinformatics}, 2021.

\bibitem{kovaka2020targeted}
Sam Kovaka, Yunfan Fan, Bohan Ni, Winston Timp, and Michael~C Schatz.
\newblock {Targeted Nanopore Sequencing by Real-time Mapping of Raw Electrical
  Signal with UNCALLED}.
\newblock {\em Nature Biotechnology}, 2020.

\bibitem{afshinnekoo2015geospatial}
Ebrahim Afshinnekoo, Cem Meydan, Shanin Chowdhury, Dyala Jaroudi, Collin Boyer,
  et~al.
\newblock {Geospatial Resolution of Human and Bacterial Diversity with
  City-scale Metagenomics}.
\newblock {\em Cell Systems}, 2015.

\bibitem{hsu2016urban}
Tiffany Hsu, Regina Joice, Jose Vallarino, Galeb Abu-Ali, Erica~M Hartmann,
  et~al.
\newblock {Urban Transit System Microbial Communities Differ by Surface Type
  and Interaction with Humans and the Environment}.
\newblock {\em Msystems}, 2016.

\bibitem{sherman2019assembly}
Rachel~M Sherman, Juliet Forman, Valentin Antonescu, Daniela Puiu, Michelle
  Daya, et~al.
\newblock {Assembly of A Pan-genome from Deep Sequencing of 910 Humans of
  African Descent}.
\newblock {\em Nature Genetics}, 2019.

\bibitem{li2021building}
Qiuhui Li, Shilin Tian, Bin Yan, Chi~Man Liu, Tak-Wah Lam, et~al.
\newblock {Building a Chinese Pan-genome of 486 Individuals}.
\newblock {\em Communications Biology}, 2021.

\bibitem{miga2021need}
Karen~H Miga and Ting Wang.
\newblock {The Need for A Human Pangenome Reference Sequence}.
\newblock {\em Annual Review of Genomics and Human Genetics}, 2021.

\bibitem{zhang2020comprehensive}
Haowen Zhang, Chirag Jain, and Srinivas Aluru.
\newblock {A Comprehensive Evaluation of Long Read Error Correction Methods}.
\newblock {\em BMC Genomics}, 2020.

\bibitem{miga2020telomere}
Karen~H Miga, Sergey Koren, Arang Rhie, Mitchell~R Vollger, Ariel Gershman,
  et~al.
\newblock {Telomere-to-telomere Assembly of A Complete Human X Chromosome}.
\newblock {\em Nature}, 2020.

\bibitem{logsdon2021structure}
Glennis~A Logsdon, Mitchell~R Vollger, PingHsun Hsieh, Yafei Mao, Mikhail~A
  Liskovykh, et~al.
\newblock {The Structure, Function and Evolution of A Complete Human Chromosome
  8}.
\newblock {\em Nature}, 2021.

\bibitem{peresini2021nanopore}
Peter Pere\u{s}\'{i}ni, Vladim\'{i}r Bo\u{z}a, Bro\u{n}a Brejov\'{a}, and
  Tom\'{a}\u{s} Vina\u{r}.
\newblock {Nanopore Base Calling on the Edge}.
\newblock {\em Bioinformatics}, 2021.

\bibitem{ahmed2021pan}
Omar Ahmed, Massimiliano Rossi, Sam Kovaka, Michael~C Schatz, Travis Gagie,
  et~al.
\newblock {Pan-genomic Matching Statistics for Targeted Nanopore Sequencing}.
\newblock {\em iScience}, 2021.

\bibitem{pomerantz2018real}
Aaron Pomerantz, Nicol{\'a}s Pe{\~n}afiel, Alejandro Arteaga, Lucas Bustamante,
  Frank Pichardo, et~al.
\newblock {Real-time DNA Barcoding in a Rainforest Using Nanopore Sequencing:
  Opportunities for Rapid Biodiversity Assessments and Local Capacity
  Building}.
\newblock {\em GigaScience}, 2018.

\bibitem{sunagawa2015structure}
Shinichi Sunagawa, Luis~Pedro Coelho, Samuel Chaffron, Jens~Roat Kultima,
  Karine Labadie, et~al.
\newblock {Structure and Function of the Global Ocean Microbiome}.
\newblock {\em Science}, 2015.

\bibitem{lax2014longitudinal}
Simon Lax, Daniel~P Smith, Jarrad Hampton-Marcell, Sarah~M Owens, Kim~M
  Handley, et~al.
\newblock {Longitudinal Analysis of Microbial Interaction between Humans and
  the Indoor Environment}.
\newblock {\em Science}, 2014.

\end{thebibliography}

\endgroup

\end{document}